\documentclass[1p]{elsarticle}
\biboptions{sort&compress}
\usepackage{amsmath,amssymb,amsfonts}
\usepackage[latin1]{inputenc}
\usepackage{mathrsfs}
\usepackage{times}
\usepackage{float}
\usepackage[vcentermath]{youngtab}
\usepackage[colorlinks,bookmarks=true,citecolor=blue,linkcolor=blue,urlcolor=blue, breaklinks=true]{hyperref}
\graphicspath{{./figures/}}

\newcommand{\bolS}{\boldsymbol{S}}
\newcommand{\bolT}{\boldsymbol{T}}

\newcommand{\bolr}{\mathrm{\bf r}}

\newcommand{\be}{\mathrm{e}}
\newcommand{\VEV}[1]{\langle #1 \rangle}  

\begin{document}
\title{%
Phases of one-dimensional SU($N$) cold atomic Fermi gases \\--- from molecular Luttinger liquids to topological phases}
\author[UPS]{S.\ Capponi\corref{cor1}}
\ead{capponi@irsamc.ups-tlse.fr}

\author[Cergy]{P. Lecheminant}

\author[YITP]{K. Totsuka}

\cortext[cor1]{Corresponding author}
\address[UPS]{Laboratoire de Physique Th\'eorique, CNRS UMR 5152, 
Universit\'e Paul Sabatier, F-31062 Toulouse, France}
\address[Cergy]{Laboratoire de Physique Th\'eorique et
Mod\'elisation, CNRS UMR 8089,
Universit\'e de Cergy-Pontoise, Site de Saint-Martin,
F-95300 Cergy-Pontoise Cedex, France}
\address[YITP]{Yukawa Institute for Theoretical Physics, 
Kyoto University, Kitashirakawa Oiwake-Cho, Kyoto 606-8502, Japan}
\begin{abstract}
Alkaline-earth and ytterbium cold atomic gases make it possible to simulate SU($N$)-symmetric fermionic systems 
in a very controlled fashion. Such a
high symmetry is expected to give rise to a variety of novel
phenomena ranging from molecular Luttinger liquids to (symmetry-protected) topological phases.  
We review some of the phases that can be stabilized in a
one dimensional lattice. The physics of this multi-component Fermi gas
turns out to be much richer and more exotic than in the standard SU(2) case. 
For $N>2$, the phase diagram is quite rich already in the case of the single-band model,  
including a molecular Luttinger liquid (with dominant superfluid instability in the $N$-particle channel) 
for incommensurate fillings, as well as various Mott-insulating phases occurring at commensurate fillings.  
Particular attention will be paid to the cases with additional orbital degree of freedom 
(which is accessible experimentally either by taking into account two atomic states 
or by putting atoms in the $p$-band levels). 
We introduce two microscopic models which are relevant for these cases and discuss 
their symmetries and strong coupling limits.  
More intriguing phase diagrams are then presented including, for instance, symmetry protected topological 
phases characterized by non-trivial edge states.
\end{abstract}
\maketitle
\section{Introduction}
\label{sec:introduction}
Symmetry lies at the heart of physics, since it is a powerful concept
to classify conventional phases of matter (with broken symmetries) as
well as phase transitions within the Ginzburg-Landau paradigm \cite{Landau-Lifshitz-StatPhys}. 
It has also been often used to idealize the actual physical situation, or simply to make analytical progress 
feasible as in the large-$N$ expansion. 
In high-energy physics, continuous symmetries based on the SU($N$) unitary
group play fundamental roles in the standard model of
particle physics.  For instance, an approximate SU($N$) symmetry, 
where $N$ is the number of species of quarks, or
flavors, underlies the description of hadrons.  

In realistic condensed-matter experiments, such a high continuous symmetry is rarely realized 
[in stark contrast to SU(2) associated with, e.g., the spin-rotational symmetry] 
since it is not guaranteed by any fundamental principles of physics 
and usually requires some sort of fine-tuning of the parameters.  
There are, however, notable exceptions 
where relatively simple physical situations enable the system to fine-tune itself and realize a higher symmetry group.
For instance, the interplay between spin and orbital degrees of freedom can lead to the realization 
of an SU(4) Kondo effect using semiconductor quantum dots \cite{Keller14} 
or an SU(4) symmetry in strongly correlated electrons with orbital degeneracy~\cite{Kugel2015}.  
Another important example is electrons in graphene which have four flavors associated 
with the low-energy spin and valley degrees of freedom;   
the fractional quantum Hall effect in graphene is governed by the Coulomb interactions which are
invariant under rotations of these four flavors and thus an SU(4) symmetry could also be realized 
\cite{Arovas-K-L-99,Nomura-M-06}.   
Though, weak atomic-range valley-dependent interactions might explicitly break this symmetry, 
it has been shown recently that an SO(5) symmetry arises in the $\nu =0$ quantum Hall regime 
of grapheme~\cite{Wu2014}.

On top of these examples, ultracold fermions loaded into optical lattices might be ideal systems
to simulate strongly correlated electrons with a high symmetry, thanks to their great 
tunability (for reviews on recent development in many-body physics of cold atoms in optical lattices, 
see, e.g., Refs.~\cite{Bloch-D-Z-08,Bloch-D-N-12,Lewensteinbook}). 
In principle, ultracold atomic gases with alkali atoms with hyperfine spin $F>1/2$ can explore the physics with
SO(5) and SU(3) symmetries~\cite{Wu2003,HonerkampH04,Lecheminant-B-A-05,Wu06,RappZHH07,AzariaCL09}.
However, alkaline-earth atoms and related ones, like ytterbium atoms, have a peculiar energy spectrum 
associated with the two-valence outer electrons which make them the best candidates for experimental realizations of
SU($N$) many-body physics~\cite{Cazalilla-H-U-09,Gorshkov-et-al-10,Cazalilla-R-14}. 
The ground state (``$g$'' state) is a long-lived singlet state $^1S_0$ and moreover 
the spectrum contains a metastable triplet excited state (``$e$'' state) $^3P_0$.  
The $g$ and $e$ states have therefore zero electronic angular momentum, so that the nuclear spin $I$ 
is almost decoupled from the electronic spin.  From perturbation theory, 
the nuclear-spin-dependent variation of the scattering lengths is estimated to be smaller than $\sim 10^{-9}$ 
($\sim 10^{-3}$) for the $g$ (the $e$) states \cite{Gorshkov-et-al-10}.  
As will be reviewed in Sec.~\ref{sec:SUN-in-alkaline-earth}, this decoupling of the electronic spin from the nuclear one 
in two-body collisions paves the way to the experimental realization of fermions with an SU($N$) symmetry 
where $N= 2 I +1$ ($I$ being the nuclear spin) is the number of nuclear-spin states.

The cooling of fermionic isotopes of these atoms below the quantum degeneracy has been 
achieved for strontium atoms $^{87}$Sr with $I= 9/2$ \cite{DeSalvo-Y-M-M-K-10,Tey-S-G-S-10} 
(see Ref.~\cite{Stellmer-S-K-14} for a review) 
and for ytterbium atoms $^{171}$Yb and $^{173}$Yb respectively with $I= 1/2$ 
and $5/2$~\cite{Fukuhara-T-K-T-07,Taie-etal-PRL-10} (see Ref.~\cite{Sugawa-T-E-T-Yb-review-13} for a review). 
These atoms enable the experimental exploration of the physics of fermions with an SU($N$) symmetry where $N$
can be as large as 10. Several experiments have been done recently with these fermionic atoms loaded into optical lattices.
An SU(6) Mott insulator has been explored with $^{173}$Yb  atoms loaded into a  three-dimensional (3D) optical lattice \cite{Taie-Y-S-T-12}.
The experimental proof of the presence of the  SU($N$) symmetry has been provided in Refs. 
\cite{Zhang-et-al-14} and \cite{Scazza-et-al-14} by using $^{87}$Sr and $^{173}$Yb, respectively,  
in a 2D ($^{87}$Sr) and 3D ($^{173}$Yb) optical lattices.  
Furthermore, the specific form of the interactions between the $g$ and $e$ states was determined in 
these works. Quite recently also, a coherent spin-orbital exchange interactions has been observed using $^{173}$Yb loaded into a 3D optical lattice~\cite{Cappellini-et-al-14}. 

All these remarkable experiments give us confidence that there will be many more upcoming breakthroughs 
in the near future. 
Motivated by recent experiments on $^{173}$Yb in the one-dimensional (1D) regime \cite{Pagano-et-al-14}, 
in this review, we will be focusing only on the 1D case in the presence of an optical lattice 
and will not discuss many other interesting results in two and three dimensions.  
Therefore, as a complement to our review, 
we refer the readers to Ref.~\cite{Cazalilla-R-14} for an recent extensive review about the realization of 
SU($N$) symmetric fermionic gases and its physics in 2D or 3D.  

More specifically, we will first discuss the zero-temperature phases that can occur 
in the SU($N$) alkaline-earth fermions when only the atoms in the $g$-state are 
loaded in one-dimensional optical lattices.    
As will be seen in Sec.~\ref{sec:singleband}, the low-energy properties of alkaline-earth atoms 
in the $g$-state are described by the single-band SU($N$) Fermi-Hubbard model with
the on-site interaction $U$ that depends on the $s$-wave scattering length between two $g$ atoms.  
When $N>2$, rich physics emerges already in this simple setting, which is very different from that 
in the standard SU(2) case.  
At a filling of one atom per site ($1/N$-filling), which best avoids three-body losses, 
the repulsive SU($N$) Fermi-Hubbard model simplifies, for large repulsive interaction, 
to the SU($N$) antiferromagnetic Heisenberg spin chain with the fundamental representation at each site.  
The latter model, the so-called Sutherland model \cite{Sutherland1975}, has been quite extensively 
studied in the context of quantum magnetism \cite{Affleck-NP86,Affleck-88,Alcaraz-M-SUN-89} 
and the underlying gapless Mott-insulating phase can be directly investigated in ongoing experiments 
with 1D ytterbium cold fermions.  
While low temperatures are difficult to achieve for fermionic gases, there is some advantage at working with
several species (large $N$) since some short-range features might be
observed at accessible temperatures due to some entropic effect, as was computed numerically in one~\cite{Bonnes2012,Messio2012} and two dimensions~\cite{Cai2013}.  
The Mott insulating phase at $1/N$-filling has been already realized experimentally with 
${}^{173}$Yb ($N=6$) loaded in three-dimensional optical lattices \cite{Taie-Y-S-T-12}.   

In sharp contrast to the case of SU(2), the SU($N$) Fermi-Hubbard model displays {\em fully gapped} 
Mott-insulating phases.  
However, these phases exhibits bond-ordering and spontaneously break translation symmetry. 
Therefore, the realization of some featureless gapped exotic phases of matter requires 
going beyond the single-band model.  
In this respect, it would be interesting to consider various two-band models which, on top of the SU($N$), 
have an additional `orbital' degree of freedom. 
The origin of the two orbitals could be either the two atomic states $g$ and $e$ ($g$-$e$ model), 
or the $p_x$ and $p_y$ orbital states ($p$-band model) of the harmonic trap.  

In the two-orbital cases, the interplay between the orbital and SU($N$) nuclear-spin
degrees of freedom will be shown to give rise to several interesting
phases, including symmetry-protected topological (SPT)
phases~\cite{Gu-W-09,Chen-G-L-W-12}.  The latter refer to non-degenerate
fully gapped phases which do not break any symmetry and defy 
the characterization with local order parameters.  Since any gapful phases in
one dimension necessarily have short-range entanglement \cite{Chen-G-W-10} and can be reduced to a trivial 
product state by a series of local unitary transformations, the presence of a
symmetry (which restricts the type of possible local unitary) is necessary to protect the properties 
of that 1D topological phase, in particular the existence of non-trivial edge
states~\cite{Chen-G-W-10,Chen-G-L-W-12}.

The outline of this review is as follows. 
In Sec.~\ref{sec:singleband}, we will provide a minimal introduction to SU($N$) symmetry in alkaline-earth systems 
and then describe the known results for the single-band Hubbard model in 1D.  
We consider both the cases with commensurate fillings, where various Mott phases occur, and 
those with incommensurate fillings where low-energy properties can be described 
using the Luttinger-liquid theory \cite{Gogolin-N-T-book,Giamarchi-book}.  

In Sec.~\ref{sec:twoband}, we will move on to the case of two-`orbital' fermions.  
After introducing two relevant microscopic models ($g$-$e$ model and $p$-band model), which are directly
relevant to experiments~\cite{Zhang-et-al-14,Scazza-et-al-14}, we will discuss their strong-coupling limits,  
i.e., the SU($N$) spin chains in various representations, for which several results are known 
or conjectured.  In particular, we will sketch the known classification of the so-called SU($N$) SPT phases 
allowing us to understand the formation of topological states in our SU($N$) fermion system 
as generalizations of the well-known Haldane phases in spin systems.  
Finally, we will present some typical numerical results to provide complete phase diagrams 
as well as clear signatures of the SPT phases through the direct measurements of nontrivial edge states. 
Conclusions will be given in Sec.~\ref{sec:conclusion}.

\section{Single-band Fermi-Hubbard model} 
\label{sec:singleband}
\subsection{Definition of the model}
\subsubsection{SU(N) symmetry in cold-atom systems}
\label{sec:SUN-in-alkaline-earth}
Like the SU(3)-symmetry in the quantum chromodynamics for the strong interactions, 
in high-energy physics, 
the existence of the multiplet of $N$ particles and the SU($N$)-symmetry that governs 
the dynamics of these particles have nothing to do with spins and are built in the theory 
from the outset.  
In condensed-matter physics, on the other hand, such internal symmetries independent of spin may originate from,  
e.g., sublattice symmetry and normally fine-tuning is necessary to realize high symmetries as discussed in
the introduction.  
In order to understand how SU($N$) symmetry arises among the multiplet which originates from 
the angular momentum {\em without any fine-tuning},  
consider two spin-$F$ fermionic atoms interacting with a contact interaction.  
Due to the rotational invariance, the two-body interaction depends {\em only} on the total spin $f$.  
Among all the possible values $f=0,1,2,\ldots, 2F$, only the $(2F+1)/2$ antisymmetric combinations 
$f=0,2,\ldots, 2F-1$ ($F$ is assumed to be half-odd integer) are allowed by fermionic statistics.  
Therefore, the most general form of rotationally [i.e., SU(2)] invariant interaction between two spin-$F$ objects may be 
written as \cite{Ho-98}\footnote{%
A similar analysis has been done in Ref.~\cite{Ohmi-M-98} for the $F=1$ bosons.}
\begin{equation}
V_{F,F}(\bolr,\bolr^{\prime})  = \delta(\bolr-\bolr^{\prime}) \sum_{f=0,2,\ldots}^{2F-1} g_{f} P_{f}  \; ,
\label{eqn:F-F-scattering}
\end{equation}
where $P_{f}$ denotes the projection operator onto the total spin-$f$ sector and 
the coupling constants $g_{f}$ are determined by the corresponding $s$-wave scattering lengths 
$a_{f}$ as ($M$ being the mass of the atoms in question)
\begin{equation}
g_{f} = \frac{4\pi \hbar^{2}}{M} a_{f} \; .
\end{equation}  
In the ground state ${}^{1}S_{0}$ of ${}^{173}$Yb, for instance, $I=F=5/2$ and we have, {\em in principle}, 
three independent coupling constants $g_{0}$, $g_{2}$, and $g_{4}$.  

In the second-quantized formulation, the above projection operators can be simply given as 
\begin{equation} 
 \hat{P}_{f} =  \sum_{m= -f}^{f} \hat{P}_{f,m} 
\equiv  \sum_{m= -f}^{f} A^{\dagger}_{f,m}(\bolr)A_{f,m}(\bolr) ,
\end{equation}
with the `pairing operators' defined by \cite{Ho-98}
\begin{equation}
\begin{split}
& A^{\dagger}_{f,m}(\bolr) \equiv \sum_{\alpha,\beta= -F}^{F}
 \langle F,\alpha ; F, \beta | f,m \rangle 
c^{\dagger}_{\alpha}(\bolr) c^{\dagger}_{\beta}(\bolr)  \\
& A_{f,m}(\bolr) \equiv \sum_{\alpha,\beta= -F}^{F}
 \langle f,m | F,\alpha ; F, \beta \rangle 
c_{\beta}(\bolr) c_{\alpha}(\bolr) \; .
\end{split}
\end{equation}
Using the definition of the pairing operators [i.e., creation/annihilation operators for 
a pair with a particular value of the hyperfine spin $(f,m)$],  
the interaction may be rewritten as \cite{Ho-98,Ho1999}
\begin{equation}
\begin{split}
& \frac{1}{2} \sum_{f=0,2,\ldots}^{2F-1} g_{f}  \hat{P}_{f}  \\
& = \frac{1}{2} \sum_{f=0,2,\ldots}^{2F-1} g_{f} \sum_{m= -f}^{f} 
\sum_{\alpha_1,\beta_1= -F}^{F}  \sum_{\alpha_2,\beta_2= -F}^{F}
 \langle F,\alpha_1 ; F, \beta_1 | f,m \rangle  \langle f,m | F,\alpha_2 ; F, \beta_2 \rangle \\
& \phantom{=}
\times c^{\dagger}_{\alpha_1}(\bolr) c^{\dagger}_{\beta_1}(\bolr)c_{\beta_2}(\bolr) c_{\alpha_2}(\bolr) \; .
\end{split}
\end{equation}
When the coupling constants (or, the scattering lengths) do not depend on $f$
(i.e., $g_{f}=g$), the above simplifies to the following expression
\begin{equation}
\begin{split}
& \frac{1}{2} g \sum_{f=0,2,\ldots}^{2F-1} \hat{P}_{f} = 
\frac{1}{2} g \sum_{f=0,2,\ldots}^{2F-1} \sum_{m=-f}^{f} A^{\dagger}_{f,m}(\bolr)A_{f,m}(\bolr)  \\
& = \frac{1}{2} g  \sum_{\alpha_1,\beta_1= -F}^{F}  \sum_{\alpha_2,\beta_2= -F}^{F}\\
& \phantom{=} \times
\underbrace{%
\left\{
\sum_{f=0,2,\ldots}^{2F-1} \sum_{m=-f}^{f}
 \langle F,\alpha_1 ; F, \beta_1 | f,m \rangle  \langle f,m | F,\alpha_2 ; F, \beta_2 \rangle
 \right\}
 }_{=\delta_{\alpha_1,\alpha_2}\delta_{\beta_1,\beta_2}} \\
& \phantom{=}
\times c^{\dagger}_{\alpha_1}(\bolr) c^{\dagger}_{\beta_1}(\bolr)c_{\beta_2}(\bolr) c_{\alpha_2}(\bolr) \\
& = \frac{1}{2}  g  \sum_{\alpha,\beta = -F}^{F} 
c^{\dagger}_{\alpha}(\bolr) c^{\dagger}_{\beta}(\bolr)c_{\beta}(\bolr) c_{\alpha}(\bolr) 
= \frac{1}{2} g : n(\bolr) n(\bolr) :
\; ,
\end{split}
\label{eqn:SUN-inv-interaction-1-band}
\end{equation}
where $n(\bolr) = \sum_{\alpha}  c^{\dagger}_{\alpha}(\bolr) c_{\alpha}(\bolr)$ is the density operator which 
is invariant under  SU($N$) symmetry with $N=2F+1$. 
Therefore, when $g_{f}$ do not depend on $f$, i.e., when the fine-tuning of the scattering lengths 
$a_{0}=a_{2}=\cdots=a_{2F-1}$ occurs, 
the original SU(2)-symmetry of the atom-atom interaction gets enlarged to SU($N$) with 
$N=2F+1$\footnote{%
Even when the fine-tuning is incomplete, we may have other extended symmetries. 
See Ref.~\cite{Nonne-L-C-R-B-11} for the discussion of the phase structure of these cases.}. 

A natural question to ask then is when and in which system this fine tuning happens?  
The collision among neutral atoms are mainly governed by short-range van der Waals interaction 
that depends only on the electronic wave functions of the two colliding atoms.  
The nuclear-spin degrees of freedom can participate in the collisional processes  
only through the hyperfine interaction.  
In alkali cold atoms with hyperfine $F=3/2$ spin, an enlarged continuous $\text{Sp}(4) \sim \text{SO}(5)$ symmetry
arises without fine-tuning\footnote{%
In order to understand this, we first identify the two quartets $\mathbf{4}$ of the two colliding $F=3/2$ 
atoms with two 4-dimensional spinor representations ($\mathbf{4}_{\text{SO(5)}}$) of SO(5).  
Then, the states of the two atoms may be decomposed, in terms of SO(5), as
\[ (F=3/2)\otimes(F=3/2) \Leftrightarrow 
\mathbf{4}_{\text{SO(5)}}\otimes \mathbf{4}_{\text{SO(5)}} \simeq 
\mathbf{1}_{\text{SO(5)}}\oplus \mathbf{5}_{\text{SO(5)}} \oplus \mathbf{10}_{\text{SO(5)}} \; , \]
among which the last one (symmetric $\mathbf{10}_{\text{SO(5)}}$) must be discarded by the Fermi statistics 
(for the $s$-wave scattering).  The remaining two ($\mathbf{1}_{\text{SO(5)}}$ and 
$\mathbf{5}_{\text{SO(5)}}$) may be identified with the singlet ($f=0$) and the quintet ($f=2$) of SU(2) 
[see Eq.~\eqref{eqn:F-F-scattering}].  
Therefore, the scattering of two {\em fermionic} $F=3/2$ atoms can be rephrased in terms of SO(5).
 }
, i.e., with two independent scattering lengths $a_{0}, a_{2}$. \cite{Wu2003,Wu06}
The physical properties of this model in one dimension have been extensively investigated over the years by means 
of various analytical and numerical approaches 
\cite{Lecheminant-B-A-05,Wu2005,Controzzi2006,Capponi2007,Lecheminant-A-B-08,Roux-C-L-A-09,%
Nonne-L-C-R-B-10,Rodriguez2010,Nonne-L-C-R-B-11,Barcza2012,Szirmai-N-14,Barcza2015,Jiang2015}.
The fine-tuning of the scattering lengths $a_{0}= a_{2}$ to achieve a higher SU(4) symmetry is not
easy to realize experimentally with these alkali atoms. 
In contrast, if the hyperfine interaction is quenched for some reasons, the scattering processes  
are independent of the nuclear spin $f$: $g_{0}=g_{2}=\cdots=g_{2f-1}$.   
As the total electron angular momentum vanishes (and so does the hyperfine interaction) 
in the ground state ${}^{1}S_{0}$ of alkaline-earth and Yb atoms,  
these atoms are the best candidates of systems that realize SU($N$) symmetry without any fine-tuning
\cite{Gorshkov-et-al-10,Cazalilla-H-U-09,Cazalilla-R-14}. 

\subsubsection{SU($N$) Hubbard model}
\label{sec:deriv-1-band-SUN-Hubbard}
In order to derive a lattice Hamiltonian that describes low-energy physics 
of an $N$-component Fermi gas with the SU($N$) symmetry moving on 1D optical lattices, 
let us begin with the single-particle problem:
\begin{equation}
\mathcal{H}_{0} = \left\{
- \frac{\hbar^{2}}{2M}\partial_{z}^{2} + V_{\text{per}}(z) 
\right\}   \equiv  \mathcal{H}_{/\!/}(z)
 \; ,
 \label{eqn:single-particle-Ham-g-e}
 \end{equation}
 where $V_{\text{per}}(z)$ is a periodic potential that introduces a lattice structure along the chain 
 (i.e. $z$) direction.  Also we have assumed that the confining potential in the other directions (i.e., 
 $x$ and $y$) is so strong that we are able to neglect the motion in the $xy$-direction and set $x=y=0$.  
If the chain is translationally invariant in the $z$-direction, the single-particle state is given by the Bloch function 
$\varphi^{(n)}_{k_z}(z)$ ($n$ being the band index) satisfying:
\begin{equation}
 \mathcal{H}_{/\!/}(z) \varphi^{(n)}_{k_z}(z) = \varepsilon^{(n)}(k_z) \varphi^{(n)}_{k_z}(z)  \; .
 \end{equation}
 To derive an effective Hubbard-like Hamiltonian \cite{Auerbach,Jaksch-Z-05}, it is convenient to move from 
the Bloch function $\varphi^{(n)}_{k_z}(z)$ to the Wannier function defined by
\begin{equation}
w^{(n)}_{R}(z) \equiv 
\frac{1}{\sqrt{N_{\text{cell}}}} \sum_{k_z} \be^{- i k_z R} \varphi^{(n)}_{k_z}(z)  
\label{eqn:Wannier1-pxpy}
\end{equation}
(the index $R$ labels the center of the Wannier function and $N_{\text{cell}}$ is the number of unit cells in 
the $z$-direction) 
and introduce the fermion operators in the Wannier basis $\{c^{(n)}_{\alpha,R} \}$:
\begin{equation}
\begin{split}
& c_{\alpha}(z)=
\sum_{R}\sum_{n}
w^{(n)}_{R}(z)c^{(n)}_{\alpha,R}  \; , \quad  
c_{\alpha}^{\dagger}(z)=
\sum_{R}\sum_{n}
w^{(n)\, \ast}_{R}(z)c^{(n)\, \dagger}_{\alpha,R}  \\
& \quad (\alpha=1,\ldots,N)   \; .
\end{split}
\end{equation}
We use these operators to rewrite the SU($N$)-invariant 
two-body interaction \eqref{eqn:SUN-inv-interaction-1-band} as
\begin{equation}
\begin{split}
& \frac{1}{2} g  \sum_{\alpha,\beta = -F}^{F} 
\int\!dz  \,c^{\dagger}_{\alpha}(z) c^{\dagger}_{\beta}(z)c_{\beta}(z) c_{\alpha}(z) \\
&= 
\frac{1}{2}\sum_{\alpha,\beta=1}^{N}\sum_{\{R_{i},n_i\}}
V
({}_{R_{1},R_{2};R_{3},R_{4}}^{n_1,n_2;n_3,n_4}) \, 
c^{(n_1)\dagger}_{\alpha,R_1}c^{(n_2)\dagger}_{\beta,R_2}
c^{(n_3)}_{\beta,R_3}c^{(n_4)}_{\alpha,R_4}  \; ,
\end{split}
\end{equation}
where 
\begin{equation}
V({}_{{R}_{1},{R}_{2};{R}_{3},{R}_{4}}^{n_1,n_2;n_3,n_4}) 
\equiv 
g \int\!dz \, 
w^{(n_1)\ast}_{{R}_1}(z) w^{(n_2)\ast}_{{R}_2}(z)
w^{(n_3)}_{{R}_3}(z)w^{(n_4)}_{{R}_4}(z) 
\label{eqn:1-band-V-by-Wannier}
\end{equation}
and we have re-labelled $\alpha=-F,\ldots, F \Rightarrow \alpha=1,\ldots, N$ ($N=2F+1$)\footnote{%
In fact, $N$ can take any number $N \leq 2F+1$ as one can load only a subset of the multiplet 
by optical pumping.}.    
As usual \cite{Auerbach,Jaksch-Z-05}, we keep only the most relevant band (denoted by 
$n_1=n_2=n_3=n_4=n_0$) and the on-site term 
${R}_1={R}_2={R}_3={R}_4={R}_i$ to obtain 
the so-called Hubbard interaction:
\begin{equation}
\begin{split}
& \frac{1}{2}
V({}_{{R}_{i},{R}_{i};{R}_{i},{R}_{i}}^{n_0,n_0;n_0,n_0}) 
 \sum_{\alpha,\beta=1}^{N} \sum_{i} 
c^{\dagger}_{\alpha,i}c^{\dagger}_{\beta,i} c_{\beta,i}c_{\alpha,i} \\
& \equiv \frac{1}{2}U \sum_{\alpha,\beta=1}^{N} \sum_{i} 
c^{\dagger}_{\alpha,i}c^{\dagger}_{\beta,i} c_{\beta,i}c_{\alpha,i} 
= \frac{1}{2}U \sum_{i} n_{i}(n_{i}-1) 
\; .
\end{split}
\label{eqn:1-band-Hubbard-U}
\end{equation}
In the above, we have introduced a short-hand notations
\begin{equation}
c_{\alpha,i} \equiv c_{\alpha,{R}_{i}} \; , \quad 
\sum_{i} \equiv \sum_{{R}_{i}} \; ,
\end{equation}
and $n_{i} = \sum_{ \alpha}  c_{\alpha,\,i}^\dag c_{\alpha,\,i}$ is the
density operator on site $i$.

When second-quantized, the single-particle part reads as
\begin{equation}
\begin{split}
& \sum_{\alpha=1}^{N}
\int\! dz \, 
c^{\dagger}_{\alpha}(z)\mathcal{H}_{/\!/}(z) c_{\alpha}(z)  \\
& = \sum_{\alpha=1}^{N}\sum_{\{R_i,n_i\}} 
\left\{%
\int\! dz \, 
w_{R_1}^{(n_1)\ast}(z)\mathcal{H}_{/\!/}(z) w_{R_2}^{(n_2)}(z)
\right\}
c_{\alpha,R_1}^{(n_1)\dagger}c_{\alpha,R_2}^{(n_2)}  .
\end{split}
\end{equation}
This may be further rewritten using the hopping amplitudes:
\begin{equation}
\begin{split}
& 
\int\! dz \, 
w_{R_1}^{(n_1)\ast}(z)\mathcal{H}_{/\!/}(z) w_{R_2}^{(n_2)}(z) \\
& =  \frac{1}{N_{\text{cell}}} \sum_{k_z}\sum_{k_z^{\prime}}
\be^{ i k_z R_1} \be^{- i k^{\prime}_z R_2} 
\left\{%
\int\!dz \, 
\varphi^{(n_1)\ast}_{k_z}(z)
\mathcal{H}_{/\!/}(z)
\varphi^{(n_2)}_{k_z^{\prime}}(z)  
\right\} \\
& = \delta_{n_1 n_2}
\left\{ 
\frac{1}{N_{\text{cell}}} \sum_{k_z}
\epsilon^{(n_1)}_{k_z} \be^{ i k_z (R_1-R_2)} 
\right\}
\equiv  -  \delta_{n_1 n_2} t^{(n_1)}(R_1-R_2)  \; ,
\end{split}
\end{equation}
and we finally obtain the kinetic term
\begin{equation}
- \sum_{\alpha=1}^{N} \sum_{n\in\text{bands}}\sum_{R_1,R_2} t^{(n)}(R_1-R_2)
c_{\alpha,R_1}^{(n)\dagger}c_{\alpha,R_2}^{(n)}  \; .
\end{equation}
Retaining only the terms with $n=n_0$ and $|R_1 - R_2|=1$ in the above and combining it with 
Eq.~\eqref{eqn:1-band-Hubbard-U}, we arrive at 
the SU($N$) generalization of the famous Fermi-Hubbard model:
\begin{equation}
{\cal H}_{\text{SU($N$)}} = - t \sum_{i}\sum_{\alpha=1}^{N}  \left(c_{\alpha,\,i}^\dag c_{\alpha,\,i+1}  + \text{H.c.}\right)
+ \frac{U}{2}  \sum_{i} n_{i}(n_{i}-1) \; .
\label{HubbardSUN}
\end{equation} 
The model (\ref{HubbardSUN}) is invariant under the global U(1) symmetry: 
$c_{\alpha,\,i} \mapsto e^{i \theta} c_{\alpha,\,i}$, which implies the conservation
of the total number of atoms. 
In the rest of this paper, we frequently use the terminology `charge' to denote the degree of freedom associated with 
this symmetry, although we are dealing with neutral atoms {\em without} electric charge.  
All the parameters in the model (\ref{HubbardSUN}) are independent from the nuclear-spin states 
($\alpha=1,\ldots,N$) and
an extended SU($N$) symmetry arises: $c_{\alpha,\,i} \mapsto \sum_{\beta} U_{\alpha \beta}  c_{\beta,\,i}$, $U$ being an SU($N$) matrix. The actual continuous symmetry group of the Hamiltonian (\ref{HubbardSUN}) 
is then U($N$) $=$ U(1) $\times$
SU($N$) but the model ${\cal H}_{\text{SU($N$)}}$ 
is often called the SU($N$) Fermi-Hubbard model to put the emphasis 
on its non-trivial SU($N$) hyperfine-spin rotational invariance. 
\subsection{Sutherland model and its low-energy physics}
\label{sec:sutherland}
The model (\ref{HubbardSUN}) describes alkaline-earth atoms in the $g$ state (i.e., ground state 
${}^{1}S_{0}$) loaded into the lowest band of the optical lattice.  
The interaction parameter $U$ is directly related to the $s$-wave scattering length associated with
the collision between two atoms in the $g$ state [see Eqs.~\eqref{eqn:1-band-V-by-Wannier} and 
\eqref{eqn:1-band-Hubbard-U}]. 
When $N=2$,  the model (\ref{HubbardSUN}) is the usual 
SU(2) Hubbard chain which is exactly solvable by means of the Bethe ansatz \cite{Lieb1968}.
The physical properties of the model have been discussed in great detail over the years and are reviewed in the book
\cite{Essler-book}. However, for $N>2$, the Hamiltonian  (\ref{HubbardSUN}) is not integrable for arbitrary $U$ and
filling $n$. Although it is possible to formally generalize the Lieb-Wu Bethe ansatz equation \cite{Lieb1968} 
to fermions with internal SU($N$) symmetry, it is believed that the corresponding model 
describes a {\em non-local} variant  of the SU($N$) Hubbard model \cite{Schlottmann1997}.  
In the absence of a lattice, the model is again integrable 
and its properties have been described in a recent review \cite{Guan2013}.

The situation becomes much simpler in the limit of large repulsive $U$ for a filling $n=1/N$  
with one atom per site which best avoids the three-body losses. In that case, 
the model (\ref{HubbardSUN}) reduces to the SU($N$)
Heisenberg antiferromagnetic spin chain with the SU($N$) fundamental representation 
(represented by the Young diagram ${\tiny \yng(1)}$; for a pedagogical explanation of the representation 
theory of SU($N$) and the Young diagrams, see, e.g., Ref.~\cite{Georgi-book-99}) on each site ({\em Sutherland 
model} \cite{Sutherland1975}):
\begin{equation}
{\cal H} = J \sum_i P_{i,i+1} ,
\label{sutherland}
\end{equation}
where $J = 2 t^2/U$ is the antiferromagnetic spin exchange 
and $P_{i,i+1}$ is the operator which permutes the SU($N$) hyperfine states on the sites $i$ and $i+1$.
For the fundamental representation (${\tiny \yng(1)}$), $P_{i,i+1}$ is compactly written, in terms of 
the SU($N$) generators that are normalized to be $\text{Tr}\,\mathcal{S}_{i}^{A}\mathcal{S}_{i}^{B} = \delta^{AB}$, 
as 
\begin{equation}
P_{i,i+1} = \frac{1}{N} + \sum_{A=1}^{N^{2}-1}\mathcal{S}_{i}^{A}\mathcal{S}_{i+1}^{A} 
\end{equation}
[SU($N$) generalization of the {\em Dirac identity\/}].  
It is known that the `spin' model \eqref{sutherland} well describes the low-energy sector of 
the original Hubbard model \eqref{HubbardSUN} for $U/t \gtrsim 12$ \cite{Manmana-H-C-F-R-2011}.  

In contrast to the original fermionic model \eqref{HubbardSUN}, the large-$U$ effective Hamiltonian \eqref{sutherland} 
can be solved exactly by the Bethe ansatz \cite{Sutherland1975}.
The low-energy spectrum is gapless with $N-1$ relativistic modes with the same velocity 
$v_{\text{s}} = 2 \pi J/N$.
The critical theory has been identified by Affleck \cite{Affleck-NP86,Affleck-88} as described  
by the level-1 SU($N$) Wess-Zumino-Witten [SU($N$)$_1$ WZW] conformal field theory 
(CFT) \cite{Knizhnik-Z-84,Witten1984,DiFrancesco-M-S-book}.
This CFT has a central charge $c= N-1$ and the low-temperature specific heat (per volume) 
scales as \cite{Affleck1986,Bloete1986}: 
$C(T) \simeq {k_{\text{B}}}^{2} N(N-1) T/(6 \hbar J)$ (with $k_{\text{B}}$ denoting the Boltzmann constant).  
The latter result has been confirmed by the thermodynamic Bethe ansatz \cite{lee94} 
(see Ref.~\cite{Alcaraz-M-SUN-89} for the determination of $c$ by finite-size corrections)  
and quantum Monte-Carlo calculations (QMC) \cite{Frischmuth-M-T-99,Messio2012}.  
It is known \cite{Affleck-88} that the SU($N$)$_1$ WZW CFT corresponds to the stable fixed point of 
generic 1D gapless systems with SU($N$) symmetry [as the SU(2)$_1$ WZW CFT describes generic 
gapless SU(2)-invariant spin chains].  

The gapless behavior in the SU($N$) Mott-insulating phase with one atom per site manifests itself in 
the spin-spin correlation functions which exhibit a universal power-law decay in
the long-distance limit at zero temperature~\cite{Affleck-NP86,Affleck-88}:
\begin{equation}
\langle \mathcal{S}^{A} (\tau, x)  \mathcal{S}^{B} (0,0)  \rangle \sim \delta^{AB} 
\cos \left\{ \frac{2\pi}{N} \left( \frac{x}{a_0} \right) \right\} 
\frac{\log^{2/N^2}\left(x^2 + v^2_{\text{s}} \tau^2 \right)}{\left(x^2 + v^2_{\text{s}} \tau^2 \right)^{1 - 1/N}},
\label{spinspincorrsutherland}
\end{equation}
where $\tau$ is the Euclidean time,
$a_0$ is the lattice spacing, and $A,B = 1, \ldots, N^2 -1$ are the components of the SU($N$) spin operator. 
The logarithmic corrections in 
Eq.~\eqref{spinspincorrsutherland} stem from the existence of a marginal operator 
in the low-energy effective field theory describing the model (\ref{sutherland}) \cite{Itoi1997,Majumdar2002}. 
When this marginal operator becomes marginally relevant by adding, e.g., further neighbor interactions, 
the system may spontaneously develop the $N$-merization [i.e., the formation of $N$-site clusters 
in SU($N$)-singlet]  
and enter gapped phases with broken translation symmetry 
\cite{Haldane-NNN-82,Lecheminant-T-06-SU4,Corboz-L-T-T-07}.  

The thermodynamics properties of the Sutherland model (\ref{sutherland}) have been investigated numerically
in Refs. \cite{Messio2012,Bonnes2012}.  
In particular, it has been shown that characteristic short-range correlations develop at
low temperature as a precursor of the algebraic correlations (\ref{spinspincorrsutherland}) in the ground state. 
The first sign of short-range order appears at an entropy per particle which increases with $N$,
leading to observable qualitative effects in ongoing ultracold atom experiments with alkaline-earth fermions.

\subsection{Phase structure}
\label{sec:phase-structure}
Since the SU($N$) Fermi-Hubbard model (\ref{HubbardSUN}) with $N>2$ is, in general, not exactly solvable, 
one has to resort to approximate but powerful techniques available in one dimension 
to map out its zero-temperature phase diagram: field-theoretical, strong-coupling, and numerical approaches.
In this section, we sketch the phase diagram of the model \eqref{HubbardSUN} for 
both incommensurate (Sec.~\ref{sec:incomm}) 
and commensurate fillings (Secs.~\ref{sec:mott}--\ref{sec:other-commensurate-fillings}).  
The main results of Sec.~\ref{sec:phase-structure} are summarized in Table~\ref{tab:phases-SUN}.    
\subsubsection{Incommensurate fillings}
\label{sec:incomm}
\label{sec:incommensurate}
We first consider the SU($N$) Fermi-Hubbard model (\ref{HubbardSUN})  for incommensurate fillings.
In this respect, there are no umklapp processes which open a gap for the charge degrees of freedom 
[or, the U(1) sector corresponding to the continuous symmetry of Eq.~\eqref{HubbardSUN}]. A metallic state is then formed whose nature strongly depends
on the sign of the coupling constant $U$.

When $U>0$, all modes are gapless and a metallic Luttinger-liquid phase emerges \cite{Gogolin-N-T-book,Giamarchi-book}.
The hallmark of this $N$-component Luttinger liquid phase is the $2k_{\text{F}}$ oscillations in the density-density and 
the spin-spin correlation functions decaying with non-universal power-law exponents 
\cite{Affleck-88,Assaraf-A-C-L-99,Manmana-H-C-F-R-2011}. 
For instance, the leading asymptotics of the SU($N$) spin-spin correlation functions (\ref{spinspincorrsutherland})
reads now in the metallic phase as:
\begin{equation}
\langle \mathcal{S}^{A} (\tau, x)  \mathcal{S}^{B} (0,0)  \rangle \sim \delta^{AB}  \frac{\cos \left( 2 k_{\text{F}} x\right)}{
\left(x^2 + v^2_{\text{c}} \tau^2 \right)^{K/N} \left(x^2 + v^2_{\text{s}} \tau^2 \right)^{1 - 1/N}},
\label{spinspinLuttinger}
\end{equation}
where $K$ and $v_{\text{c}}$ are respectively the Luttinger parameter and the characteristic velocity of 
the charge excitation which depend
on the interaction and density \cite{Gogolin-N-T-book,Giamarchi-book}. In particular,
$K$ determines the singularity in the momentum
distribution around the Fermi point $k_{\text{F}}$:
$n(k) \sim n(k_{\text{F}}) + {\rm const.} \; {\rm sign} \left(k -k_{\text{F}}\right) |k -k_{\text{F}} |^{\alpha}$
with $\alpha = (1 - K)^2/(2NK)$  \cite{Assaraf-A-C-L-99}.
This power-law singularity at the Fermi level is inherent in Luttinger liquids 
unlike in the standard Fermi liquid \cite{Gogolin-N-T-book,Giamarchi-book}.
Similarly, the single-particle density of states also has an anomalous power-law behavior for any finite value of $N$:
$\rho(\omega) \sim  |\omega|^{\alpha}$.

\begin{figure}[!htb]
\begin{center}
\includegraphics[width=0.5\textwidth]{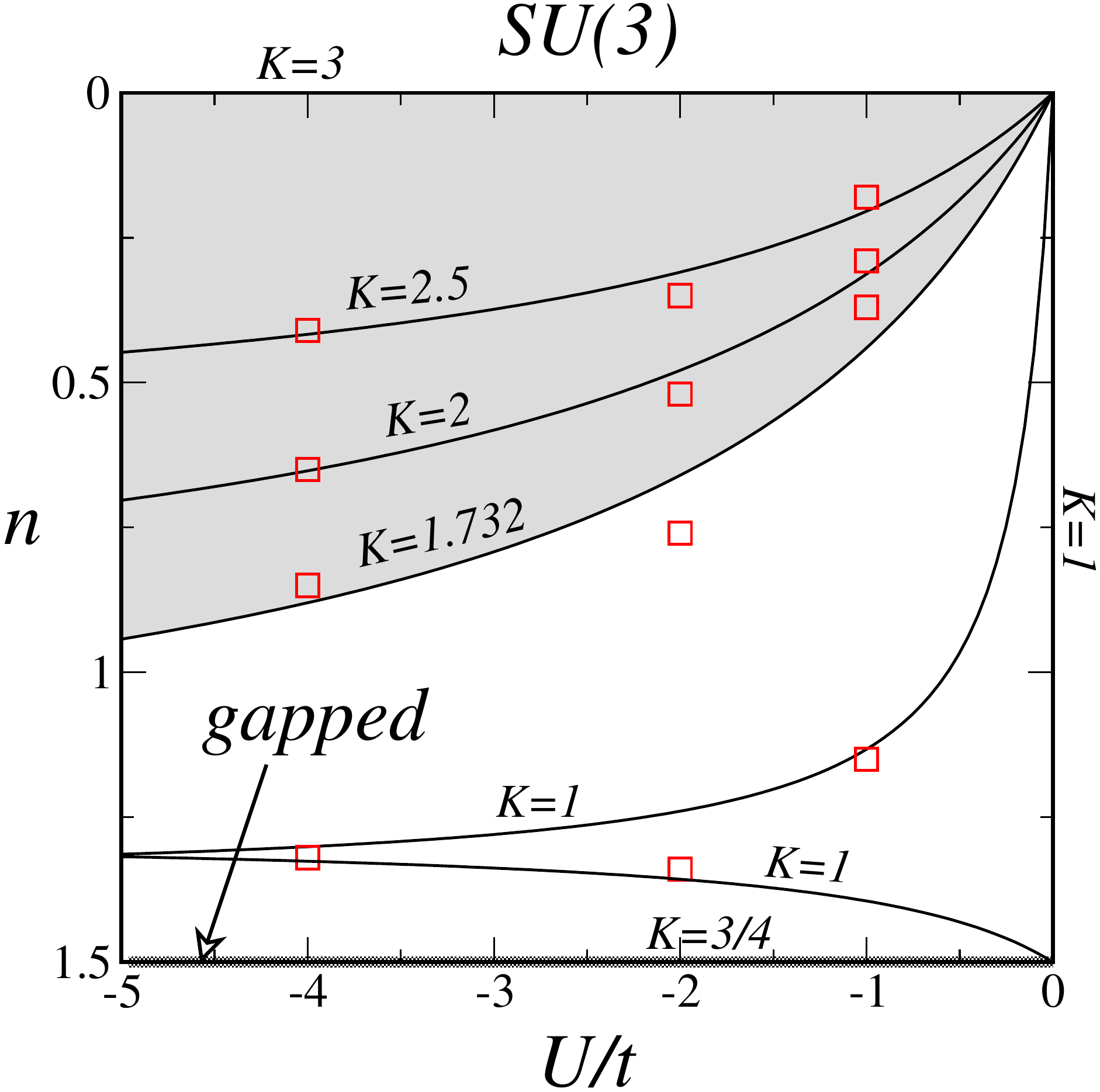}
\caption{(Color online) The Luttinger parameter $K$ as function
of $U$ and the density $n$ for the SU(3) Fermi-Hubbard model with attractive interaction ($U<0$). Grey region marks
the onset of the MS phase. The $n=3/2$ line corresponds to the half-filled case where a fully gapped Mott-insulating
phase occurs as discussed in Sec. \ref{sec:halfilling}. From Ref.~\cite{Capponi2008}.
\label{fig:SU3Kc}}
\end{center}
\end{figure}

On the other hand, 
the physics is very different in the attractive case ($U<0$); there is a spin gap for the SU($N$) degrees of freedom
and the only remaining gapless mode is the charge one. The metallic phase is then characterized by a CFT with
the central charge $c=1$ reflecting a single bosonic gapless mode. The resulting spin-gap phase is 
the Luther-Emery phase \cite{Luther-E-74,Gogolin-N-T-book,Giamarchi-book}, 
which turns out to be very exotic when $N>2$ at sufficiently low density \cite{Capponi2008,Roux-C-L-A-09}.
In the $N=2$ case, the Luther-Emery phase describes the competition between a charge-density wave (CDW) instability
and a superconducting one \cite{Gogolin-N-T-book,Giamarchi-book}.  
In the attractive SU(2) Hubbard model, the leading instability is the
superconducting one. 
When $N>2$, on the other hand, the SU($N$) symmetry plays an important role in one dimension by preventing any 
pairing between fermions: there is no way to form an SU($N$) singlet with only {\em two} fermions.  
When $N>2$, the usual pairing instability is then completely suppressed with exponential-decaying correlation functions 
in stark contrast to the $N=2$ case.
The only possible gapless fluctuation corresponding to the superfluid  instability is a molecular one 
where $N$ fermions form SU($N$) singlet (an analog of baryons in high-energy physics): 
$\mathcal{M}_i = c_{1,\,i}^\dag c_{2,\,i}^\dag \ldots c_{N,\,i}^\dag$, i.e, {\em trionic} ($N=3$) 
and {\em quartetting} ($N=4$) superfluid instabilities.  
The Luther-Emery phase of the model  (\ref{HubbardSUN}) with $U<0$ is
then governed by the competition between the instability toward this molecular superfluid (MS) 
and the one toward CDW.  
Among these competing order parameters, the one that exhibits the slowest
power-law-decaying correlations at zero temperature corresponds to the leading instability.  
The equal-time correlation functions of these order parameters have been determined 
by means of bosonization \cite{Lecheminant-B-A-05,Lecheminant-A-B-08,Capponi2008} as:
\begin{equation}
\begin{split}
& \langle n_{i}  n_{i+x} \rangle \sim \cos \left( 2 k_{\text{F}} x\right)  x^{- 2 K/N}  \\
& \langle \mathcal{M}_{i}  \mathcal{M}^{\dag}_{i+x} \rangle 
\sim 
\begin{cases}
x^{- N/ 2K}   \quad  & \text{for } N   \text{ even}  \\  
\sin \left(k_{\text{F}} x\right)  x^{- (K + N^2/ K)/2N}   \quad & \text{for }  N    \text{ odd} .
\end{cases}
\end{split}
\label{MScorr}
\end{equation}
Either CDW or MS instability thus dominates depending on the value of the Luttinger parameter $K$; 
dominant MS instability requires $K > N/\sqrt 3$ ($K > N/2$)
when $N$ is odd (even). At issue is the value of the the Luttinger parameter $K$. Its 
expression as a function of $U$ and the density $n$ has been numerically determined in Ref.~\cite{Capponi2008} and is reproduced in Fig.~\ref{fig:SU3Kc} for $N=3$. 
In the low-density regime, the MS phase, characterized by the bound states made of $N$ fermions, 
exists for a wide range of attractive $U$ (the shaded region in Fig.~\ref{fig:SU3Kc}).   
The latter phase might also be viewed as a `molecular' Luttinger liquid 
with the molecules of $N$ atoms and, as has been mentioned above, is characterized by 
the suppression of the usual Cooper pairs.  
In the high-density regime (the lower part of Fig.~\ref{fig:SU3Kc}), on the other hand, 
the dominant instability for $U<0$ is the standard CDW. 
\subsubsection{Mott transition}
\label{sec:mott}
For commensurate fillings, umklapp processes might become strongly relevant perturbations and open a spectral
gap for the charge degrees of freedom. The nature of the resulting Mott-insulating phase depends crucially on 
the filling $n$ \cite{Szirmai-L-S-08}. In the case of one atom per site (i.e., $n=1/N$ filling), 
the Mott-insulating phase is gapless in the large-$U$ limit and is described 
by the Sutherland model (\ref{sutherland}) and its physical properties were discussed already in Sec. \ref{sec:sutherland}.
In the $N=2$ case, it is well-known from the exact Bethe-ansatz solution that there is no (finite-$U$) 
Mott transition and a charge gap opens exponentially as soon as positive $U$ is switched on \cite{Lieb1968}. 
A gapless Mott-insulating phase with central charge $c=1$, 
described by the SU(2) Heisenberg spin model, emerges thus for all $U>0$.  

Again, the situation might be very different when $N>2$ as advocated in Ref. \cite{Assaraf-A-C-L-99}.
In the weak-coupling ($U \ll t$) limit, the umklapp operator appearing in the low-energy effective Hamiltonian 
has the scaling dimension $\Delta = N K$ \cite{Assaraf-A-C-L-99},    
which is relevant when the Luttinger parameter
is sufficiently small: $K < 2/N$. When $N=2$, this is always the case since $K<1$ 
as soon as a repulsive interaction is switched on \cite{Gogolin-N-T-book,Giamarchi-book}.  
In stark contrast, the umklapp operator is always \emph{irrelevant} when $N>2$ for sufficiently small $U$ 
and a Mott-transition at finite $U$ is thus expected toward the large-$U$ insulating phase 
with the gapless SU($N$) sector 
described by the SU($N$) Sutherland model \eqref{sutherland} \cite{Assaraf-A-C-L-99}.  
This problem has been investigated numerically by various different techniques \cite{Assaraf-A-C-L-99,Szirmai-S-05,Buchta2007,Manmana-H-C-F-R-2011}.
The determination of the position of the Mott-transition turned out to be a difficult numerical problem due to the
smallness of the charge gap. The QMC simulations of Ref.~\cite{Assaraf-A-C-L-99} found a critical value 
$U_{\text{c}} \simeq 2.2 t$ ($N=3$) and $U_{\text{c}} \simeq 2.8 t$ ($N=4$).  
Density matrix renormalization group (DMRG) ~\cite{White-92} calculations done in Ref.~\cite{Manmana-H-C-F-R-2011} reported smaller values: $U_{\text{c}} \simeq 1.1 t$ ($N=3$) and $U_{\text{c}} \simeq 2.1 t$ ($N=4$) while those
in Ref.~\cite{Buchta2007} concluded the same value $U_{\text{c}} = 0$ for all $N \geq 2$.
The physical properties of the metallic phase when $ U < U_{\text{c}}$ 
are described by the multi-component Luttinger liquid with $N$ gapless channels (see Sec.~\ref{sec:incommensurate}).  
Above the critical value $U >U_{\text{c}}$, there is a charge gap and we are  left with 
$N-1$ gapless spin modes whose physical properties are governed by the Sutherland model (\ref{sutherland}).  
The Mott-transition is argued \cite{Assaraf-A-C-L-99} to be of the Berezinskii-Kosterlitz-Thouless 
universality class \cite{Berezinskii-71,Kosterlitz-T-73}.

\subsubsection{Half-filled case}
\label{sec:halfilling}
We now consider the half-filled case with $k_{\text{F}} = \pi/2a_0$ ($N/2$ fermions per site; $n=1/2$). 
For $N=2$, it corresponds to the situation
where we have one atom per site and reduces to the case considered already in Sec.~\ref{sec:sutherland}; 
the physics for $U>0$ is governed by the Heisenberg model with a gapless $c=1$ behavior 
[corresponding to the level-1 SU(2) WZW CFT].  
On the attractive side $U<0$, we can apply a transformation ({\em Shiba transformation} \cite{Shiba-72,Emery-76}; 
See section 2.2.4 of Ref.~\cite{Essler-book} for a detailed discussion of the Shiba transformation), 
that interchanges spin and charge while flipping the sign of $U$, to 
show that now a gap opens in the spin sector while the charge sector remains gapless (Luther-Emery liquid).  

In contrast, when $N>2$,  all degrees of freedom are fully gapped for {\em any} values 
of $U$ (whether positive or negative) 
due to the absence of spin-charge separation in the low-energy limit \cite{Buchta2007,Nonne-L-C-R-B-11}.  
The resulting fully gapped Mott-insulating phase is two-fold degenerate 
as the result of the spontaneous breakdown of the one-site translation symmetry \cite{Nonne-L-C-R-B-11}.  
The physical nature of the Mott-insulating phase depends crucially on the sign of $U$. 

\begin{figure}[!htb]
\begin{center}
\centerline{
\includegraphics[width=0.5\textwidth]{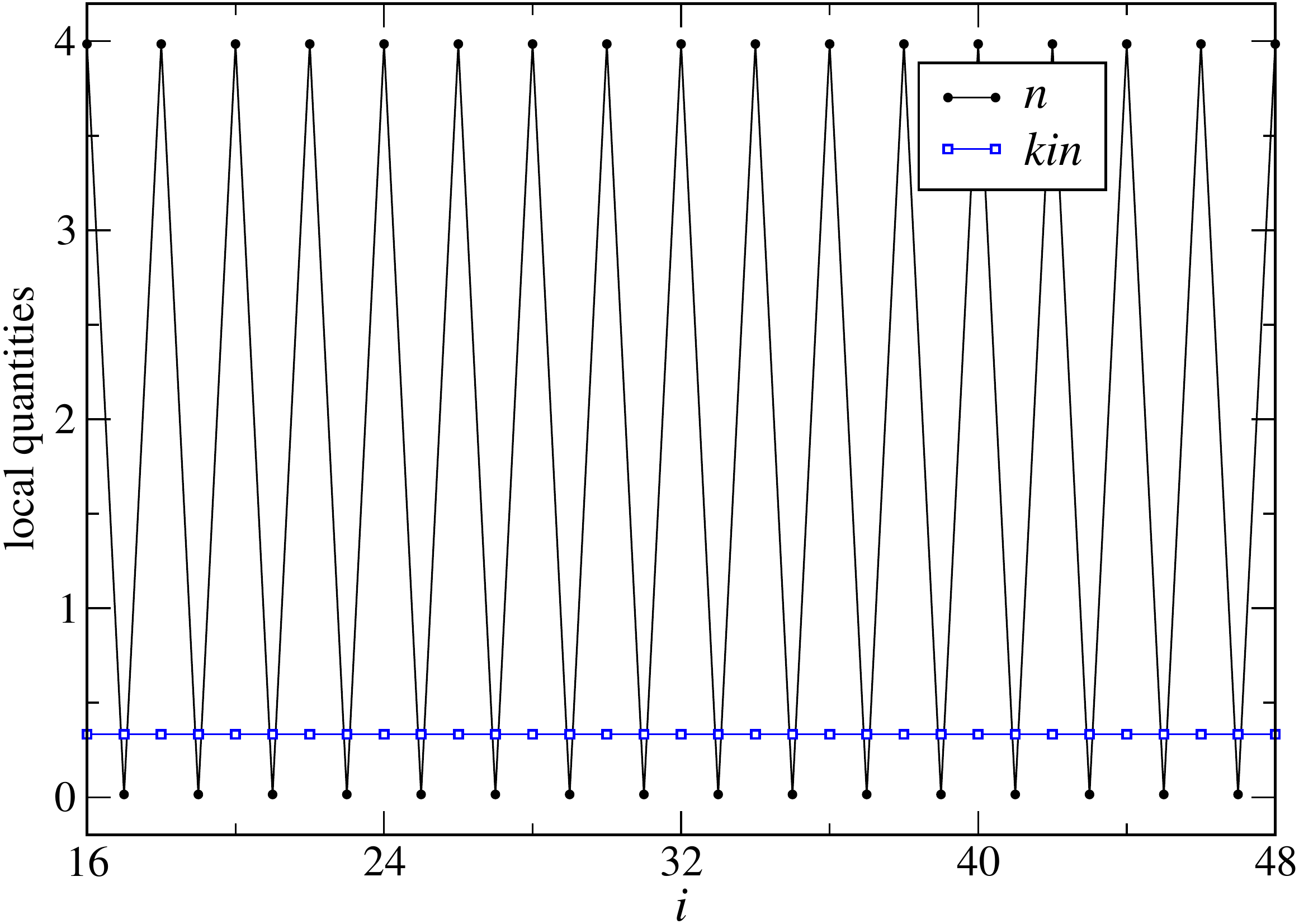}
\hspace{3mm}
\includegraphics[width=0.5\textwidth]{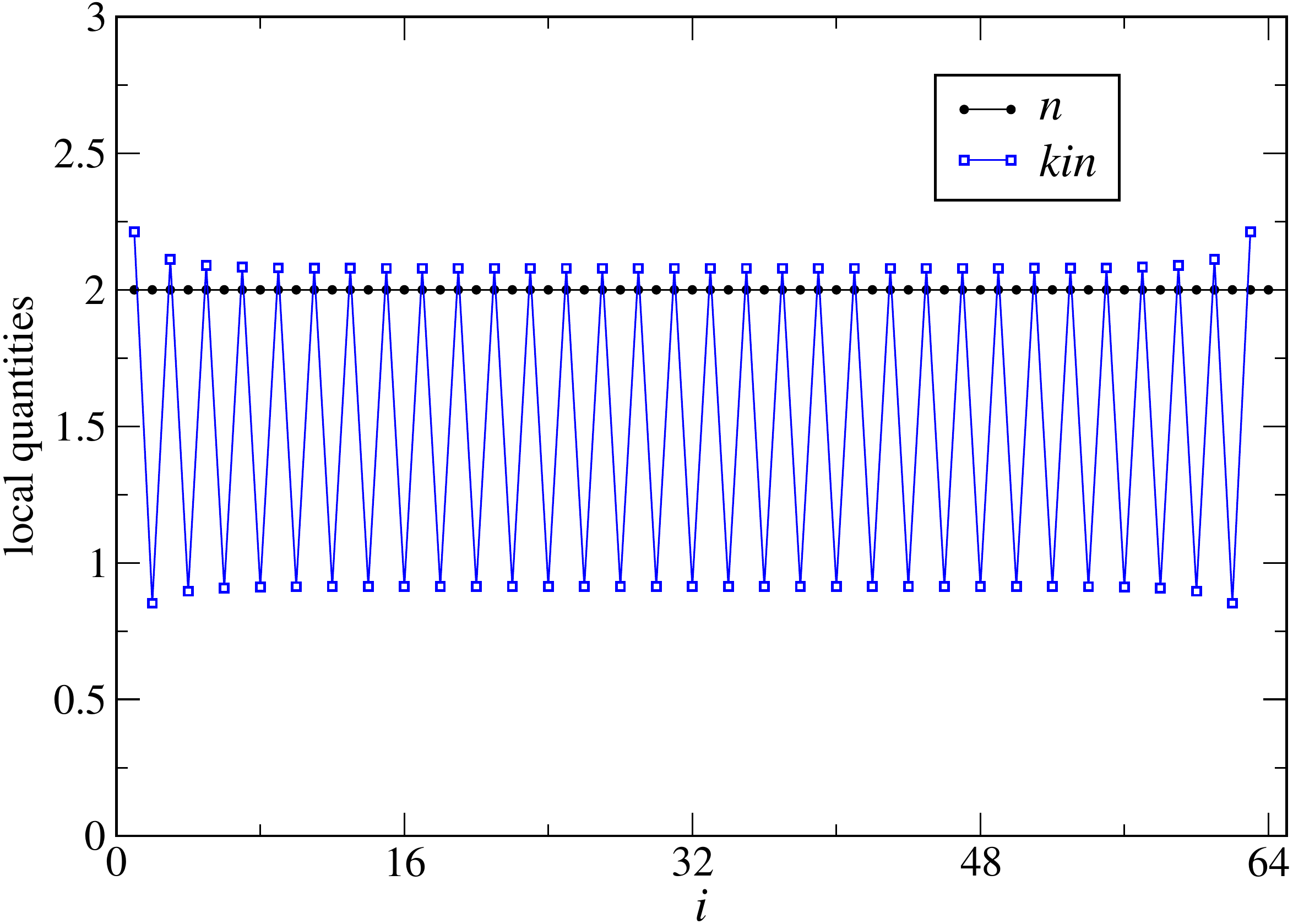}
}
\caption{(Color online) 
Local density $n_i$ and kinetic bond energy for the fermionic SU(4) model with two particles per site 
at $U/t=-8$ (left, data are only shown in the bulk) and $U/t=8$ (right) on a $L=64$ chain 
computed by DMRG. Data strongly indicate CDW and SP phase respectively. 
\label{fig:SU4}}
\end{center}
\end{figure}

\begin{figure}[htb]
\begin{center}
\includegraphics[scale=1.2]{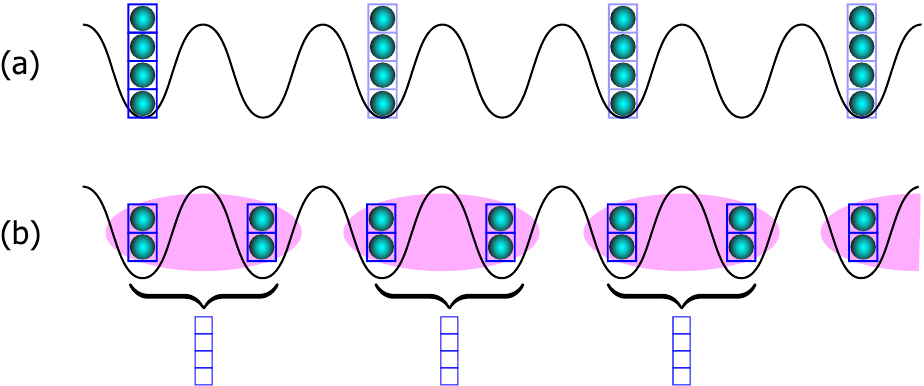}
\caption{(Color online) (a) Charge density wave (CDW) and (b) spin-Peierls (SP) phases of SU(4) Hubbard model.    
In both SU(4)-singlet phases, translation symmetry is broken.  
In CDW [(a)], $N$-mers ($N=4$, here), that are each SU($N$)-singlet, form (period-2) crystalline structures.   
In SP [(b)], a six-dimensional representation 
($\mathbf{6}$) is formed at each site and then it is combined with another $\mathbf{6}$ 
on the adjacent site to form an SU(4)-singlet. 
\label{fig:SU4-SP-CDW}}
\end{center}
\end{figure}

In the attractive case ($U<0$), 
long-range ordering of period-2 CDW emerges \cite{Zhao-U-W-06,Zhao-U-W-07,Nonne-L-C-R-B-11}.  
In the strong-coupling region ($|U| \gg t$), the picture of the CDW formation is simple; 
SU($N$)-singlet molecules of $N$ atoms ($N$-mers) are formed first 
and then the preformed $N$-mers organize themselves into period-2 crystalline structures 
in such a way that they optimize the $O(t^{2}/|U|)$ repulsive 
interaction generated by virtual hopping \cite{Zhao-U-W-06,Zhao-U-W-07}.   
To illustrate the crystalline pattern of $N$-mers, we show, in the left panel of Fig.~\ref{fig:SU4},  
the spatial profile of physical quantities 
(local fermion density and kinetic-energy density) obtained by DMRG for the half-filled SU(4) Hubbard 
model at $U/t= -8$.  
The local fermion density clearly shows period-two oscillation indicative of CDW of tetramers 
(note that the maxima of the density is close to 4), while the kinetic energy does not exhibit any special feature.  
A typical wave function of the CDW phase is shown in Fig. \ref{fig:SU4-SP-CDW}(a).

When $U>0$, on the other hand, it has been shown that a gapful dimerized [or spin-Peierls (SP)] phase 
with bond ordering [see Fig.~\ref{fig:SU4-SP-CDW}(b) for an intuitive picture of the ground state]  
appears upon switching on the weak repulsive interaction~\cite{Buchta2007,Nonne-L-C-R-B-11}.
In this respect, the situation is very different from the $N=2$ case where, at half-filling (one atom per site),  
a Mott-insulating phase with gapless spin excitations is stabilized when $U>0$. 
It has been shown numerically (by means of 
QMC and DMRG) for $N=4$ that there is adiabatic continuity from weak to strong coupling and that 
the SP phase occurs for all $U>0$  \cite{Nonne-L-C-R-B-11,Assaraf-A-B-C-L-04}. 
In the large-$U$ limit, the existence of this SP phase can be simply understood from the fact that
the half-filled SU(4) Fermi-Hubbard model reduces to the SU(4) Heisenberg
spin chain in the antisymmetric self-conjugate representation (${\tiny \yng(1,1)}$) of SU(4); 
the latter model is known to display, at zero-temperature, a dimerized phase with two-fold ground-state degeneracy  \cite{Marston-A-89,OnufrievM99,Paramekanti2007}.  
In the right panel of Fig.~\ref{fig:SU4}, we also present the plot of the local fermion density and the kinetic-energy density 
for the SP phase obtained by direct DMRG simulations of the SU(4) Hubbard model at $U/t=8$.   
As can be clearly seen, there is a period-two oscillation in the profile of the kinetic energy, 
whereas the local-density profile is completely flat.\footnote{
Clearly, this mechanism does not work in the case of $N=\text{odd}$ where a simple uniform Mott insulator 
with $N/2$ particles at each site is impossible.  However, preliminary DMRG simulations showed that 
we still have an SU($N$)-singlet dimerized phase with uniform charge distribution.}

\subsubsection{Other commensurate fillings}
\label{sec:other-commensurate-fillings}
\label{sec:commensurate}
The nature of SU($N$) Mott-insulating phases for general commensurate fillings $n = p/q$ ($p$ and $q$ 
being relatively primes) has been investigated with combined use of bosonization and DMRG 
simulations \cite{Szirmai-L-S-08}.
For $N=2$, there is no Mott transition and the charge sector remains gapless for any commensurate fillings 
other than half-filling \cite{Giamarchi-book} . 

\begin{figure}[!htb]
\begin{center}
\includegraphics[width=0.7\textwidth,clip]{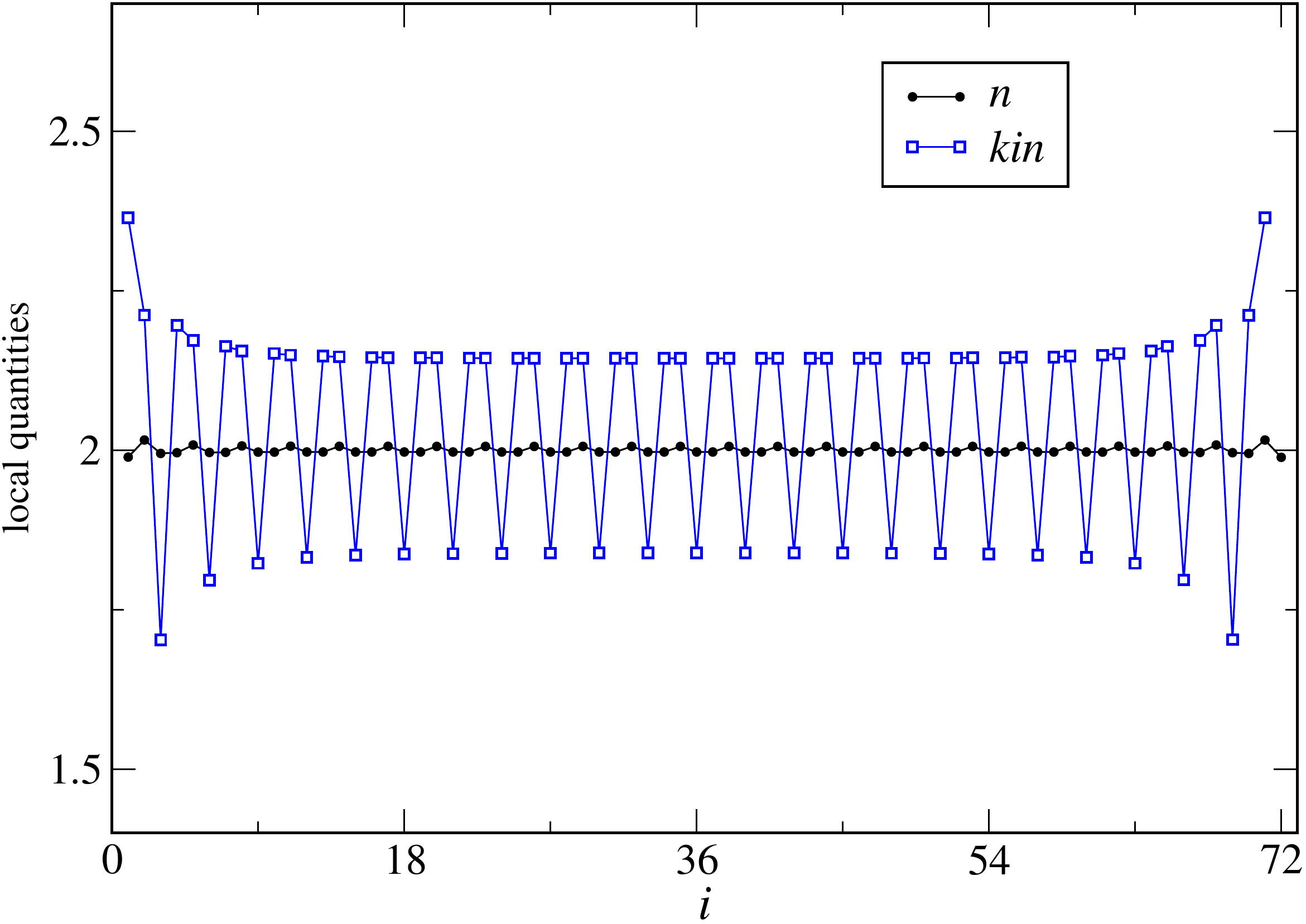}
\caption{(Color online) Local density $n_i$ and kinetic bond energy for the fermionic SU(6) model with two particles per site at $U/t=8$ on a $L=72$ chain computed by DMRG. There is a strong trimerization pattern with a three-periodicity.
\label{fig:SU6}}
\end{center}
\end{figure}

Again, for $N>2$, the physics turns out to be much richer.
If $q>N$, umklapp processes are irrelevant and a metallic $N$-component Luttinger-liquid phase
is stabilized, which has $N$ gapless degrees of freedom [1 for charge and $N-1$ for SU($N$) spin] 
and hence the central charge $c=N$.
When $q=N$, spin-charge separation occurs and a charge gap opens for finite $U>0$;    
a gapless Mott-insulating phase emerges with $N-1$ bosonic modes.  
The physics is then quite similar to the one discussed above (Sec~.\ref{sec:sutherland}) 
for the Sutherland model for the $1/N$-filling (i.e., one atom per site).  
Last, when $q < N$, umklapp processes are strongly relevant and couple the charge and the spin 
degrees of freedom.  As a consequence,
fully gapped Mott-insulating phases are formed which spontaneously break the one-site translation symmetry. In particular, 
bond-ordered (dimerized, trimerized, or tetramerized) phases are found depending on the filling  \cite{Szirmai-L-S-08}.
For instance, a trimerized phase with a three-fold ground-state degeneracy 
can be stabilized in the SU(6) Hubbard model with two particles per site ($n=1/3$).  In Fig.~\ref{fig:SU6} we provide complementary DMRG data showing the emergence of this phase. 

For filling $n=m/N$ ($m=1,\ldots,N-1$), a Mott insulator with $m$ atoms per site is formed 
in the large-$U$ limit.  
Then, one can perform a strong-coupling expansion  to derive an effective spin model 
for the remaining SU($N$) low-energy degrees of freedom.  
The resulting magnet takes the form of the SU($N$) antiferromagnetic
Heisenberg chain with the Hamiltonian \cite{Affleck-88}:
\begin{equation}
{\cal H} = J \sum_i \sum_{A=1}^{N^2-1} {\cal S}^{A}_{i} {\cal S}^{A}_{i+1} ,
\label{sunheisenbergantisym}
\end{equation}
where ${\cal S}^{A}_{i}$ is the SU($N$) spin operators at site $i$ which transform 
in the antisymmetric $m$-tensor representation of SU($N$):
\begin{equation}
\text{\scriptsize $m$} \left\{ 
\yng(1,1,1,1)
 \right.  \; .
\end{equation}
 For $m=1$ ($N$-dimensional fundamental representation) and $m=N-1$ (its conjugate), 
 one recovers the Sutherland model (\ref{sutherland}) with
a gapless behavior described by SU($N$)$_1$ WZW CFT \cite{Affleck-88}.  
The physical properties of the model (\ref{sunheisenbergantisym}) for other values of $m$
have been investigated by a CFT approach \cite{Affleck-88,lecheminant-T2015} and variational QMC calculations
\cite{Dufour2015}; when $m$ and $N$ have no common divisor, a gapless SU($N$)$_1$ WZW quantum 
criticality emerges.\footnote{%
From the effective-field-theory point of view, 
whether $N$ and $m$ have a common divisor or not affects the selection rule for the relevant perturbations 
allowed at the level-1 SU($N$) WZW fixed point and thereby governs the ground-state properties \cite{Affleck-88}.} 
If $N$ is divisible by $m$ (i.e., $N = mp$), on the other hand, 
either a fully gapped bond-ordered phase with $p$-fold ground-state 
degeneracy or the SU($N$)$_1$  quantum critical phase realizes 
depending on whether the underlying umklapp operator is relevant or (marginally) irrelevant \cite{Affleck-88,Dufour2015}.
The latter case corresponds to situations when $N(m-1) \ge 2 m^2$  \cite{Affleck-88}.
These predictions have been checked with a great accuracy recently by variational QMC calculations \cite{Dufour2015}.  
When $m$ and $N$ have a common divisor, one observes that these results are not in 
agreement with the SU($N$) generalization of the Haldane conjecture proposed in Ref. \cite{Rachel-T-F-S-G-09} 
(see Sec.~\ref{sec:LSM} for more details of the conjecture). 
\begin{table}[!htb]
\begin{center}
\begin{tabular}{lcc}
\hline\hline
filling $n$  &  $N=2$ & $N \geq 3$ \\
\hline
incommensurate & $\text{C}1\text{S}1$ 
& 
\begin{minipage}{15em}
$\text{C}1\text{S}(N-1)$ for $U>0$,  \\
$\text{C}1\text{S}0$ for $U<0$ (MS or CDW)
\end{minipage}
\\
\hline
$1/N$ & 
\begin{minipage}{8em}
$\text{C}0\text{S}1$ for $U>0$, \\  
$\text{C}1\text{S}0$ for $U<0$  
\end{minipage}
& 
\begin{minipage}{15em}
$\text{C}1\text{S}(N-1)$ for $0<U\leq U_{\text{c}}$, \\ 
$\text{C}0\text{S}(N-1)$ for $U_{\text{c}}< U$ 
\end{minipage}
\\
\hline
$1/2$ (half-filling) & same as $n=1/N$ 
& 
\begin{minipage}{15em} 
$\text{C}0\text{S}0$ 
[dimerized (SP) for $U>0$,  \\
period-2 CDW for $U<0$]
\end{minipage}
\\
\hline
$m/N$ ($m=1,\ldots,N-1$)& 
same as $n=1/N$ 
& 
\begin{minipage}{15em}
$\text{C}0\text{S}(N-1)$ or $p$-merization \\
when $N=mp$, \\ 
$\text{C}0\text{S}(N-1)$
when $N$ and $m$ coprime 
\end{minipage}
\\
\hline
\begin{minipage}{10em} 
generic $p/q$ \\ 
($p$, $q$: coprime) 
\end{minipage}
& $\text{C}1\text{S}1$
&
\begin{minipage}{15em} 
$\text{C}1\text{S}(N-1)$ when $q>N$, \\
$\text{C}0\text{S}(N-1)$ when $q=N$ ($U>U_{\text{c}}$),  \\
$\text{C}0\text{S}0$ with broken translation \\
or $\text{C}0\text{S}(N-1)$ when $q<N$
\end{minipage}
 \\
\hline\hline
\end{tabular}
\caption{Phases of single-band SU($N$) Hubbard chain \eqref{HubbardSUN} for various fillings. 
The notation $\text{C}m_{\text{c}}\text{S}m_{\text{s}}$ ($m_{\text{c}}=0,1$, $m_{\text{s}}=0,\ldots,N-1$) denotes 
a phase with $m_{\text{c}}$ ($m_{\text{s}}$) gapless 
charge [SU($N$)] degrees of freedom.  
For incommensurate fillings, we have only one gapless charge degree of freedom ($\text{C}1\text{S}0$) 
and leading instability is molecular superfluid (MS) of $N$-mers at low densities or $2k_{\text{F}}$-CDW 
for higher densities (see, e.g., Fig.~\ref{fig:SU3Kc}).  
The low-energy physics of the phase $\text{C}0\text{S}(N-1)$ is described by 
the Sutherland model \eqref{sutherland} or level-1 SU($N$) WZW CFT.  
The `SP' and `period-2 CDW' states at half-filling are illustrated in Fig.~\ref{fig:SU4-SP-CDW}. %
\label{tab:phases-SUN}}
\end{center}
\end{table}

\section{Two-orbital fermionic Hubbard models}
\label{sec:twoband}
So far, we have described the phases of the SU($N$) fermions that appear when 
no additional degrees of freedom (e.g., the ground and the excited atomic states 
of alkaline-earth fermionic atoms) are taken into account.   
However, as we have mentioned in Sec.~\ref{sec:introduction}, we can incorporate 
an additional degrees of freedom (``orbital'') into the system 
by taking into account another set of states with SU($N$) symmetry 
(e.g., the metastable ${}^{3}P_{0}$ state `$e$' of alkaline-earth atoms).  
Due to the interplay of SU($N$) and orbital, we have much richer phase diagrams.  
In this section, we consider the one-dimensional SU($N$) fermions 
having {\em two} orbital degrees of freedom and map out the ground-state phases.  
\subsection{Various physical realizations}
In this review, we describe two different ways to introduce additional ``orbital'' degree of freedom. 
One is to use two atomic states, the ground state `$g$' (${}^{1}S_{0}$) and a metastable excited state 
`$e$' (${}^{3}P_{0}$) of the alkaline-earth fermions\cite{Gorshkov-et-al-10}.   
Although we use the terminology `orbital' here, it is in fact related to an additional internal 
degree of freedom ($g$ and $e$) and has nothing to do with the real orbital.  The highly suppressed internal conversion 
${}^{1}S_{0} \leftrightarrow {}^{3}P_{0}$ guarantees separate conservation of the number of fermions 
in the $g$ and $e$ states leading to nearly perfect U(1) symmetry in the orbital sector.  

The other uses the two degenerate $p$-bands of a 1D optical lattice. 
Let us consider a 1D optical lattice (running in the $z$-direction) 
with moderate strength of the harmonic confining potential 
$V_{\perp}(x,y)=\frac{1}{2}m\omega_{xy}^{2}(x^{2}+y^{2})$ in the direction (i.e. $xy$) 
perpendicular to the chain.  The two degenerate $p$-bands are formed by the two degenerate 
first excited states of the above two-dimensional harmonic oscillator and we use them to introduce 
the orbital degree of freedom.  In this scheme, the two orbital states originate from the 
two symmetry-related excited states with specific spatial structures.  For this reason, 
we obtain a slightly different effective Hamiltonian for the latter case (in particular, we have much less 
degree of freedom in the effective Hamiltonian).    
\subsubsection{$g$-$e$ model}
\label{sec:g-e-model}
Since both $g$ ($^{1}S_{0}$) and $e$ ($^{3}P_{0}$) states have vanishing total electron angular momentum, 
the same mechanism as in Sec.~\ref{sec:SUN-in-alkaline-earth} leads to the SU($N$)-symmetry 
in the two-body scattering processes (i.e., for all combinations $g$-$g$, $e$-$e$, and $g$-$e$). 
Even if the scattering lengths for the two atoms in a given fixed combination do not depend on the total $f$, 
they may be different for different combinations of the atomic states (e.g., $g$-$g$ and $e$-$e$).  
In general, the scattering length may differ for the following four combinations of the two colliding particles:
\begin{equation}
\begin{split}
& |gg\rangle \equiv |g\rangle_{1} |g\rangle_{2} \; , \;\;
|ee\rangle \equiv |e\rangle_{1} |e\rangle_{2}  \; , \;\; \\
& |ge^{\pm}\rangle \equiv  \frac{1}{\sqrt{2}}(|g\rangle_{1} |e\rangle_{2}\pm |e\rangle_{1}|g\rangle_{2})
\;  .
\end{split}
\end{equation} 
Then, the strength of the interaction is determined by the four $s$-wave scattering lengths
$a_{X}$ ($X = gg$, $ee$, $ge^{+}$, $ge^{-}$) as
\begin{equation}
g_{X} = \frac{4\pi \hbar^{2}}{M} a_{X} \; .
\label{eqn:g-by-a-2}
\end{equation}
The scattering lengths $a_{gg}$, 
$a_{ee}$ and $a_{ge}^{\pm}$ are for two atoms in the electronic states $|gg\rangle$, 
$|ee\rangle$ and $|ge^{\pm}\rangle$, respectively. 
The known values of the scattering length $a_X$ for ${}^{173}\text{Yb}$ and ${}^{87}\text{Sr}$ 
are summarized in Table~\ref{tab:scat-length}.  In stark contrast to alkali-metal atoms, 
magnetic Feshbach resonance cannot be used, due to the vanishing electron spin, 
to tune these scattering lengths in alkaline-earth atoms.  
However, quite interestingly, orbital Feshbach resonances have been proposed theoretically~\cite{Zhang2015} 
and observed experimentally in Yb~\cite{Hoefer2015,Pagano2015}, thus allowing 
to change $a_{ge}^{\pm}$ with a magnetic field. 

\begin{table}[!htb]
\begin{center}
\begin{tabular}{lll}
\hline\hline
  &  ${}^{173}\text{Yb}$ $(I=5/2)$ & ${}^{87}\text{Sr}$ $(I=9/2)$ \\
\hline
$a_{gg}$ & $10.55\text{ [nm]}$ (Ref.~\cite{Kitagawa-et-al-PRA-08}) 
& $5.09\text{ [nm]}$ (Refs.~\cite{Escobar-et-al-08,Stein-K-T-10}) \\
$a_{ee}$ & $16.2 \pm 0.55 \text{ [nm]}$ (Ref.~\cite{Scazza-et-al-14}) & 
$9.31 \pm 0.58 \text{ [nm]}$ (Ref.~\cite{Zhang-et-al-14})  \\
$a_{ge}^{+}$ & 
$
\begin{cases}
 175 \pm 16 \text{ [nm]} \;\; & (\text{Ref.~\cite{Cappellini-et-al-14}})\\
  115 \pm 10 \text{ [nm]} \;\; & (\text{Ref.~\cite{Scazza-et-al-14}})  \\
  99 \pm 2 \text{ [nm]} \;\; & (\text{Ref.~\cite{Hoefer2015}})   \; 
\end{cases} 
$
& $8.94 \pm 0.42 \text{ [nm]}$ (Ref.~\cite{Zhang-et-al-14}) \\
$a_{ge}^{-} $ 
&
$11.6 \pm 0.1 \text{ [nm]}$ (Refs.~\cite{Scazza-et-al-14,Hoefer2015})
&
$3.60 \pm 1.16 \text{ [nm]}$ (Ref.~\cite{Zhang-et-al-14}) \\
$a_{ge}^{+}+a_{ge}^{-} $  & $111\text{ [nm]}$ (Ref.~\cite{Pagano2015}) 
&  \\
\hline\hline
\end{tabular}
\caption{Values of scattering length $a_X$ ($X = gg$, $ee$, $ge^{+}$, $ge^{-}$) known from 
experiments.%
\label{tab:scat-length}}
\end{center}
\end{table}

When the two-body interaction $\hat{V}$ is independent of the nuclear spin (this is the case 
to a good approximation in alkaline-earth atoms), 
the most general form of $\hat{V}$ may be given by the following contact interaction:
\begin{equation}
\begin{split}
& \hat{V}(\bolr-\bolr^{\prime}) \\
& = 
\left\{
g_{gg}|gg\rangle\langle gg| + g_{ee}|ee\rangle\langle ee|
+ g_{ge}^{+}|ge^{+}\rangle\langle ge^{+}| + g_{ge}^{-}|ge^{-}\rangle\langle ge^{-}|
\right\} \delta(\bolr-\bolr^{\prime}) \; .
\end{split}
\label{eqn:orbital-part-g-e}
\end{equation}
Note that all the four couplings 
$g_{gg}$, $g_{ee}$, and $g^{\pm}_{ge}$ are independent of the nuclear-spin states of 
the colliding atoms (see the discussion of Sec.~\ref{sec:SUN-in-alkaline-earth}).  
From \eqref{eqn:orbital-part-g-e}, the orbital-dependent part of the matrix elements is calculated easily:
\begin{equation}
\begin{split}
\langle a,b|\hat{V}|m,n\rangle = &
\biggr\{
g_{gg} \delta_{a,g}\delta_{b,g}\delta_{m,g}\delta_{n,g}
+ g_{ee} \delta_{a,e}\delta_{b,e}\delta_{m,e}\delta_{n,e}  \\
&  + \frac{1}{2} \left(g^{+}_{ge} + g^{-}_{ge}\right)
\left( \delta_{a,g}\delta_{b,e}\delta_{m,g}\delta_{n,e} 
+\delta_{a,e}\delta_{b,g}\delta_{m,e}\delta_{n,g} \right)  \\
&  + \frac{1}{2} \left(g^{+}_{ge} - g^{-}_{ge}\right) 
\left( \delta_{a,g}\delta_{b,e}\delta_{m,e}\delta_{n,g} 
- \delta_{a,e}\delta_{b,g}\delta_{m,g}\delta_{n,e} \right) 
\biggr\} \delta(\bolr - \bolr^{\prime})  \\
& \quad 
(a,b,m,n=e,g) \; .
\end{split}
\label{eqn:ge-2-body-scattering}
\end{equation}

The derivation of the lattice Hamiltonian for the case of two orbitals 
closely follows that described in Sec.~\ref{sec:deriv-1-band-SUN-Hubbard} for the single-band case 
except that now we have two species of fermions and use the corresponding Wannier functions: 
\begin{equation}
\begin{split}
& c_{m\alpha}(z)=
\sum_{R}\sum_{n\in\text{bands}}
w^{(n)}_{m,R}(z)c^{(n)}_{m\alpha,R}  \; , \quad  
c_{m\alpha}^{\dagger}(z)=
\sum_{R}\sum_{n}
w^{(n)\, \ast}_{m,R}(z)c^{(n)\, \dagger}_{m\alpha,R}  \\
& w^{(n)}_{m,R}(z) \equiv 
\frac{1}{\sqrt{N_{\text{cell}}}} \sum_{k_z} \be^{- i k_z R} \varphi^{(n)}_{m,k_z}(z)  
\quad (m=e,g; \; \alpha=1,\ldots,N)   \; .
\end{split}
\end{equation}
We plug these operators to rewrite the two-body interaction 
\begin{equation}
\frac{1}{2} \sum_{\alpha,\beta=1}^{N}\sum_{a,b,m,n=e,g}
\int\!dz \int\!dz^{\prime}
c^{\dagger}_{a\alpha}(z)c^{\dagger}_{b\beta}(z^{\prime}) 
\langle a,b|\hat{V}|m,n\rangle 
c_{n\beta}(z^{\prime})c_{m\alpha}(z)  \; .
\end{equation}
In contrast to the single-band case where we have obtained only the Hubbard-$U$ interactions,  
we now have the following four different types of interactions:
\begin{equation}
\begin{split}
& \frac{1}{2}\sum_{\alpha,\beta=1}^{N}\sum_{\{R_{i},n_i\}}
V_{gg} 
({}_{R_{1},R_{2};R_{3},R_{4}}^{n_1,n_2;n_3,n_4}) \, 
c^{(n_1)\dagger}_{g\alpha,R_1}c^{(n_2)\dagger}_{g\beta,R_2}
c^{(n_3)}_{g\beta,R_3}c^{(n_4)}_{g\alpha,R_4} \\
&  + \frac{1}{2}\sum_{\alpha,\beta=1}^{N}\sum_{\{R_{i},n_i\}}
V_{ee}
({}_{R_{1},R_{2};R_{3},R_{4}}^{n_1,n_2;n_3,n_4}) \, 
c^{(n_1)\dagger}_{e\alpha,R_1}c^{(n_2)\dagger}_{e\beta,R_2}
c^{(n_3)}_{e\beta,R_3}c^{(n_4)}_{e\alpha,R_4}  \\
&  + \frac{1}{2}\sum_{\alpha,\beta=1}^{N}\sum_{\{R_{i},n_i\}}
V^{+}_{ge}
({}_{R_{1},R_{2};R_{3},R_{4}}^{n_1,n_2;n_3,n_4}) \, 
c^{(n_1)\dagger}_{g\alpha,R_1}c^{(n_2)\dagger}_{e\beta,R_2}
c^{(n_3)}_{e\beta,R_3}c^{(n_4)}_{g\alpha,R_4}  \\
& + \frac{1}{2}\sum_{\alpha,\beta=1}^{N}\sum_{\{R_{i},n_i\}}
V^{-}_{ge}
({}_{R_{1},R_{2};R_{3},R_{4}}^{n_1,n_2;n_3,n_4})\, 
c^{(n_1)\dagger}_{g\alpha,R_1}c^{(n_2)\dagger}_{e\beta,R_2}
c^{(n_3)}_{g\beta,R_3}c^{(n_4)}_{e\alpha,R_4}  \; .
\end{split}
\label{eqn:ineraction-g-e-scheme}
\end{equation}
In the above, the interactions are given in terms of the Wannier basis as
\begin{equation}
\begin{split}
& V_{mm} 
({}_{R_{1},R_{2};R_{3},R_{4}}^{n_1,n_2;n_3,n_4}) 
\equiv 
g_{aa} \int\!dz \, 
w^{(n_1)\ast}_{a,R_1}(z) w^{(n_2)\ast}_{a,R_2}(z)
w^{(n_3)}_{a,R_3}(z)w^{(n_4)}_{a,R_4}(z) \quad 
(m=g,e) \\
& V^{+}_{ge}
({}_{R_{1},R_{2};R_{3},R_{4}}^{n_1,n_2;n_3,n_4}) 
\equiv 
(g_{ge}^{+} + g_{ge}^{-})
\int\!dz \, 
w^{(n_1)\ast}_{g,R_1}(z)w^{(n_2)\ast}_{e,R_2}(z)
w^{(n_3)}_{e,R_3}(z)w^{(n_4)}_{g,R_4}(z) \\
& V^{-}_{ge}
({}_{R_{1},R_{2};R_{3},R_{4}}^{n_1,n_2;n_3,n_4}) 
\equiv 
(g_{ge}^{+} - g_{ge}^{-})
\int\!dz \, 
w^{(n_1)\ast}_{g,R_1}(z)w^{(n_2)\ast}_{e,R_2}(z)
w^{(n_3)}_{g,R_3}(z)w^{(n_4)}_{e,R_4}(z) ,
\end{split}
\label{eqn:V-by-Wannier}
\end{equation}
and the coupling constants $g_{gg}$, $g_{ee}$, and $g_{ge}^{\pm}$ 
are calculated from the scattering lengths as Eq.~\eqref{eqn:g-by-a-2}.  
The first two terms in Eq.~\eqref{eqn:ineraction-g-e-scheme} describe the density-density interactions 
between the fermions in the same orbital, while 
the third and the fourth ones correspond to fermions from {\em different} orbitals. 
Specifically, the third one is just the density-density interaction of  a pair of fermions 
on different orbitals and the last one is the orbital exchange interaction (or, the Hund coupling).   
Here it should be noted that there is no special relation among the four couplings 
$V_{gg}$, $V_{ee}$, $V^{\pm}_{ge}$ as the Wannier functions for the two orbitals 
$w^{(n)}_{g/e,R_{i}}(z)$ are {\em not} symmetry-related to each other\footnote{This is not the case
for the $p$-band model where the two orbitals are related to each other by $C_{4}$-symmetry.}.     
As before, we restrict ourselves only to the same band $n_1=n_2=n_3=n_4=n_0$ and keep only the onsite 
terms $R_{1}=R_{2}=R_{3}=R_{4}= i$ 
to obtain the following Hamiltonian ($g$-$e$ model)~\cite{Gorshkov-et-al-10}: 
\begin{equation}
\begin{split}
 \mathcal{H}_{g\text{-}e} =& 
  - \sum_{m=g,e}  t_{m} \sum_{i} \sum_{\alpha=1}^{N} 
   \left(c_{m\alpha,\,i}^\dag c_{m\alpha,\,i+1}  + \text{H.c.}\right) \\
&  -\sum_{m=g,e}\mu^{(m)} \sum_i n_{m,i}  
 + \sum_{m=g,e} \frac{U_{mm}}{2} \sum_{i} n_{m,\,i}(n_{m,\,i}-1) \\
&  +V \sum_i n_{g,\,i} n_{e,\,i} 
  + V_{\text{ex}}^{g\text{-}e} \sum_{i,\alpha \beta} 
  c_{g\alpha,\,i}^\dag c_{e\beta,\,i}^\dag 
  c_{g\beta ,\,i} c_{e\alpha,\,i} ,
  \end{split}
\label{eqn:Gorshkov-Ham}
\end{equation} 
where the index $\alpha$ labels the nuclear-spin multiplet 
and the orbital indices $m=g$ and $e$ label the two atomic states 
${}^{1}S_{0}$ and ${}^{3}P_{0}$, respectively.  
We have also introduced the number of the fermion $m$ at each site 
\begin{equation}
n_{m,i} \equiv \sum_{\alpha=1}^{N}c^{\dagger}_{m\alpha,i}c_{m\alpha,i} \quad (m=g,e) 
\end{equation}
and suppressed the common band index $n_0$.  
The coupling constants are given in terms of $V_{gg}$, $V_{ee}$, and $V^{\pm}_{ge}$ as 
[see Eq.~\eqref{eqn:V-by-Wannier}]
\begin{subequations}
\begin{equation}
U_{mm} = V_{mm} 
({}_{R_i,R_i;R_i,R_i}^{n_0,n_0;n_0,n_0}) , \;\;
V = V^{+}_{ge}
({}_{R_i,R_i;R_i,R_i}^{n_0,n_0;n_0,n_0}) , \;\;
V_{\text{ex}}^{g\text{-}e} = V^{-}_{ge}
({}_{R_i,R_i;R_i,R_i}^{n_0,n_0;n_0,n_0})  \; ,
\end{equation}
while the hopping $t_{m}$ and the chemical potential $\mu^{(m)}$ are 
given by the Bloch energy as
\begin{equation}
\begin{split}
& t_{m} \equiv t_{m}(1) \, , \;\; \mu^{(m)} \equiv t_{m}(0) \\
& t_{m}(j - j^{\prime}) \equiv 
- \frac{1}{N_{\text{cell}}}  \sum_{k_z}
\varepsilon_{m}^{(n)}(k_z)\be^{ i k_{z}(j - j^{\prime})} 
\quad (m=g,e) \; .
\end{split}
\end{equation}
\end{subequations}

\begin{figure}[!htb]
\begin{center}
\includegraphics[scale=0.6]{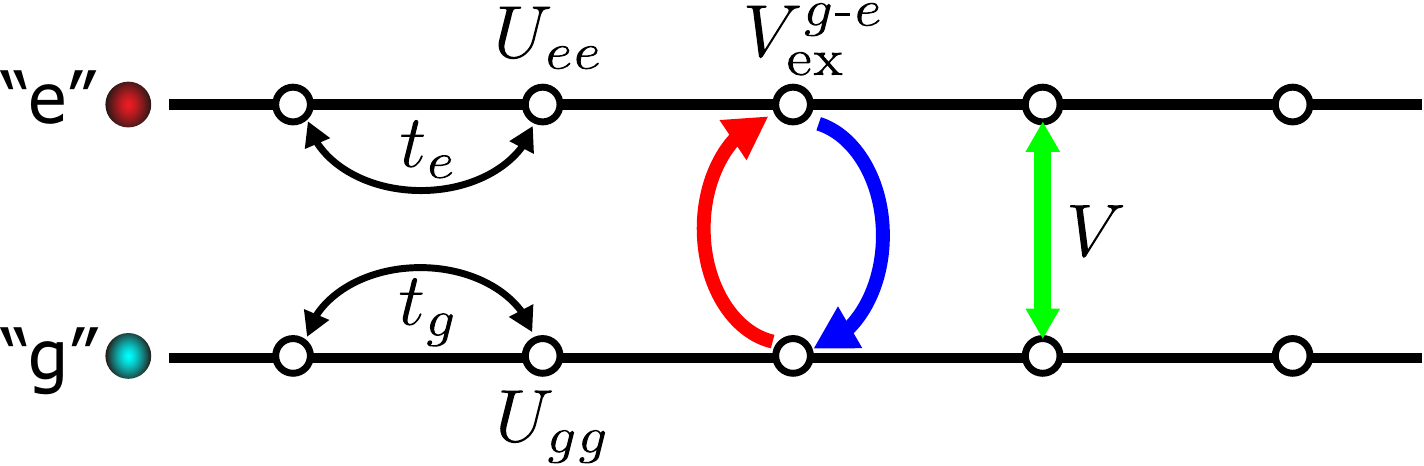}
\caption{(Color online) Two-leg ladder representation of the $g$-$e$ Hamiltonian \eqref{eqn:Gorshkov-Ham}.     
\label{fig:alkaline-2leg}}
\end{center}
\end{figure}

In order to understand the physical processes contained in this Hamiltonian, it is helpful to 
represent it as two coupled (single-band) SU($N$) Hubbard chains (see Fig.~\ref{fig:alkaline-2leg}). 
On each chain, we have the standard hopping $t_{m}$ {\em along} each chain and the Hubbard-type 
interaction $U_{mm}$, and the two are coupled to each other by the nearest-neighbor 
Coulomb interaction $V$ and the $g$-$e$ exchange process $V_{\text{ex}}^{g\text{-}e}$.  
The `hopping' between the chains does not exist as the transition $g \leftrightarrow e$ is strongly 
suppressed.  
Therefore, on top of the obvious SU($N$)-symmetry, the Hamiltonian is invariant under the following U(1)-symmetry
\begin{equation}
c_{g\alpha,\,i} \mapsto \be^{i \theta_{\text{o}}} c_{g\alpha,\,i} \, , \; 
c_{e\alpha,\,i} \mapsto \be^{-i \theta_{\text{o}}} c_{e\alpha,\,i}  \; .
\label{eqn:orbital-U1}
\end{equation}
This is a consequence of the fact that the total fermion numbers for $g$ and $e$ are 
conserved {\em separately}.\footnote{%
This breaks down when there is transition (i.e., `hopping') between the two atomic states $g$ and $e$.}  

To understand the global phase structure, it is useful to rewrite the exchange interaction 
$V_{\text{ex}}^{g\text{-}e}$ in two different ways.  
First, we introduce the second-quantized SU($N$) generators of each orbital as
\begin{equation}
\hat{\mathcal{S}}_{m,i}^{A} 
= \sum_{\alpha,\beta=1}^{N}c^{\dagger}_{m\alpha,i}(\mathcal{S}^{A})_{\alpha\beta}c_{m\beta,i}  \quad (m=g,e) \; .
\end{equation} 
If we normalize the SU($N$) generators $\{S^{A}\}$ as\footnote{%
This corresponds to, e.g., using the SU(2) generators $\sigma^{a}/\sqrt{2}$ ($a=x,y,z$) instead of 
the standard ones $\sigma^{a}/2$.}
\begin{equation}
\text{Tr}\,(\mathcal{S}^{A}\mathcal{S}^{B}) = \delta^{AB} \; ,
\label{eqn:normalization-SUN-gen}
\end{equation}
they satisfy the following identity:
\begin{equation}
\sum_{A=1}^{N^{2}-1} (\mathcal{S}^A)_{\alpha\beta} (\mathcal{S}^A)_{\gamma \delta}  =  \left(
\delta_{\alpha \delta} \delta_{\beta \gamma} - \frac{1}{N} 
\delta_{\alpha \beta} \delta_{\gamma \delta} \right)  \; .
\label{eqn:SASA-tensor}
\end{equation}
Then, it is straightforward to show that the orbital-exchange interaction $V_{\text{ex}}^{g\text{-}e}$ 
can be written as the `Hund coupling' between the SU($N$) `spins' of the two orbitals
\begin{equation}
\sum_{i,\alpha \beta} 
  c_{g\alpha,\,i}^\dag c_{e\beta,\,i}^\dag 
  c_{g\beta ,\,i} c_{e\alpha,\,i} 
= - \sum_{i}\left( 
\sum_{A=1}^{N^{2}-1}\hat{\mathcal{S}}_{g,i}^{A}\hat{\mathcal{S}}_{e,i}^{A} 
\right)
- \frac{1}{N} \sum_{i}n_{g,i}n_{e,i} \; .
\label{eqn:exchange-to-Hund}
\end{equation}
The fermionic anti-commutation is crucial in obtaining the minus sign in front of 
the Hund coupling.  
The above expression enables us to rewrite the original $g$-$e$ Hamiltonian \eqref{eqn:Gorshkov-Ham} 
as \cite{Bois-C-L-M-T-15}
\begin{equation}
\begin{split}
  \mathcal{H}_{g\text{-}e} =& 
  -  \sum_{i}\sum_{m=g,e} t_{m} \sum_{\alpha=1}^{N} 
   \left(c_{m\alpha,\,i}^\dag c_{m\alpha,\,i+1}  + \text{H.c.}\right) \\
&  -\sum_{m=g,e}\mu^{(m)} \sum_i n_{m,i}  
  +\sum_{i}\sum_{m=g,e} \frac{U_{mm}}{2} n_{m,\,i}(n_{m,\,i}-1)  \\
& +\left(V - \frac{1}{N}V_{\text{ex}}^{g\text{-}e}\right) \sum_i n_{g,\,i} n_{e,\,i} 
- V_{\text{ex}}^{g\text{-}e}
\underbrace{%
\sum_{i}\left( 
\sum_{A=1}^{N^{2}-1}\hat{\mathcal{S}}_{g,i}^{A}\hat{\mathcal{S}}_{e,i}^{A} 
\right) 
}_{\text{Hund}} \; .
\end{split}
\label{eqn:Gorshkov-Ham-Hund}
\end{equation} 
From this, one readily sees that positive $V_{\text{ex}}^{g\text{-}e}$ leads to 
{\em ferromagnetic} coupling between the two SU($N$) spins on the $g$ and $e$ orbitals 
thereby maximizing the SU($N$) `spin' at each site.  

To derive the third form, we first introduce the {\em orbital pseudo-spin}
\begin{equation}
\begin{split}
& \hat{T}_i^a 
= \frac{1}{2}\sum_{\alpha=1}^{N}\sum_{m,n=e,g} c_{m \alpha,\,i}^\dag \sigma^a_{m n} c_{n \alpha,\,i} 
\equiv \sum_{\alpha=1}^{N} \hat{T}_{\alpha,i}^{a}  \\
& (a=x,y,z; \; m,n=g,e; \; \sigma^a \textrm{ :Pauli matrices})  \; .
\end{split}
\label{eqn:1pseudospinoperator}
\end{equation}
Clearly, the U(1)-symmetry \eqref{eqn:orbital-U1} is generated by $\sum_{i}\hat{T}^{z}_{i}$ and, 
in the following, we call it U(1)$_{\text{o}}$.    
Using these pseudo-spin operators, we obtain the third form of the $g$-$e$ Hamiltonian \cite{Bois-C-L-M-T-15}
\begin{equation}
\begin{split}
  \mathcal{H}_{g\text{-}e} =& 
  -  \sum_{i}\sum_{m=g,e} t_{m} \sum_{\alpha=1}^{N} 
   \left(c_{m\alpha,\,i}^\dag c_{m\alpha,\,i+1}  + \text{H.c.}\right)  \\
&  -\sum_{m=g,e}\left( \mu^{(m)}+3V_{\text{ex}}^{g\text{-}e}/4 \right) \sum_i n_{m,i}   \\
&  +\sum_{i}\sum_{m=g,e} \frac{U_{mm}-V_{\text{ex}}^{g\text{-}e}/2}{2} n_{m,\,i}(n_{m,\,i}-1) \\
&  +\left(V + V_{\text{ex}}^{g\text{-}e}/2 \right) \sum_i n_{g,\,i} n_{e,\,i} 
  + V_{\text{ex}}^{g\text{-}e} 
\underbrace{%
\sum_{i}  ( \hat{\bolT}_{i})^{2} 
 }_{\text{Hund}}  \; .
  \end{split}
  \label{eqn:Gorshkov-Ham-Hund-2}
\end{equation} 
This expression, which is equivalent to \eqref{eqn:Gorshkov-Ham-Hund}, represents 
the exchange interaction $V_{\text{ex}}^{g\text{-}e}$ in terms of the orbital SU(2) pseudo-spin $\bolT$.  
It is important to note that the sign of $V_{\text{ex}}^{g\text{-}e}$ is opposite to the one in 
Eq.~\eqref{eqn:Gorshkov-Ham-Hund}; positive (negative) $V_{\text{ex}}^{g\text{-}e}$ tends to 
quench (maximize) the orbital pseudo-spin and maximize (quench) the SU($N$) spin.   
This dual nature of the orbital pseudo-spin $\bolT$ and the SU($N$) spin is the key to understand the structure 
of the phase diagram and we will come back to this point in Sec.~\ref{sec:low-energy-dof}.  
\begin{figure}[!htb]
\begin{center}
\includegraphics[scale=0.6]{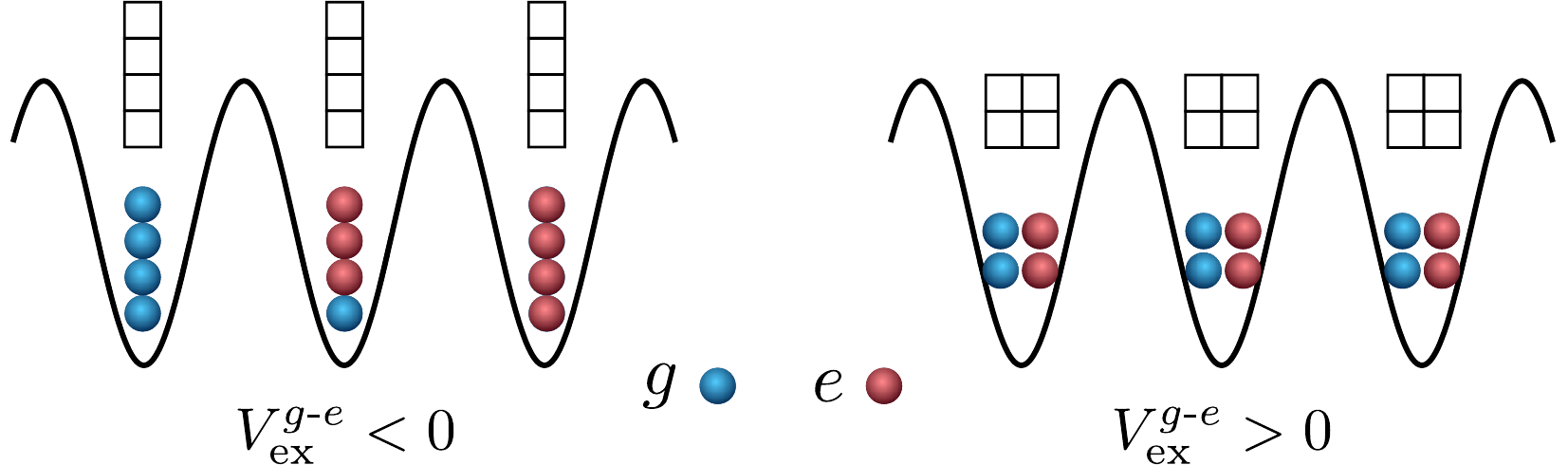}
\caption{(Color online) Typical strong-coupling ground states for SU(4) $g$-$e$ Hamiltonian 
\eqref{eqn:Gorshkov-Ham-Hund} with $t_m=0$. 
The SU($N$) `magnetic moment' that appears at each site in the Mott-insulating limit is shown by Young diagrams.  
For $V_{\text{ex}}^{g\text{-}e}<0$, SU($N$)-spin at each site is quenched 
[see Eq.~\eqref{eqn:Gorshkov-Ham-Hund}],  
while orbital pseudo-spin $\hat{\boldsymbol{T}}$ is quenched 
(and hence SU($N$) spin is maximized) for $V_{\text{ex}}^{g\text{-}e}>0$.   
\label{fig:strong-coupling-SU4}}
\end{center}
\end{figure}

Before concluding this subsection, let us give another useful form of the $g$-$e$ 
Hamiltonian \cite{Bois-C-L-M-T-15}: 
\begin{equation}
\begin{split}
\mathcal{H}_{g\text{-}e} = & - \, \sum_{i}\sum_{m=g,e} \sum_{\alpha=1}^{N} 
t_{m}  \left( c_{m\alpha,\,i}^\dag c_{m\alpha,\,i+1}+  \text{H.c.}  \right) \\
& -\frac{1}{2}\left(\mu_e + \mu_g\right) \sum_{i} n_{i} +\frac{U}{2} \sum_i  n_i^2 \\
&  +J \sum_i \left\{ (\hat{T}_i^x)^2 + (\hat{T}_i^y)^2\right\} + J_z \sum_i (\hat{T}_i^z)^2 
 - \left(\mu_g -\mu_e\right)\sum_{i} \hat{T}^{z}_{i}  \\
& + U_{\text{diff}} \sum_i  \hat{T}^{z}_{i} n_{i}
\; ,
\end{split}
\label{alkamodel-2b}
\end{equation}
where a set of the new coupling constants are given in terms of those in Eq.~\eqref{eqn:Gorshkov-Ham}
\begin{equation}
\begin{split}
& U=\frac{1}{4} (U_{gg}+ U_{ee} +2 V), \;\; 
U_{\text{diff}} = \frac{1}{2}(U_{gg} - U_{ee}) , \\
& J=V^{g\text{-}e}_{\text{ex}} , \;\;
J_{z} = \frac{1}{2}(U_{gg}+ U_{ee} -2 V), \\
& \mu_{g} = \frac{1}{2} (2 \mu^{(g)}+ U_{gg}+V^{g\text{-}e}_{\text{ex}}), \;\; 
\mu_{e} = \frac{1}{2} (2 \mu^{(e)}+ U_{ee} + V^{g\text{-}e}_{\text{ex}})  \; .
\end{split}
\label{eqn:Hund-by-Gorshkov}
\end{equation}
The model \eqref{alkamodel-2b} with $U_{gg}=U_{ee}$, $\mu^{(g)}=\mu^{(e)}$ is dubbed 
the generalized Hund model and has been studied extensively for $N=2$ in the cold-fermion 
context \cite{Nonne-B-C-L-10,Nonne-B-C-L-11}.  
It is obvious that, when $t_g=t_e$, $J=J_{z}=U_{\text{diff}}=0$, $\mu_{g}=\mu_{e}$, 
the Hamiltonian $\mathcal{H}_{g\text{-}e}$ is U($2N$)-invariant and the orbital part ($J$ and $J_{z}$) 
breaks it down to the generic symmetry 
$\text{U(1)}_{\text{c}}{\times}\text{SU($N$)}_{\text{s}}{\times}\text{U(1)}_{\text{o}}$:
\begin{equation}
\begin{split}
\text{U($2N$)}  & \xrightarrow[U_{\text{diff}}=0]{J= J_{z}(\neq 0)} 
\text{U(1)}_{\text{c}}{\times}\text{SU($N$)}_{\text{s}}{\times}\text{SU(2)}_{\text{o}} \\
& \xrightarrow[\text{or }U_{\text{diff}}\neq 0]{J\neq J_{z}} 
\text{U(1)}_{\text{c}}{\times}\text{SU($N$)}_{\text{s}}{\times}\text{U(1)}_{\text{o}} \;.
\end{split}
\label{eqn:symmetries-g-e}
\end{equation} 
In Sec.~\ref{sec:phasediagrams}, we use the form \eqref{alkamodel-2b} of the $g$-$e$ model 
and the parametrization 
$t_g=t_e$, $J=J_{z}=U_{\text{diff}}=0$, $\mu_{g}=\mu_{e}$ to map out its phases.  
\subsubsection{$p$-band model}
\label{sec:p-band-hamiltonian}
\begin{figure}[!htb]
\begin{center}
\includegraphics[scale=0.9]{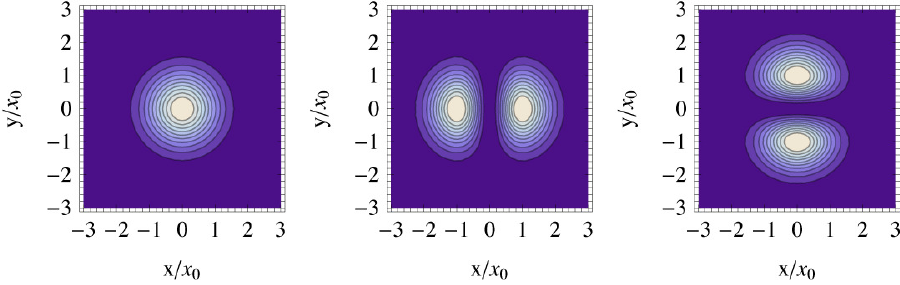}
\caption{(Color online) Contour plots of squared wave functions $| \phi_{n_x,n_y}|^{2}$ 
for three orbitals $(n_x,n_y)=(0,0)$, $(1,0)$ and $(0,1)$.   From Ref.~\cite{Bois-C-L-M-T-15}  
\label{fig:pxpy-orbitals}}
\end{center}
\end{figure}

In the second scheme, we use only the $g$-state (${}^{1}S_{1}$) and, to implement the orbital degree 
of freedom, introduce the two degenerate $p$-bands of a 1D optical lattice. 
Let us consider a 1D optical lattice (running in the $z$-direction) 
with moderate strength of (harmonic) confining potential $V_{\perp}(x,y)=\frac{1}{2}M\omega_{xy}^{2}(x^{2}+y^{2})$ 
in the direction (i.e. $xy$) 
perpendicular to the chain.  Then, the single-particle part of the Hamiltonian reads as
\begin{equation}
\begin{split}
\mathcal{H}_{0} &= \left\{
- \frac{\hbar^{2}}{2M}\partial_{z}^{2} + V_{\text{per}}(z) 
\right\}  +
\left\{ 
- \frac{\hbar^{2}}{2M}\left( \partial_{x}^{2} + \partial_{y}^{2} \right) 
 + V_{\perp}(x,y) 
 \right\}  \\
& \equiv \mathcal{H}_{/\!/}(z) + \mathcal{H}_{\perp}(x,y) 
 \; ,
 \end{split}
 \label{eqn:single-particle-Ham}
 \end{equation}
where $V_{\text{per}}(z)$ is a periodic potential that introduces a lattice structure along the chain 
 (i.e. $z$) direction.  
If the chain is infinite in the $z$-direction, the single-particle state is given by the following Bloch function:
\begin{equation} 
\psi^{(n)}_{n_x,n_y,k_z}(x,y,z) = \phi_{(n_x,n_y)}(x,y) \varphi^{(n)}_{k_z}(z) \; ,
\end{equation}
where the two functions $\varphi^{(n)}_{k_z}(z)$ and $\phi_{n_x,n_y}(x,y)$ respectively are 
the eigenfunctions of $\mathcal{H}_{/\!/}(z)$ and $\mathcal{H}_{\perp}(x,y)$ 
(see Fig.~\ref{fig:pxpy-orbitals}):
\begin{subequations}
\begin{align}
& \mathcal{H}_{/\!/}(z) \varphi^{(n)}_{k_z}(z) = \varepsilon^{(n)}(k_z) \varphi^{(n)}_{k_z}(z)  \\
& \mathcal{H}_{\perp}(x,y) \phi_{(n_x,n_y)}(x,y)  = \epsilon_{(n_x,n_y)}  \phi_{(n_x,n_y)}(x,y)  \; .
\label{eqn:Schroedinger-transverse-part}
\end{align}
with
\begin{equation}
\epsilon_{(n_x,n_y)} = \left( n_x + n_y + 1 \right) \hbar \omega_{xy}  \quad 
(n_x, n_y = 0,1,2, \ldots ) \; .
\end{equation}
\end{subequations}
This implies that due to the motion perpendicular to the chain, each one-dimensional Bloch band specified by $n$ 
splits into subbands labeled by $(n_x,n_y)$: 
\begin{equation}
E^{(n)}_{(n_x,n_y)}(k_z) =  \varepsilon^{(n)}(k_z) +  \epsilon_{(n_x,n_y)}  \; .
\label{eqn:en-subbands}
\end{equation} 
We call, for a given main band index $n$, 
the subbands with $(n_x,n_y)=(0,0)$, $(1,0)$, and $(0,1)$ as `$s$', `$p_x$' and `$p_y$', 
respectively.   

Now let us suppose that only the $n=0$ bands are occupied,  
and that, among them, the lowest one (the $s$-band) is completely filled.  
Then, it is legitimate to keep the next two bands $p_x$ and $p_y$ in the effective Hamiltonian 
in describing the low-energy physics \cite{Kobayashi-O-O-Y-M-12,Kobayashi-O-O-Y-M-14}.      
The argument goes almost in the same way except that now we have to take into account 
the motion in the perpendicular (i.e., $xy$) directions.     
Now the Wannier function
\begin{equation}
w^{(n)}_{(n_x,n_y);R}(x,y,z) \equiv 
\frac{1}{\sqrt{N_{\text{cell}}}} \phi_{(n_x,n_y)}(x,y) \sum_{k_z}
\be^{- i k_z R} \varphi^{(n)}_{k_z}(z)
\label{eqn:Wannier-pxpy}
\end{equation}
($R$ labels the center of the Wannier function and $N_{\text{cell}}$ is the number of unit cells) 
is used instead to expand the Fermi operators:
\begin{equation}
c_{\alpha}(\bolr) 
= \sum_{R}\sum_{n=\text{bands}}\sum_{(n_x,n_y)} w_{(n_x,n_y);R}^{(n)}(x,y,z) 
c_{(n_x,n_y),\alpha,R}^{(n)} \quad (\alpha=1,\ldots,N)  \; .
\end{equation}   
We plug them into the two-body (contact) interaction for atoms in the $g$ state 
\begin{equation}
\frac{1}{2} \sum_{\alpha,\beta=1}^{N}
\int\!d\bolr \int\!d\bolr^{\prime}
c^{\dagger}_{\alpha}(\bolr)c^{\dagger}_{\beta}(\bolr^{\prime}) 
 V(\bolr - \bolr^{\prime})
c_{\beta}(\bolr^{\prime})c_{\alpha}(\bolr)   
\end{equation}
with 
\begin{equation}
V(\bolr - \bolr^{\prime}) = g_{gg} \delta(\bolr - \bolr^{\prime})  \; .
\end{equation}
Retaining only the terms with $R_1=R_2=R_3=R_4=i$, $n_1=n_2=n_3=n_4=0$, 
$(n_x,n_y)=(1,0),(0,1)$, we obtain
\begin{equation}
\begin{split}
& \frac{1}{2} \sum_{R}\sum_{\alpha,\beta=1}^{N} \Biggl\{ 
\sum_{a=p_x,p_y} U_{aaaa} \,
c_{a,\alpha,R}^{\dagger} c_{a,\beta,R}^{\dagger}c_{a,\beta,R}c_{a,\alpha,R}  \\
& + \sum_{\substack{a\neq b\\=p_x,p_y}} 
U_{aabb} \, 
c_{a,\alpha,R}^{\dagger} c_{a,\beta,R}^{\dagger}c_{b,\beta,R}c_{b,\alpha,R}
+ \sum_{\substack{a\neq b\\=p_x,p_y}} 
U_{abba} \, 
c_{a,\alpha,R}^{\dagger} c_{b,\beta,R}^{\dagger}c_{b,\beta,R}c_{a,\alpha,R} \\
& + \sum_{\substack{a\neq b\\=p_x,p_y}} 
U_{abab} \, 
c_{a,\alpha,R}^{\dagger} c_{b,\beta,R}^{\dagger}c_{a,\beta,R}c_{b,\alpha,R}  \Biggr\} \; ,
\end{split}
\end{equation}
where the superscript `$(0)$' for the fermion operators of the lowest Bloch band has been suppressed.   
We have also introduced a short-hand notation $m=p_{x},p_{y}$ with 
$p_{x}=(n_x,n_y)=(1,0)$ and $p_{y}=(n_x,n_y)=(0,1)$.  

As the Wannier functions $w_{a;R}^{(0)}$ are related to each other by the $C_{4}$-symmetry, 
the coupling constants $U_{abcd}$ defined by
\begin{equation}
U_{abcd} \equiv g_{gg} 
\int\! d\bolr \, 
w_{a;R}^{(0)\ast}(\bolr)w_{b;R}^{(0)\ast}(\bolr)
w_{c;R}^{(0)}(\bolr)w_{d;R}^{(0)}(\bolr) \quad 
(a,b,c,d=p_{x}, p_{y}) 
\label{eqn:def-Uabcd-pxpy}
\end{equation}
obey the following relation:
\begin{equation}
\begin{split}
&  U_{p_x p_x p_x p_x} = U_{p_y p_y p_y p_y} \equiv U_{1}  \\
& U_{p_x p_x p_y p_y} = U_{p_y p_y p_x p_x}
= U_{p_x p_y p_y p_x}= U_{p_y p_x p_x p_y}
= U_{p_x p_y p_x p_y}= U_{p_y p_x p_y p_x}  \\
& \equiv U_{2} \; .
\end{split}
\label{eqn:def-U1-U2}
\end{equation}
Therefore, in contrast to the case of the $g$-$e$ model \eqref{eqn:Gorshkov-Ham}, 
we have only two independent couplings $U_1$ and $U_2$ for any $C_{4}$-symmetric potentials.  
In fact, further simplification occurs for axially-symmetric potentials like the harmonic one used 
here.  In these cases, $\phi_{n_x,n_y}(x,y)$ in Eq.~\eqref{eqn:Wannier-pxpy} is replaced by 
\begin{equation}
f(r)\cos \theta \quad \text{for }p_x \, , \;\; 
f(r)\sin \theta \quad \text{for }p_y  \quad (r=\sqrt{x^{2}+y^{2}})
\end{equation}
with some real-valued function $f(r)$ depending the actual choice of the potential.   
Carrying out the integration in Eq.~\eqref{eqn:def-Uabcd-pxpy}, we obtain
\begin{equation}
U_{1} = \frac{3\pi}{4}\mathcal{I}_{r,z} \, , \;\;
U_{2} = \frac{\pi}{4}\mathcal{I}_{r,z} \; ,
\label{eqn:U1-U2-axially-symmetric}
\end{equation}
where $\mathcal{I}_{r,z}$ is the result of the integration over $r$ and $z$ that is common to 
all $U_{abcd}$.   This implies that $(U_1,U_2)$ are bound to satisfy the relation 
\begin{equation}
U_{1}=3 U_{2}
\label{eqn:constraint-axial-sym}
\end{equation}
for {\em any} axially-symmetric potentials $V_{\perp}(x,y)=f(\sqrt{x^2+y^2})$.  
That is, as far as we use axially-symmetric traps, there is only {\em one} free parameter 
for the interactions.  
One possible way to deviate from the line $U_1 = 3U_2$ is to use such ($C_{4}$-symmetric) 
anharmonic potentials as 
\cite{Bois-C-L-M-T-15}:
\begin{equation}
V_{\perp}(x,y) = \frac{1}{2}M {\omega^2_{xy}} (x^{2}+y^{2}) 
+ \frac{1}{2}\beta (x^{4}+y^{4}) \quad (\beta \geq 0) 
\label{eqn:anharmonic-potential}
\end{equation} 
(see Fig.~3 of Ref.~\cite{Bois-C-L-M-T-15} for the ratio $U_1/U_2$ obtained for the above $V_{\perp}$).   

\begin{figure}[htb]
\begin{center}
\includegraphics[scale=0.5]{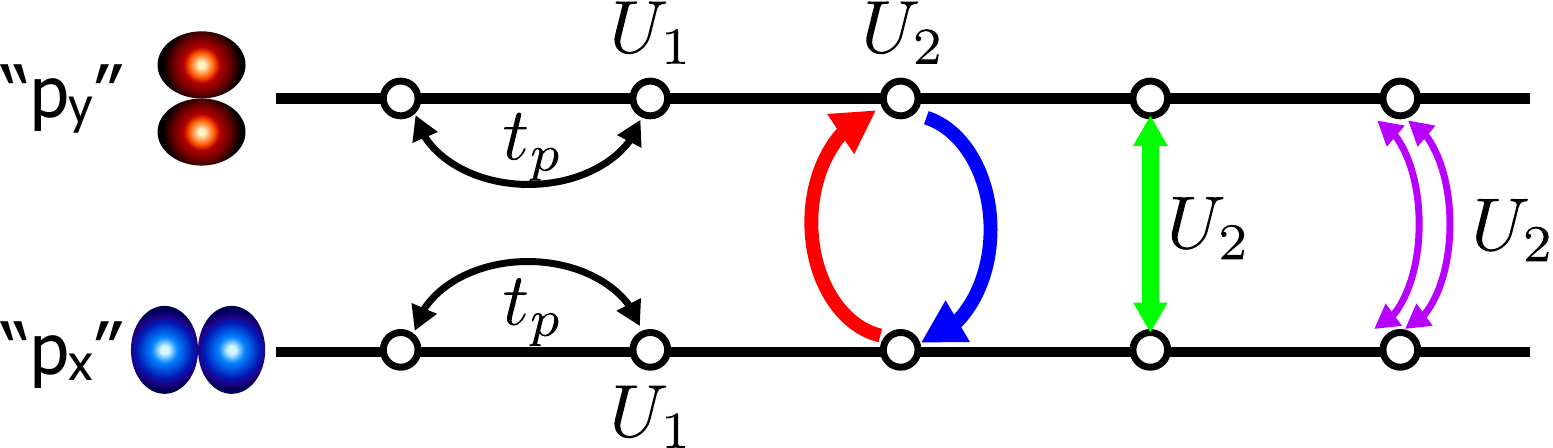}
\caption{(Color online) The two-leg ladder representation of the $p$-band model \eqref{eqn:p-band}. 
On top of the interactions included already in the $g$-$e$ model, pair-hopping processes between 
the two orbitals are allowed.   
\label{fig:p-band-2leg}}
\end{center}
\end{figure}

As the single-particle energy $E^{(n)}_{(n_x,n_y)}(k_z)$ [Eq.~\eqref{eqn:en-subbands}]  
is the same for the two $p$-bands, we have the same hopping amplitude and the chemical potential 
for $p_x$ and $p_y$.  Summing up all these, we obtain the following form 
of the $p$-band model \cite{Kobayashi-O-O-Y-M-12,Kobayashi-O-O-Y-M-14,Bois-C-L-M-T-15}:
\begin{equation}
\begin{split}
\mathcal{H}_{p\text{-band}} 
=& - t_{p} \sum_{i} \sum_{m=p_x,p_y} ( c_{m\alpha,i}^{\dagger}c_{m\alpha,i+1}  + \text{H.c.} ) \\
& - \mu_{p} \sum_{i} \sum_{m=p_x,p_y} n_{m,i} 
+  \frac{1}{2}U_{1}  \sum_{i}\sum_{m=p_x,p_y} n_{m,i}(n_{m,i}-1) \\
& +  U_{2}\,  \sum_{i}n_{p_x,i}n_{p_y,i} 
+ U_{2} \sum_{i}
c_{p_x\alpha,i}^{\dagger} c_{p_y\beta,i}^{\dagger}c_{p_x\beta,i}c_{p_y\alpha,i} \\
& + \frac{1}{2}U_{2}\sum_{i} \sum_{\alpha,\beta=1}^{N} \sum_{\substack{m\neq n\\=p_x,p_y}} 
c_{m\alpha,i}^{\dagger} c_{m\beta,i}^{\dagger}c_{n\beta,i}c_{n\alpha,i}  \; ,
\end{split}
\label{eqn:p-band}
\end{equation}
where the hopping $t_p$ and the chemical potential $\mu_{p}$ are given by the single-particle energy 
\begin{equation}
t_{p} = - \frac{1}{N_{\text{cell}}} \sum_{k_z}
E^{(n)}_{(1,0)}(k_z) \be^{ i k_z } \, , \;\; 
\mu_{p} = - \frac{1}{N_{\text{cell}}} \sum_{k_z} E^{(n)}_{(1,0)}(k_z)  \; .
\end{equation}
The last term of Eq.~\eqref{eqn:p-band} comes from the pair-hopping between the two orbitals, which is not allowed 
for the setting of the $g$-$e$ model, and breaks 
U(1)$_{\text{o}}$-symmetry in general, while the other five terms already existed 
in the $g$-$e$ model \eqref{eqn:Gorshkov-Ham}.  
In fact, except for the last term, $\mathcal{H}_{p\text{-band}}$ coincides with the Hamiltonian 
$\mathcal{H}_{g\text{-}e}$ [\eqref{eqn:Gorshkov-Ham}] after the following identification
(see Fig.~\ref{fig:p-band-2leg})
\begin{equation}
\begin{split}
& t_{g}= t_{e}= t_{p} \, , \; \mu^{(g)}=\mu^{(e)}=\mu_{p} \, , \\
& U_{gg}=U_{ee} = U_{1} \, , \; V = U_{2}  \, , \; V_{\text{ex}}^{g\text{-}e} = U_{2}  \; .
\end{split}
\label{eqn:g-e-vs-p-band}
\end{equation}
Due to the pair-hopping between the two orbitals, the $p$-band Hamiltonian 
\eqref{eqn:p-band} appears to break the U(1)$_{\text{o}}$-symmetry.    
To see this, we rewrite \eqref{eqn:p-band} using the orbital pseudo-spin:
\begin{equation}
\begin{split}
\mathcal{H}_{p\text{-band}} = & -t_p \,  \sum_{i}\sum_{m=p_x,p_y} 
  \left( c_{m\alpha,\,i}^\dag c_{m\alpha,\,i+1}+  \text{H.c.}  \right)  \\
& -\mu_{p} \sum_i  n_i
+\frac{1}{4}(U_1 + U_2) \sum_i  n_i^2  \\
&  +\sum_i \left\{ 2U_2 (\hat{T}_i^x)^2 + (U_1 - U_2)(\hat{T}_i^z)^2\right\}  \; ,
\end{split}
\label{eqn:p-band-simple}
\end{equation}
which is to be compared with Eq.~\eqref{alkamodel-2b}.   
As can be easily seen, the last two terms break $\text{U(1)}_{\text{o}}$ in general.   
Therefore, the generic symmetry of the $p$-band model is
\begin{equation}
\text{U(1)}_{\text{c}}{\times}\text{SU($N$)}_{\text{s}}{\times}\mathbb{Z}_{2,\text{o}} \; .
\end{equation}
However, this is not always the case 
and, in fact, there is a {\em hidden} U(1)$_{\text{o}}$ symmetry in the case of axially-symmetric traps.  
In fact, if one plugs into \eqref{eqn:p-band-simple} 
the relation $U_{1}=3 U_{2}$ that holds for {\em any} axially-symmetric $V_{\perp}(x,y)$ 
[see Eq.~\eqref{eqn:constraint-axial-sym}], the orbital part assumes 
a fully U(1)-symmetric form: $2U_2\left\{  (T_j^x)^2 +(T_j^z)^2\right\}$ and 
$\mathcal{H}_{p\text{-band}}$ reduces to a special case of $\mathcal{H}_{g\text{-}e}$ [Eq. \eqref{alkamodel-2b}] 
(with $\mu_{g}=\mu_{e}$, $U_{\text{diff}}=0$) 
after the redefinition: $T_{i}^{y} \leftrightarrow T_{i}^{z}$, $T_{i}^{z} \to -T_{i}^{y}$.\footnote{%
This is in a sense an artifact of the choice of the basis ($p_x$ and $p_y$).   
In fact, if we had chosen the orbital-angular-momentum (along the $z$-axis) basis, 
the U(1)$_{\text{o}}$-symmetry would have been explicit.}

A few remarks are in order here about the special points where great simplification or 
enhanced symmetries emerge.  
When $U_2 =0$, the $p$-band model decouples into two non-interacting copies 
of SU($N$) Hubbard chains and we can borrow the results from Sec.~\ref{sec:singleband}.  
Moreover, along the line $U_1 =U_2$,
the $p$-band model  (\ref{eqn:p-band-simple}) reduces to the above $U_2 =0$ case (two decoupled chains) 
after the redefinition $T_{i}^{x} \leftrightarrow T^{z}_{i}$, $U_1 \to U_1/2$.   
Finally, in the $N=2$ case, the $p$-band model can be recast in the following form \cite{Kobayashi-O-O-Y-M-14}:
\begin{equation}
\begin{split}
\mathcal{H}_{p\text{-band}} = & 
- t_p \sum_{\alpha=\uparrow,\downarrow} \sum_{m=p_x,p_y}\sum_{j}
( c_{m\alpha,j}^{\dagger}c_{m\alpha,j+1}  + \text{H.c.} ) \\
&-\mu_{p} \sum_i  n_i - \frac{2}{3}U_{1}  \sum_{i}\sum_{m=p_x,p_y} 
\left( \hat{\boldsymbol{S}}_{m,i} \right)^{2}  \\
& - 2 U_{2}\sum_{i} \hat{\boldsymbol{S}}_{p_x,i}{\cdot} \hat{\boldsymbol{S}}_{p_y,i}  
+ 2U_{2} \sum_{i} \hat{\boldsymbol{K}}_{p_x,i}{\cdot} \hat{\boldsymbol{K}}_{p_y,i} 
\end{split}
\end{equation}  
using the following two commuting sets of spin and pseudo-spin operators 
$\{\hat{S}^{a}\}$ and $\{\hat{K}^{a}\}$ \cite{Yang-Z-90}:
\begin{subequations}
\begin{align}
& \hat{S}^{a}_{m,i} = \frac{1}{2}c_{m\alpha,i}^{\dagger}\sigma^{a}_{\alpha\beta} c_{m\beta,i}  
\quad (a=x,y,z; \; m= p_x, p_y) \\
\begin{split}
& \hat{K}^{+}_{m,i} \equiv (-1)^{i} c_{m\uparrow,i}^{\dagger}c_{m\downarrow,i}^{\dagger} , \quad 
\hat{K}^{-}_{m,i} \equiv (-1)^{i} c_{m\downarrow,i} c_{m\uparrow,i}  , \\
& \hat{K}_{m,i}^{z} \equiv \frac{1}{2}(n_{m\uparrow,i} + n_{m\downarrow,i} -1 ) 
= \frac{1}{2}(n_{m,i}-1)  \; .
\end{split}
\label{eqn:cgarge-pseudo-spin}
\end{align}
\end{subequations}
Note that the set of operators $\hat{K}$ generates the so-called $\eta$-SU(2)-symmetry in the charge sector.  
From this, we can see that 
the $N=2$ $p$-band model at half-filling enjoys an enlarged 
$\text{SU(2)}_{\text{spin}} \times \text{SU(2)}_{\text{charge}} \sim \text{SO(4)}$ symmetry for {\em all} $U_1,U_2$ 
which stems from an additional SU(2) symmetry for the charge degrees 
of freedom at half-filling \cite{Kobayashi-O-O-Y-M-14}.    
For more details about the SO(4) symmetry in the Hubbard model, see section 2.2 of Ref.~\cite{Essler-book}. 
\subsection{Results known for Mott-insulating limits}
\label{sec:Mott}
As in the usual Hubbard Hamiltonian, when we have an integer number of particles at each site, 
the system becomes a Mott insulator in the limit 
\begin{equation}
U = \frac{1}{4}(U_{gg} + U_{ee} + 2 V) \to \infty 
\end{equation}
[see Eq.~\eqref{alkamodel-2b}].  
Depending on $N$ and the number of fermions at each site, we have a variety of SU($N$) spins 
emerging at each site in the Mott-insulating limits (see, e.g., Fig.~\ref{fig:strong-coupling-SU4}).  
\subsubsection{Low-energy degrees of freedom in Mott phases}
\label{sec:low-energy-dof}
To find the SU($N$) and the orbital SU(2) content, we first note that when $J=V^{g\text{-}e}_{\text{ex}}=0$, 
$J_z=0$, $U_{\text{diff}}=0$,  
the system attains the maximal symmetry $\text{U($2N$)}$ [see \eqref{eqn:symmetries-g-e}].  
In this limit, for a given number of fermions per site (say, $n$), we can uniquely assign the following U($2N$) 
irreducible representation:
\begin{equation}
\text{\scriptsize $n$} \left\{ 
\yng(1,1,1,1)
 \right.  \quad (0 \leq n \leq 2N) \; .
 \end{equation}
Formally, the $\frac{(2N)!}{n!(2N-n)!}$ states in the above representations may be decomposed in terms of 
the direct product of SU($N$) and the orbital SU(2).  
This problem is explicitly worked out in Ref.~\cite{Itzykson-N-66} (see also Appendix A of 
Ref.~\cite{Bois-C-L-M-T-15} for a concise explanation in the context of the SU($N$) Hubbard model).   
As a simplest example, let us consider the $n=1$ case.  Then the $2N$ singly-occupied states at each site 
are decomposed as
\begin{equation}
\underbrace{\yng(1)}_{\text{SU($2N$)}} 
\sim (\underbrace{\yng(1)}_{\text{SU}(N)},\underbrace{\yng(1)}_{\text{SU(2)}}) 
= (\mathbf{N} \; , \; T=1/2)  \; .
\end{equation}
Therefore, the low-energy physics of the highly degenerate ground-state manifold is described 
by the composite of the SU($N$) `spins' in the fundamental representation (${\tiny \yng(1)}$) and 
the orbital pseudo-spins $T=1/2$ \cite{Gorshkov-et-al-10}.  

Different results are obtained when we consider the case of half filling. 
For instance, the 70 half-filled states of the SU(4) $g$-$e$ model (i.e., $N=4$, $n=4$) may be decomposed 
as
\begin{equation}
\begin{split}
\text{\scriptsize $n=4$} \left\{ 
\yng(1,1,1,1)
 \right. 
& \sim \left( \underbrace{\yng(2,2)}_{\text{SU(4)}}\; ,\;  \underbrace{\bullet}_{\text{SU(2)}} \right) 
\oplus \left( \yng(2,1,1)\; ,\; \yng(2) \right) 
 \oplus \left( \bullet \; , \; \yng(4) \right)  \\
& = \left( \mathbf{20} , T=0 \right)\oplus \left( \mathbf{15} , T=1 \right) \oplus 
\left( \mathbf{1} , T=2 \right) 
 \label{eqn:decomp-SU2-SUN-4} 
 \end{split}
 \end{equation}
(with $\bullet$ being the singlet).   
Note that the Fermi statistics dictates the possible combinations of  SU($N$) and orbital.  
The Hund coupling $V^{g\text{-}e}_{\text{ex}}$ 
 \begin{equation}
 - V_{\text{ex}}^{g\text{-}e}
\sum_{i}\left( 
\sum_{A=1}^{N^{2}-1} \mathcal{S}_{g,i}^{A} \mathcal{S}_{e,i}^{A} 
\right) 
= +V_{\text{ex}}^{g\text{-}e} \sum_{i}  ( \bolT_{i})^{2} 
+ (\text{density-density interaction})
 \end{equation}
[see Eqs.~\eqref{eqn:Gorshkov-Ham-Hund} and \eqref{eqn:Gorshkov-Ham-Hund-2}]  
then selects one of the irreducible representations appearing 
on the right-hand side:
\begin{equation}
\begin{split}
& \left( \yng(2,2) \; ,\;  \bullet \right)   \qquad \text{for }  V_{\text{ex}}^{g\text{-}e} > 0 \\
& \left( \bullet \; , \; \yng(4) \right) \qquad \text{for }  V_{\text{ex}}^{g\text{-}e} < 0  \; .
\end{split}
\label{eqn:SUN-spin-Mott-1}
\end{equation}
That is, when the SU($N$) spin is maximized (quenched), the orbital pseudo-spin $\bolT$ is 
quenched (maximized) as is expected from the discussion in Sec.~\ref{sec:g-e-model}.  
This dual nature of the SU($N$) and the orbital is common to all SU($N$) 
$g$-$e$ models with even $N$.   

For odd-$N$, on the other hand, we have different decompositions.  For instance, for $N=3$ and $n=3$ 
(half-filling), the decomposition in terms of $\text{SU(3)}{\times}\text{SU(2)}$ reads as
\begin{equation}
\begin{split}
\text{\scriptsize $n=3$} \left\{
\yng(1,1,1)\right. 
& \sim \left( \underbrace{\yng(2,1)}_{\text{SU(3)}} \; , \; \underbrace{\yng(1)}_{\text{SU(2)}} \right) 
\oplus \left(\bullet \; , \; \yng(3) \right)   \\
&=  \left( \mathbf{8} , T=1/2 \right)\oplus \left( \mathbf{1} , T=3/2 \right)  \; .
\end{split}
\end{equation}
Therefore, the situation on the $V_{\text{ex}}^{g\text{-}e} > 0$ side is rather different for 
$N$-even and $N$-odd.   Specifically, the low-energy degrees of freedom in the half-filled Mott insulator 
for $N$-even are the {\em pure} SU($N$) spins, while those for $N$-odd are the composite of 
SU($N$) spins [with $(N+1)/2$ boxes in the first column and $(N-1)/2$ boxes in the second] and 
orbital pseudo-spins $T=1/2$.   
\subsubsection{Strong-coupling limits}
\label{sec:strong-coupling}
In the previous section, we have seen that different low-energy degrees of freedom emerge 
depending on $N$, filling $n(\in \mathbb{Z})$, and the sign of $V_{\text{ex}}^{g\text{-}e}$.   
Therefore, when the hopping amplitudes vanish (atomic limit), i.e. $t_g=t_e=0$ or $t_p=0$ depending on the model, the ground state has extensive degeneracy.  
This degeneracy is lifted when the hopping $t_m$ is taken into account.  Clearly, $t_m$ changes the fermion number $n$ at each 
site and non-trivial matrix elements require at least $t_m^{2}$ processes.  

For the clarity of the argument, we restrict ourselves to the case of half-filling 
(i.e., the number of fermions per site $n=N$) in the following.   
Let us begin with the case where $N=\text{even}$, $V_{\text{ex}}^{g\text{-}e}>0$.  
Then, in the atomic-limit ground state, we have 
\begin{equation}
\text{\scriptsize $N/2$} \left\{
\yng(2,2,2)\right. 
\label{eqn:SUN-spin-Mott-2}
\end{equation}
with the orbital pseudo-spin quenched ($\boldsymbol{T}=\mathbf{0}$).  
The resulting effective Hamiltonian is obtained by the second-order perturbation and 
reads as follows \cite{Bois-C-L-M-T-15}
\begin{equation}
\mathcal{H}_{\text{SU($N$)}}  = J_{\text{SU($N$)}} 
\sum_{A=1}^{N^{2}-1} \mathcal{S}_{i}^{A}\mathcal{S}_{i+1}^{A}  + \text{const.}  \; ,
\label{eqn:2nd-order-effective-Ham-Gorshkov}
\end{equation}
where the exchange coupling $J_{\text{SU($N$)}}$ is $N$-independent: 
\begin{equation}
J_{\text{SU($N$)}}  \equiv
\frac{1}{2} \left\{
\frac{t_{g}^{2}}{U+U_{\text{diff}}+J+\frac{J_z}{2}} 
+ \frac{t_{e}^{2}}{U-U_{\text{diff}}+J+\frac{J_z}{2}}
\right\}   
\label{eqn:exch-coupling-Gorshkov}
\end{equation}
and the SU($N$) spins $\{ \mathcal{S}_{i} \}$ transform under the representation \eqref{eqn:SUN-spin-Mott-2} 
[see Eq.~\eqref{eqn:SUN-spin-Mott-1}].  
As $\boldsymbol{T}$ is quenched, we obtain a pure `spin' Hamiltonian (as we have seen 
above, this is not the case when $N=\text{odd}$).  

In the case of $\mathcal{H}_{\text{$p$-band}}$, the condition $V_{\text{ex}}^{g\text{-}e}>0$ translates to 
$U_2>0$.  However, $T^{z}$ is not conserved in general and we cannot use 
the argument in Sec.~\ref{sec:low-energy-dof} as it is.  
However, we found that when $U_{1}>U_{2}(>0)$ the lowest-energy 
state is $T=0$ singlet that enables us to follow exactly the same steps and obtain \cite{Bois-C-L-M-T-15}
\begin{equation}
\mathcal{H}_{\text{SU($N$)}}  = \frac{t_p^{2}}{U_1 +U_2} 
\sum_{A=1}^{N^{2}-1} \mathcal{S}_{i}^{A}\mathcal{S}_{i+1}^{A}  + \text{const.}  \; .
\label{eqn:2nd-order-effective-Ham-p-band}
\end{equation}
In Sec.~\ref{sec:SPT}, we will show that the SU($N$) spin Hamiltonians 
\eqref{eqn:2nd-order-effective-Ham-Gorshkov} and \eqref{eqn:2nd-order-effective-Ham-p-band} 
host a topological phase in quite a large region of the parameter space.  

When $V_{\text{ex}}^{g\text{-}e}<0$ (or $U_2 <0$), on the other hand, the low-energy degree of freedom is 
the orbital pseudo-spin transforming like a $T=N/2$ spin [Eq.~\eqref{eqn:SUN-spin-Mott-1}]
\begin{equation}
\underbrace{\yng(3)}_{N \text{ boxes} \, (T=N/2)} 
\end{equation}
and the SU($N$) `spins' are quenched.  
The resulting effective Hamiltonian is rather different for the $g$-$e$ Hamiltonian and 
the $p$-band Hamiltonian.   For the $g$-$e$ Hamiltonian, we obtain the following 
$T=N/2$ Hamiltonian:
\begin{equation}
\begin{split}
\mathcal{H}_{\text{orbital}} = &\sum_{i}
\left\{  
\mathcal{J}_{xy} \left(
T^{x}_{i} T^{x}_{i+1} + T^{y}_{i} T^{y}_{i+1} 
\right)
+ \mathcal{J}_{z} T^{z}_{i} T^{z}_{i+1}
-(J- J_z)  (T_{i}^z)^2 \right\}  \\
& + \sum_{i} \left\{ 
N U_{\text{diff}}
- \left(\mu _g -\mu _e\right) 
\right\} T_{i}^{z}  
\end{split}
\label{eqn:eff-Ham-orbital-Haldane}
\end{equation}
with the {\em easy-axis} exchange coupling \cite{Bois-C-L-M-T-15}
\begin{equation}
\begin{split}
& \mathcal{J}_{xy} \equiv \frac{4 t_g t_{e}}{N\left\{U + |J| \left( N+\frac{1}{2} \right)\right\}}  \\
& \mathcal{J}_{z} \equiv   
\frac{2 \left\{ {t_g}^{2} + {t_{e}}^{2}\right\}}{N \left\{U + |J| \left( N+\frac{1}{2} \right)\right\}} 
\quad (\mathcal{J}_{xy} \leq  \mathcal{J}_{z})   \; .
\end{split}
\end{equation}
In the ideal situation where the two orbitals are symmetric, the model reduces to 
the spin $T=N/2$ Heisenberg model ($\mathcal{J}_{xy}=\mathcal{J}_{z}$) 
with the single-ion $D$-term, for which much is known 
(see, e.g. Refs.~\cite{Schulz-86,Chen-H-S-03,Tonegawa-O-N-S-N-K-11} and references cited therein).  
When $|J_{z}| \ll |J|$ ($J<0$), the system has {\em easy-plane} 
anisotropy and, if $N$ is even\footnote{%
When $N$ is odd, the large-$D$ limit is not very trivial and the effective Hamiltonian is given by 
the $S=1/2$ XXZ chain.} 
and $|J|$ is large enough compared with $\mathcal{J}_{xy}$ and $\mathcal{J}_{z}$, 
the ground state is in the so-called large-$D$ phase, which, in its extreme case, is a product of 
$T^{z}=0$ states.    In the fermionic language, it is a state where there are $N/2$ fermions on each 
orbital (and hence $T^{z}_{i}=0$) and these $N$ fermions form total SU($N$)-singlet at each site. 

When $J_{z}$ takes a large negative value (this is the case typically for large $V$), 
the $D$-term is of easy-axis-type and the ground state is Ising-like: 
$T^{z}=\cdots, -N/2,N/2,-N/2,N/2, \cdots$.  This is the $2k_{\text{F}}(=\pi)$ density wave of the {\em orbital} 
pseudo-spin and we call it {\em orbital density wave} (ODW).    See Sec.~\ref{sec:phasediagrams} 
for the location of the phase.  

In the case of the $p$-band model, physics of the orbital sector is even more involved. 
Since the condition $V_{\text{ex}}^{g\text{-}e}=J<0$ translates to $U_{2}<0$ in the $p$-band model 
[see Eq.~\eqref{eqn:g-e-vs-p-band}],   
the interaction is attractive $U_1 + U_2 <0$ in the physical region $U_1 \simeq 3U_2$ and we have to take into account several different values of $n_{i}$.  
Therefore, the physics here is quite different from the usual Mott physics described above.  
For instance, at $\mu=-N |U_1+U_2|$, we have two degenerate SU($N$)-singlet states 
$n_{i}=0$ ($T=0$) and $n_{i}=2N$ ($T=0$)  
which feel a repulsive interaction coming from $t^2$-processes thereby stabilizing the 
$2k_{\text{F}}$-CDW\footnote{%
The basic mechanism underlying the CDW here is the same as the CDW in the case of 
the half-filled single-band SU($N$) Hubbard chain discussed in Sec.~\ref{sec:halfilling}.} 
in a region around the line $U_1 = 3U_2(<0)$ for $N \geq 3$ 
(see, e.g., Fig.~22 of Ref.~\cite{Bois-C-L-M-T-15}).   

Last, the case with $V_{\text{ex}}^{g\text{-}e}=J=0$ is slightly different.  
This region is most conveniently investigated in the limit $U=U_{gg}=U_{ee}=V \gg t$, 
$\mu^{(g)}=\mu^{(e)}$.   Then the $g$-$e$ model \eqref{alkamodel-2b} reduces to 
the {\em single-band} SU($2N$) Hubbard model at half-filling discussed in Sec.~\ref{sec:halfilling};   
the ground state is a dimerized SP state (see Fig.~\ref{fig:SU4-SP-CDW}).  
This highlights the importance of the orbital degree of freedom and the exchange interaction 
$V_{\text{ex}}^{g\text{-}e}$ between the two orbital states in realizing topological phases.  
\subsubsection{Sigma-model mapping}
\label{sec:Read-Sachdev}
Some insights into the nature of the ground states may be gained by mapping the problems onto 
the sigma model with the $\theta$-term.  
Generalizing the semi-classical sigma-model mapping \cite{Haldane-NLsigma-88,Auerbach} 
\`{a} la Haldane of the usual spin chains, 
Read and Sachdev \cite{Read-S-PRL-89,Read-S-NP-89,Read-S-90} 
considered a family of SU($N$) `spin systems' where an irreducible representation 
$\mathcal{R}$ (specified by a rectangular Young diagram with $m$ rows and $n_{\text{c}}$ columns) 
and its conjugate $\bar{\mathcal{R}}$ are assigned respectively on the two sublattices of 
a $D$-dimensional bipartite lattice.  
\begin{equation}
\mathcal{R} = 
\left.
\vphantom{\yng(6,6,6)}
\smash{
\underbrace{ \yng(	6,6,6)}_{n_{\text{c}}}} 
 \right\} \text{\scriptsize $m$}
 \vphantom{\yng(6,6,6,6)} 
 \qquad 
\bar{ \mathcal{R}} = 
 \left.
\vphantom{\yng(6,6)}
\smash{
\underbrace{ \yng(6,6)}_{n_{\text{c}}}%
} 
 \right\} \text{\scriptsize $N-m$}
 \vphantom{\yng(6,6,6)} 
 \label{eqn:Read-Sachdev-ref-state}
\end{equation}

If we represent the state \eqref{eqn:Read-Sachdev-ref-state} in terms of $n_{\text{c}}$ copies of 
SU($N$) fermions, these states are in fact invariant under $\text{U($N-m$)}{\times}\text{U($m$)}$.  
Therefore, the coherent state, generated from the reference state \eqref{eqn:Read-Sachdev-ref-state} 
by applying any elements of U($N$), represents $\frac{\text{U($N$)}}{\text{U($N-m$)}{\times}\text{U($m$)}}$ 
\cite{Read-S-NP-89}.  
This is closely parallel to that the Bloch spin coherent state generated from the spin state polarized 
in the $z$-direction is isomorphic to 
$\frac{\text{SU(2)}}{\text{U(1)}} \simeq \frac{\text{U(2)}}{\text{U(1)}{\times}\text{U(1)}}=\mathbb{C}\text{P}^{1}
\simeq \text{S}^{2}$.  
The role of the expectation value $\langle \bolS \rangle$ of spins in the usual spin path integral is now 
played by an $N{\times}N$ hermitian matrix $Q$ ($Q^{2}=\mathbf{1}$).  
Since the number of columns $n_{\text{c}}$ controls the semi-classical limit as the spin $S$ does in 
the SU(2) case, we may expect that an SU($N$) analogue of N\'{e}el ordering occurs in the limit 
$n_{\text{c}}\to \infty$.  
The order parameter is the staggered component $\Omega$ of the matrix $Q$ and its low-energy 
fluctuations are governed by the following (finite-temperature) effective action \cite{Read-S-NP-89}:
\begin{equation}
\begin{split}
\mathcal{S}_{\text{E}} = &
\frac{1}{2}\chi_{\text{s}}   
\int_{0}^{\beta\hbar}d\tau \int\!d^{D}\bolr \, 
\text{Tr}\left\{ \partial_{\tau}\Omega(\tau,\bolr)\right\}^{2}   \\
& + \frac{1}{2} \int_{0}^{\beta\hbar}d\tau  \int\! d^{D}\bolr  
\sum_{\mu,\nu=1}^{D}
\rho_{\text{s}}(\bolr)_{\mu\nu} 
\text{Tr}\left\{\partial_{\mu}\Omega(\tau,\bolr) \, \partial_{\nu}\Omega(\tau,\bolr) \right\} \; ,
\end{split}
\label{eqn:Grassmannian-sigma}
\end{equation}
where the two phenomenological parameters $\chi_{\text{s}}$ and $\rho_{\text{s}}(\bolr)_{\mu\nu}$ 
respectively are the transverse susceptibility and the spin stiffness\footnote{%
For the nearest-neighbor Heisenberg SU($N$) magnets on a hypercubic lattice in $D$ dimensions,  
\begin{equation}
\rho_{\text{s}}(\bolr)_{\mu\nu} = \left(\frac{n_{\text{c}}}{2}\right)^{2}\frac{J}{N} a_{0}^{2-D} \delta_{\mu\nu} \, , \;\;
\chi_{\text{s}} = \frac{N}{16D J a_{0}^{D}} 
\end{equation}
(with $a_{0}$ being the lattice constant).} \cite{Chakravarty-H-N-89,Read-S-NP-89}.  
In one dimension ($D=1$), this action is supplemented by the $\theta$-term\footnote{%
In contrast to the corresponding expression for the O(3) non-linear sigma model, 
an extra factor $i$ is not necessary here as the integral itself is pure imaginary.  
In the case of the O(3) non-linear sigma model, we express the matrix field by 
a unit vector $\boldsymbol{n}$ as $\Omega=\boldsymbol{n}{\cdot}\boldsymbol{\sigma}$ to recover 
the factor $i$ in the $\theta$-term.} associated with 
$\Pi_{2}\left(\frac{\text{U($N$)}}{\text{U($N-m$)}{\times}\text{U($m$)}}\right) = \mathbb{Z}$:
\begin{equation}
\begin{split}
\mathcal{S}_{\theta} &= 
 - \frac{\Theta_{\text{top}}}{16\pi}  \int_{0}^{\beta\hbar} d\tau \int\!dx\, \epsilon_{\mu\nu}
\text{Tr}\left\{ \Omega(\tau,x) \partial_{\mu}\Omega(\tau,x)\,  \partial_{\nu} \Omega(\tau,x) \right\} \\
& \qquad  (\epsilon_{\tau x}=+1, \;  \Theta_{\text{top}}=n_{\text{c}}\pi)  \; .
\end{split}
\label{eqn:SUN-theta-term-2}
\end{equation}
Following the same line of argument as the one first used by Haldane \cite{Haldane-PLA-83,Haldane-PRL-83}, 
we may conclude, for $D=1$, that when the number of columns $n_{\text{c}}$ is even for which the $\theta$-term is 
irrelevant to bulk properties, the system is in a featureless gapped phase \cite{Read-S-NP-89,Read-S-90}.  
On the other hand, when $n_{\text{c}}$ is odd, the system spontaneously dimerizes.  
In the context of our SU($N$) Hubbard model, the case $n_{\text{c}}=1$ corresponds to the Mott-insulating phase 
discussed in Sec.~\ref{sec:halfilling} (the single-band SU($N$) Hubbard model at half-filling)\footnote{%
As we are dealing in this paper with the translationally invariant systems, $\mathcal{R}=\bar{\mathcal{R}}$, 
i.e., $m=N/2$.} 
and the model with $n_{\text{c}}=2$ describes the half-filled 2-orbital system 
(Sec.~\ref{sec:twoband}) deep in the Mott-insulating region.  
For instance, the appearance of the SP phase in the single-band model \cite{Buchta2007,Nonne-L-C-R-B-11} 
discussed in Sec.~\ref{sec:halfilling} is consistent with the above field-theory prediction.   
In Sec.~\ref{sec:SPT}, we will argue that the featureless phase predicted by the model 
\eqref{eqn:Grassmannian-sigma} is in fact topological by explicitly constructing the model wave function. 
\subsubsection{Lieb-Schultz-Mattis argument and generalized Haldane conjecture}
\label{sec:LSM}
The simple argument first used by Lieb, Schultz, and Mattis \cite{Lieb-S-M-61} is quite general 
but sometimes able to give strong constraints on the nature of the ground state and the low-energy spectrum 
over it.  
In fact, Affleck and Lieb \cite{Affleck-L-86} extended the original argument of Ref.~\cite{Lieb-S-M-61} 
to include the case of self-conjugate representations of SU($N$).  
Specifically, they showed for SU($N$) ($N=\text{even}$) chains based on the self-conjugate representation 
with $N/2$ rows and $n_{\text{c}}$ columns that 
(i) if $n_{\text{c}}=1$ [anti-symmetric $(N/2)$-tensor representation], the finite-size ground state is unique.  
Furthermore, assuming the uniqueness of the ground state, they showed that (ii) for {\em any} 
translationally invariant choice of representations\footnote{%
The proof of the existence of low-lying states works regardless of whether the ground state 
is unique or not.  However, unless the (finite-size) ground state is unique, the proof does not 
tell anything about the {\em excited} states.} (i.e., the same representation $\mathcal{R}$ is 
assigned for all sites), the SU($N$) chain 
harbors low-lying excitations whose gap is bounded by $1/L$ ($L$ being the system size) provided that 
the number of boxes $n_{\text{Y}}$ in the Young diagram representing $\mathcal{R}$ 
is {\em not} divisible by $N$.   
In other words, except for the cases of $n_{\text{Y}}=0$ (mod $N$) 
[including the one with $N/2$ ($N=\text{even})$ rows and two columns 
which is relevant to our effective spin model \eqref{eqn:2nd-order-effective-Ham-Gorshkov}], 
this statement excludes the possibility of gapped topological 
ground states.  Remarkably, this is perfectly consistent with the recent group-cohomology classification of 
the gapped topological phases in 1D\cite{Duivenvoorden-Q-13} (see Sec.~\ref{sec:SUN-SPT} for the detail).  

In the context of the SU($N$) ($N=\text{even}$) spin chains based on the representation with $N/2$ rows 
and $n_{\text{c}}$ columns (self-conjugate representation; the model \eqref{eqn:2nd-order-effective-Ham-Gorshkov} 
obtained in Sec.~\ref{sec:strong-coupling} corresponds to $n_{\text{c}}=2$), 
this means a unique gapless ground state or gapped degenerate ground states 
unless $N{\times}n_{\text{c}}/(2N)\equiv 0$ (mod $1$), 
i.e., $n_{\text{c}}=\text{even}$.  
This is consistent with the argument in the previous section since the  sigma-model 
mapping tells us that when $\Theta_{\text{top}}=n_{\text{c}}\pi\equiv 0$ (mod $2\pi$) 
the ground state is unique and gapped.\footnote{%
When $n_{\text{c}}=\text{odd}$, on the other hand, the sigma-model argument generically predicts 
dimerized ground states except for $N=2$, where we expect the gapless SU(2)$_1$ WZW criticality.}

There is an SU(4) spin model with two exact charge-conjugation (or translation) breaking ground states 
\cite{Affleck-A-M-R-91}.  According to the Lieb-Schultz-Mattis argument \cite{Affleck-L-86},  
the physical spin ${\tiny \yng(1,1)}$ ($n_{\text{Y}} \neq 0$ mod 4) at each site implies that 
the (thermodynamic-limit) ground state is either gapless or (at least two-fold) degenerate.   
The existence of the two degenerate ground states means that the second possibility realizes here.  

All the above statements concern 1D systems.  
In the SU(2) cases, the higher-dimensional extension of the Lieb-Schultz-Mattis argument 
is known \cite{Hastings-LSM-04,Nachtergaele-S-07}.  
A similar extension for the SU($N$) cases would be intriguing in view of the proposal of exotic spin 
liquids in two dimensions \cite{Hermele-G-R-09,Hermele-G-11}.  

Last, we mention a generalization of the Haldane's conjecture for SU(2) spin chains to SU($N$)  
proposed in Ref.~ \cite{Rachel-T-F-S-G-09} (see also Ref.~\cite{Greiter-R-07}). 
Three different cases (type-I, II, and III) have been introduced there depending on $N$ and   
$n_{\text{Y}}$. Type I concerns the case when $N$ and $n_{\text{Y}}$ have no common divisor and 
an SU($N$)$_1$ quantum criticality with central charge $c=N-1$ is expected. 
When $n_{\text{Y}}$ is divisible by $N$ (type-II), a Haldane-gap phase is expected  in general 
(note that this case is {\em not} 
covered by the theorem of Ref.~\cite{Affleck-L-86}). 
The last category (type-III) corresponds to 
the case when  $n_{\text{Y}}$ and $N$ have a common divisor different from $N$. For short-range interactions, 
an SU($N$)$_1$  quantum critical behavior is expected. 
In this respect, some DMRG calculations for $N=4$ and $n_{\text{Y}} =2$, i.e. type III
behavior, have found an SU(4)$_1$ criticality \cite{Rachel-T-F-S-G-09}. 
As already noticed in Sec.~\ref{sec:other-commensurate-fillings}, however, 
this conjecture is not consistent with the known results that  SU($N$) Heisenberg spin chain
in self-conjugate antisymmetric representations (i.e., type III case)  have a spectral and a ground-state degeneracy.
\subsection{Symmetry-protected topological phases}
\label{sec:SPT}
In this section, we try to characterize the nature of the ground state of the SU($N$) spin chain 
\eqref{eqn:2nd-order-effective-Ham-Gorshkov} obtained deep in the Mott-insulating phase.    
Specifically, we show that the ground state of the model \eqref{eqn:2nd-order-effective-Ham-Gorshkov} 
shares essentially the same properties with that of the solvable valence bond solids (VBS) 
models in Sec.~\ref{sec:solvable-Ham} 
and that it is in fact in one of the symmetry-protected topological (SPT) phases.   
The concept of SPT phases is a generalization of topological insulators \cite{Hasan-K-10} to 
a class of interacting states of matter that are characterized by short-range entanglement 
and are well-defined {\em only} in the presence of certain symmetries (called protecting symmetries) 
\cite{Gu-W-09,Chen-G-W-10}.  
Being topological, this class of topological phases defies the traditional characterization 
with broken symmetry and the corresponding local order parameters; two phases are distinguished not by the symmetry 
they possess but by the presence of quantum phase transitions that separate them \cite{Chen-G-W-10}.  
One way of distinguishing between topological phases from trivial ones and 
labeling the former is to use the {\em physical} edge states.  
However, this approach is not quite satisfactory in the following respects. First, even topologically trivial 
states may have certain structures around the edges of the system, as, e.g., in the spin-2 Heisenberg 
chain \cite{Nishiyama-T-H-S-95,Qin-N-S-95}.  
Second, in order to see the edge excitations, it is necessary 
to consider the excitation spectrum, while the topological properties are intrinsic to the ground state 
itself and should be seen only by examining the ground-state wave function. 

Recently, the use of the entanglement spectrum, which is defined as the logarithm of the spectrum 
of the reduced density matrix, in characterizing topological phases 
has been suggested in Ref.~\cite{Li-H-08}.  
This is based on the observation that the entanglement spectrum {\em resembles} 
the spectrum of the physical edge excitations. 
The idea has been successfully applied to various 1D systems \cite{Pollmann-T-B-O-10,Pollmann-B-T-O-12,%
Fidkowski-K-11,Turner-P-B-11,Zheng-Z-X-L-11,Tu2011,Takayoshi-T-T-15} and enabled us to characterize 
topological phases and the quantum phase transitions among them. 
In this section, we present a clear evidence from the entanglement spectrum 
that the ground state of the SU(4) Heisenberg model 
\eqref{eqn:2nd-order-effective-Ham-Gorshkov} is indeed in an SPT phase protected by SU(4) 
[projective unitary group PSU(4), precisely\footnote{%
The difference between SU($N$) and PSU($N$) may become clear below.}] symmetry. 
\subsubsection{Haldane phase --the simplest example}
\label{sec:Haldane-phase}
To understand the structure of topological phases in the case of SU($N$) symmetry, 
it is convenient to begin with the simplest prototypical case $N=2$.  
Since the Haldane's conjecture, we know that the ground-state properties of 
the spin-$S$ Heisenberg chain are qualitatively different depending on the parity 
of $2S$ \cite{Haldane-PLA-83,Haldane-PRL-83}; 
when $2S=\text{even}$, the ground state is in a featureless non-magnetic phase 
({\em Haldane phase}) with the gapped triplon ($S=1$)  excitations in the bulk,  
while, for odd $2S$, we have a gapless (i.e., algebraic) ground state with spinon ($S=1/2$) excitations.   
This conjecture has been later confirmed both by the construction of 
a rigorous example (see below) \cite{ Affleck-K-L-T-87,Affleck-K-L-T-88} 
and by extensive numerical simulations \cite{White-H-93,Schollwock-G-J-96,Todo-K-01}.  
Soon after, it has been pointed out that the featureless gapped ground state of the integer-$S$ spin chains 
may have a {\em hidden} ``topological'' order characterized 
by non-local order parameters \cite{denNijs-R-89,Girvin-A-89,Kennedy-T-92-PRB,Kennedy-T-92-CMP} 
{\em at least} when $S$ is an odd integer \cite{Oshikawa-92}.  
However, it was not until the concept of SPT phases was established 
that the true meaning of ``topological order'' in the Haldane phase was fully understood \cite{Gu-W-09}.   
Now it is realized that the gapped phases in integer-spin chains with some protecting symmetry 
(e.g., time-reversal, reflection) are further categorized into topological phases and trivial ones.  

\begin{figure}[htb]
\begin{center}
\includegraphics[scale=0.6]{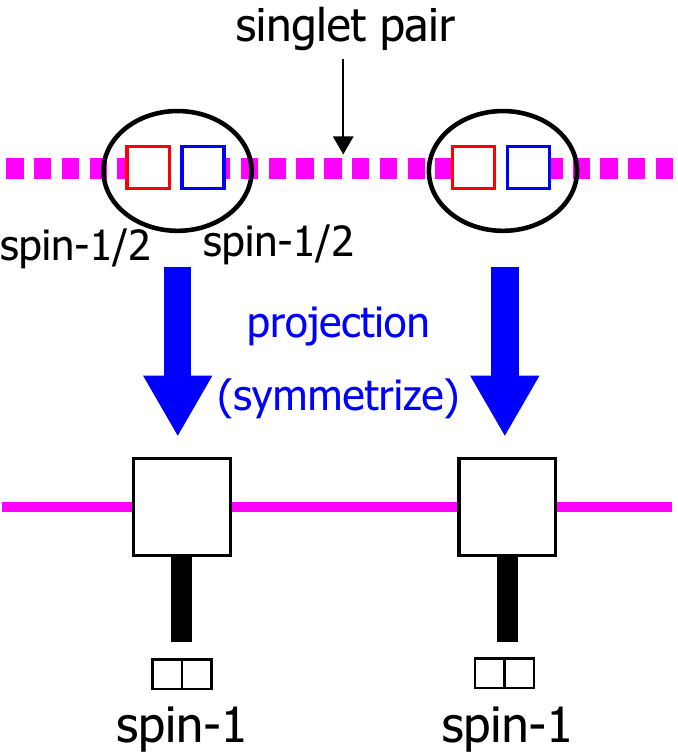}
\caption{(Color online) Valence-bond construction of the spin-1 VBS states \eqref{eqn:spin-1-VBS}.  
Two $S=1/2$ spins (ancillary qubits) at each site are symmetrized to obtain physical spin-1.  
If we replace the spin-1/2s with spin-1s and apply the anti-symmetrization at each site, 
we obtain the state \eqref{eqn:spin-1-another-VBS}.   
\label{fig:SU2-VBS}}
\end{center}
\end{figure}

To understand the properties of the Haldane phase of integer-spin antiferromagnets, 
it is convenient to consider the spin-1 VBS state introduced 
by Affleck, Kennedy, Lieb, and Tasaki \cite{Affleck-K-L-T-87,Affleck-K-L-T-88}.  
The basic idea of construction is to first decompose an $S=1$ at each site into a pair of $S=1/2$s,  
then form uniform tiling of dimer singlets (hence `valence-bond solid') among S=1/2s on neighboring sites, 
and fuse the $S=1/2$ pairs at the {\em same} site back to the original spin-1s (see Fig.~\ref{fig:SU2-VBS}).  
One of the most convenient ways of representing the VBS state is the matrix-product-state (MPS) 
representation \cite{Verstraete-M-C-08,Schollwoeck-11,Orus-review-14}\footnote{%
The symbol $\bigotimes_{i}$ implies both the matrix multiplication and the tensor product of 
the local (spin-1) Hilbert spaces.}:
\begin{equation}
\begin{split}
& |S=1\text{ VBS}\rangle_{\alpha,\beta} 
= \bigotimes_{i} \mathcal{A}_{i} \\
& = \sum_{m_i= x,y,z} \left\{ A(m_{1})\cdots A(m_i) \cdots A(m_{L}) \right\}_{\alpha,\beta}
|m_1\rangle \otimes |m_2\rangle \otimes \cdots |m_{L}\rangle \; ,
\end{split}
\label{eqn:spin-1-VBS}
\end{equation}
where $\{|x\rangle,|y\rangle,|z\rangle\}$ are defined by
\begin{subequations}
\begin{align}
& 
|x \rangle_{i} = -\frac{1}{\sqrt{2}}\left(
|+1\rangle_{i} -|-1\rangle_{i}
\right)
\label{eqn:xyz-basis1}
 \\
& 
|y \rangle_{i} = \frac{i}{\sqrt{2}}\left(
|+1\rangle_{i} +|-1\rangle_{i}
\right) 
\\
& 
|z \rangle_{i} = |0\rangle_{i} 
\label{eqn:xyz-basis2}
\end{align}
\end{subequations}
and the matrices $\mathcal{A}_{i}$ and $A(a)$ ($a=x,y,z$) are given in terms of the Pauli matrices:  
\begin{equation}
\begin{split}
& A(a) = \sigma^{a} \quad (a=x,y,z) \\
& \mathcal{A}_{i} = \sum_{a=x,y,z} A(a)|a \rangle_{i} 
= \sigma^{x}|x \rangle_{i} + \sigma^{y}|y \rangle_{i} + \sigma^{z}|z \rangle_{i}  \; .
\end{split}
\end{equation}
Remarkably, it can be shown \cite{Affleck-K-L-T-87,Affleck-K-L-T-88} that this quantum many-body state 
\eqref{eqn:spin-1-VBS} is the unique ground state\footnote{%
It is unique in a periodic system or in the thermodynamic limit.  On a finite open system, 
the ground state is four-fold degenerate because of the edge modes \cite{Affleck-K-L-T-88}.} 
of the following simple spin-1 Hamiltonian ({\em VBS model\/}):
\begin{equation}
\mathcal{H}^{N=2}_{\text{VBS}} = \sum_i \left\{
  \mathbf{S}_i \cdot \mathbf{S}_{i+1}
+   \frac{1}{3} \left(\mathbf{S}_i \cdot \mathbf{S}_{i+1}\right)^2  \right\}  \; .
\label{eqn:AKLT}
\end{equation}

It is easy to see that the VBS state \eqref{eqn:spin-1-VBS} describes a non-magnetic short-range 
phase with an excitation gap.  
To see whether the state is topologically non-trivial or not, 
it is useful to consider how the state \eqref{eqn:spin-1-VBS} transforms 
under the symmetry operation.  Being non-magnetic, {\em the bulk} does not 
respond to the symmetry operation but the edges do.  As the consequence, the symmetry operation 
gets {\em fractionalized} into two pieces; one acts on the left edge and the other on the right. 
For instance, the ground state of the spin-1 AKLT model \eqref{eqn:AKLT} $|S=1\text{ VBS}\rangle_{\alpha,\beta}$ 
hosts two {\em emergent} $S=\frac{1}{2}$ spins (i.e., $\alpha,\beta=\uparrow,\downarrow$) 
on both edges and hence transforms under the SO(3) rotation as 
\begin{equation}
|S=1\text{ VBS}\rangle_{\alpha,\beta} \xrightarrow{\text{SO(3)}} 
\sum_{\alpha^{\prime},\beta^{\prime}}U^{\dagger}_{\alpha,\alpha^{\prime}}U_{\beta,\beta^{\prime}}
|S=1\text{ VBS}\rangle_{\alpha^{\prime},\beta^{\prime}} \; ,
\label{eqn:S1-VBS-adjoint}
\end{equation}
where $U$ is the $S=\frac{1}{2}$ rotation matrix of SU(2).  
Putting it another way, $U$ serves as the mathematical labeling of the physical edge states. 
It is important to note that $U$ appearing in Eq.~\eqref{eqn:S1-VBS-adjoint}, in general, 
may be a projective representation of SO(3) as both $U^{\dagger}$ and $U$ appear in the equation. 
Since in the VBS state \eqref{eqn:spin-1-VBS} $U$ belongs to a non-trivial projective representation 
that is intrinsically different from any irreducible (integer-spin) representations of the original SO(3), one sees 
that $|S=1\text{ VBS}\rangle_{\alpha,\beta}$ is in a non-trivial topological phase with emergent edge states.  

On the other hand, one can construct another model state of a spin-1 chain: 
\begin{equation}
|S=1\text{ VBS-II} \rangle_{\alpha,\beta} 
= \bigotimes_{i} \mathcal{B}_{i} 
\label{eqn:spin-1-another-VBS}
\end{equation}
with the following $3{\times}3$ matrices $\mathcal{B}_{i}$:
\begin{equation}
\begin{split}
& \mathcal{B}_{i} = S^{x}|x\rangle_i + S^{y}|y\rangle_i +S^{z}|z\rangle_i \\
& S^{x}= 
\begin{pmatrix}
0 & 0 & 0 \\ 0 & 0 & -i \\
0 & i & 0 
\end{pmatrix}
 \; , \;\;
S^{y}= 
\begin{pmatrix}
0 & 0 & i \\ 0 & 0 & 0 \\
-i & 0 & 0 
\end{pmatrix}
 \; , \;\;
S^{z}= 
\begin{pmatrix}
0 & -i & 0 \\ i & 0 & 0 \\
0 & 0 & 0 
\end{pmatrix}
 \; .
 \end{split}
\end{equation} 
By construction, it is obvious that the above state also exhibits edge states with spin-1, and one may 
suspect that it describes a SPT state.  
Using $[S^{a}]_{bc}= -i \epsilon^{abc}$, one can easily see that 
the state \eqref{eqn:spin-1-another-VBS} transforms as before [see Eq.~\eqref{eqn:S1-VBS-adjoint}] 
but with $U$ now belonging to the spin-1 representation.  Since the spin-1 representation is trivial 
in the sense of projective representation of SO(3), one can eliminate the would-be edge states by continuously 
deforming the Hamiltonian \cite{Pollmann-B-T-O-12} and this ground state is in a trivial phase.  
This reasoning may be readily generalized; when $U$ transforms like a half-odd-integer spin, 
the phase is topological, while when $U$ transforms in an integer-spin representation [i.e., linear representation 
of SO(3)], the system is in a trivial phase.   

For later convenience, we rephrase the situation in terms of Young diagrams.  
The spin-$S$ representation of SU(2) is represented by the following Young diagram made of $2S$ boxes:
\begin{equation}
\underbrace{%
\yng(2) \cdots \yng(1)
}_{\text{$2S$ boxes}} \; .
\end{equation}
Then, the above result may be summarized as follows; when $U$ belongs to the representations 
\begin{equation}
\yng(1) \, , \; \yng(3) \, , \ldots \; , 
\end{equation}
the state represented by the corresponding MPS is topologically non-trivial
as we cannot annihilate these ``emergent'' edge spins by fusing physical integer spins on 
neighboring sites.  

On the other hand, the phase is trivial when 
\begin{equation}
\;\;
\yng(2)\, , \; \yng(4) \, , \ldots \; .
\end{equation}
That is, the number of boxes (mod 2) in the Young diagram for the representation 
to which $U$ belongs labels the topological classes protected by SO(3) and leads to 
the $\mathbb{Z}_{2}$ classification of the SO(3) SPT phases \cite{Chen-G-W-11,Chen-G-L-W-13}\footnote{%
In the argument presented here, the Haldane phase is protected by the on-site symmetry SO(3) [not SU(2)].  
However, it is known that other discrete symmetries, e.g., time-reversal and 
$\mathbb{Z}_{2}{\times}\mathbb{Z}_{2}$ can also protect 
the Haldane phase \cite{Pollmann-T-B-O-10,Pollmann-B-T-O-12}.}.  
As will be seen in Sec.~\ref{sec:phasediagrams}, 
the $N=2$ models (both $g$-$e$ and $p$-band) host several different Haldane phases 
(in the spin, orbital, and charge sectors) in their phase diagrams.   
\subsubsection{SU($N$) topological phases}
\label{sec:SUN-SPT}
Using the MPS representation \cite{Garcia-V-W-C-07} of the gapped ground state in one dimension, 
the above ``physical'' idea can be generalized and made mathematically precise.  
In fact, when a given ground state that is represented by an MPS
\begin{equation}
\sum_{\{m_{i}\}}A(m_1)A(m_2) \cdots A(m_L)|m_1\rangle{\otimes}\cdots {\otimes} |m_{L}\rangle
\label{eqn:MPS-general}
\end{equation}
is invariant under some symmetry $G$, 
a $D$-dimensional unitary matrix $U_{g}$ ($g \in G$) exists such that \cite{Garcia-W-S-V-C-08} 
\begin{equation}
A(m_i) \xrightarrow{G} \be^{i\phi_{g}} U^{\dagger}_{g} A(m_i) U_{g} \; ,
\label{eqn:MPS-UAU}
\end{equation}
where $A(m_i)$ denotes the $D{\times}D$ MPS matrices corresponding to the local state $|m_i\rangle$ and 
$\be^{i\phi_{g}}$ is a phase that depends on $g \in G$.   
As has been mentioned above, the unitary matrix $U_{g}$ is in fact a projective representation 
of the symmetry $G$, that corresponds to the physical edge states \cite{Pollmann-T-B-O-10}.  
Therefore, the enumeration of topologically stable phases in the presence of symmetry $G$ boils down to 
counting the possible (non-trivial) projective representations of $G$ \cite{Chen-G-W-11}.  
This problem was solved for SU($N$) and other Lie groups 
in Ref.~\cite{Duivenvoorden-Q-13} and the picture 
in the previous section basically generalizes to the case of SU($N$) with some mathematical complications.  

Now the role of SO(3) in the previous section is played by 
$\text{PSU($N$)} \simeq \text{SU($N$)}/\mathbb{Z}_{N}$ [note $\text{SO(3)}\simeq \text{PSU(2)}$].  
Considering $\text{PSU($N$)}$, instead of SU($N$), amounts to restricting ourselves only to 
the irreducible representations of SU($N$) specified by Young diagrams with the number of boxes 
$n_{\text{Y}}$ divisible by $N$ [i.e., $n_{\text{Y}}=N k$ ($k=0,1,\ldots$)]\footnote{%
We refer the readers who want to know more about the mathematical details to 
Sections II and III of Ref.~\cite{Duivenvoorden-Q-13}.}.   
This subset of irreducible representations roughly corresponds to the integer-spin ones in the SU(2) case.  
In view of the results of the Lieb-Schultz-Mattis argument presented in Sec~\ref{sec:LSM}, 
considering the symmetry PSU($N$) is quite reasonable for the hunt for gapped topological phases 
in one dimension, where no genuine topological phase with topologically degenerate ground states 
exists \cite{Chen-G-W-10}.   

As in the previous section, 
the topological class of a given ground state (typically written as an MPS) is determined by 
looking at to which projective representation the unitary $U_{g}$ of the state belongs.  
Since inequivalent projective representations of PSU($N$) are labeled 
by $n_{\text{Y}}$ (mod $N$) \cite{Duivenvoorden-Q-13}, there are $N-1$ non-trivial topological 
classes specified by the $\mathbb{Z}_{N}$ label $n_{\text{top}}=n_{\text{Y}}$ (mod $N$).   
In the following, we use the name ``class-$n_{\text{top}}$'' for the topological phase 
with $n_{\text{top}}= n_{\text{Y}}$ (the class-0 corresponds to trivial phases).   

We can think of other protecting symmetries associated with SU($N$) \cite{Duivenvoorden-Q-13}.  
For instance, when we take $\text{SU(4)}/\mathbb{Z}_{2} \simeq \text{SO(6)}$, 
we consider only the SU(4) irreducible representations with $n_{\text{Y}}=0$ (mod $2$), 
which may be viewed as the linear representations of SO(6) (containing no spinor representation).    
In this case, we are led to a $\mathbb{Z}_{2}$-classification. 

\begin{figure}[htb]
\begin{center}
\includegraphics[scale=0.7]{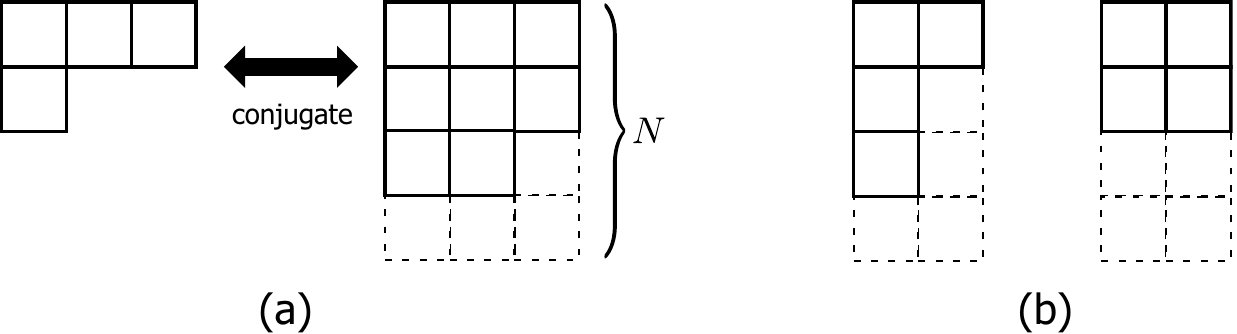}
\caption{(a) Young diagrams for an SU($N$) representation 
and its conjugate for $N=4$. 
(b) Examples of self-conjugate representations ($N=4$).   
\label{fig:Young-conjugation}}
\end{center}
\end{figure}

A remark is in order about the definition of the topological class. 
In contrast to the SU(2) case where all the irreducible representations are self-conjugate, 
we must distinguish between an irreducible representation and its conjugate in SU($N$) 
(see Fig.~\ref{fig:Young-conjugation}).  
The relation \eqref{eqn:MPS-UAU} suggests that if we have the edge state transforming under 
the projective representation $\mathcal{R}$ [of PSU($N$)] on the right edge, we necessarily have 
its conjugate $\bar{\mathcal{R}}$ on the left.   This means that when we talk about 
the topological class we must first fix which edge state we use to label the topological phases.  
Throughout this paper, 
we define the topological class by the {\em right edge state} [i.e., by $U_{g}$ acting {\em from the right} 
in Eq.~\eqref{eqn:MPS-UAU}].  
We will see, in the next section, that the SU($N$) VBS states to be discussed in Sec.~\ref{sec:solvable-Ham} 
belongs to class-$N/2$.  
\subsubsection{VBS construction of the model wave functions}
\label{sec:solvable-Ham}
In the previous section, we have seen that there are $N$ topologically distinct phases 
in the presence of $\text{PSU($N$)} (\simeq \text{SU($N$)}/\mathbb{Z}_{N})$ symmetry 
\cite{Duivenvoorden-Q-13}.  To understand their physical properties, it is convenient to analyze 
the corresponding `fixed-point' states that are easy to analyze.  
In the Haldane phase of integer-spin  antiferromagnets, the spin-1 VBS state \eqref{eqn:spin-1-VBS} 
discussed in Sec.~\ref{sec:Haldane-phase} does the job.  

\begin{figure}[htb]
\begin{center}
\includegraphics[scale=0.6]{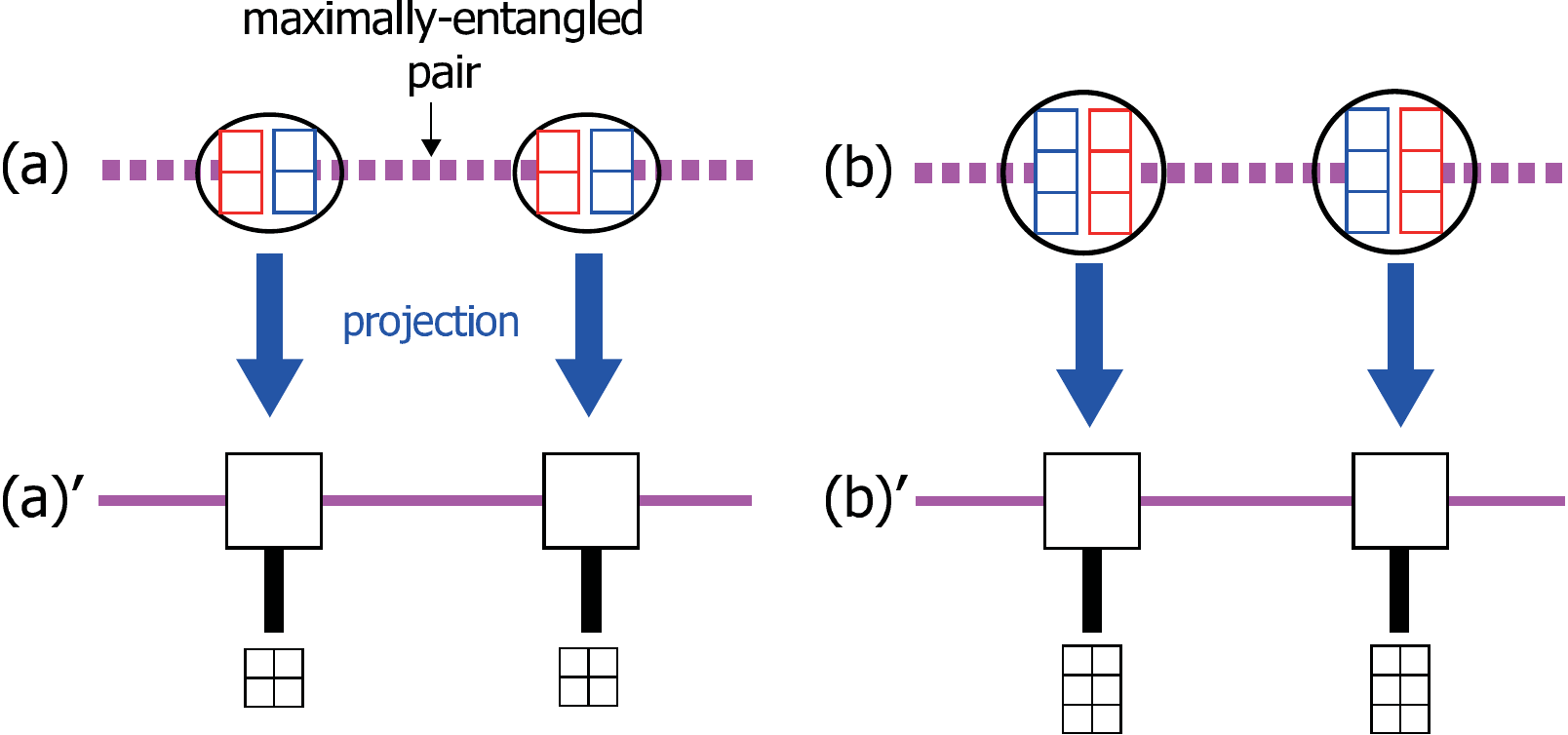}
\caption{(Color online) Valence-bond construction of the topological VBS states.  
(a) and (a)': class-2 SU(4) VBS state.  (b) and (b)': class-3 SU(6) VBS state.  From Ref.~\cite{Bois-C-L-M-T-15}.
\label{fig:VB-construction}}
\end{center}
\end{figure}

A similar strategy applies to SU($N$) ($N$: even, $N \geq 4$; see Fig.~\ref{fig:VB-construction}) 
to construct the SU($N$) generalization of the VBS state.   
For instance, we use a pair of six-dimensional representations (${\tiny \yng(1,1)}$) to 
build the SU(4) VBS state shown in Fig.~\ref{fig:VB-construction}(a) \cite{Nonne-M-C-L-T-13,Bois-C-L-M-T-15}.   
At the last stage of the construction, we project the tensor product ${\tiny \yng(1,1)}{\otimes}{\tiny \yng(1,1)}$ 
onto the 20-dimensional representation ${\tiny \yng(2,2)}$ to obtain the physical Hilbert space. 
The resulting state may be conveniently expressed in the form of the MPS \eqref{eqn:MPS-general} 
using $6\times 6$ matrices $A(m)$ \cite{Tanimoto-T-14}.  
The parent Hamiltonian of this SU(4) VBS state reads as\footnote{%
In fact, the parent Hamiltonian contains two free parameters aside from the overall factor.  
Requiring that $\left(\mathcal{S}_i \cdot \mathcal{S}_{i+1}\right)^4$ and 
$\left(\mathcal{S}_i \cdot \mathcal{S}_{i+1}\right)^5$ should not appear, we obtain Eq.~\eqref{eqn:SU4-VBS}.} 
\cite{Nonne-M-C-L-T-13,Bois-C-L-M-T-15}
\begin{equation}
\mathcal{H}^{N=4}_{\text{VBS}}  = \sum_i \left\{
  \mathcal{S}_i \cdot \mathcal{S}_{i+1}
+   \frac{13}{108} \left(\mathcal{S}_i \cdot \mathcal{S}_{i+1}\right)^2 
+ \frac{1}{216}\left(\mathcal{S}_i \cdot \mathcal{S}_{i+1}\right)^3  \right\} \;  ,
\label{eqn:SU4-VBS}
\end{equation}
where we have introduced a short-hand notation 
$\mathcal{S}_i \cdot \mathcal{S}_{i+1} \equiv \sum_{A=1}^{N^{2}-1}\mathcal{S}^{A}_{i}\mathcal{S}^{A}_{i+1}$.  
Similarly, from a pair of 20-dimensional representations ${\tiny \yng(1,1,1)}$ of SU(6), 
we can construct the VBS ground state of the following Hamiltonian 
[with $\mathcal{S}_i$ belonging to the 175-dimensional representation ${\tiny \yng(2,2,2)}$; 
see Fig.~\ref{fig:VB-construction}(b)] \cite{Tanimoto-T-14}\footnote{%
After completing this paper, we were informed that T.~Quella et al. had independently obtained similar results 
in their unpublished work \cite{Quella-unpublished}.}
\begin{equation}
\begin{split}
\mathcal{H}^{N=6}_{\text{VBS}} =&  \sum_{i} \biggl\{
\mathcal{S}_i \cdot \mathcal{S}_{i+1} 
+ \frac{47}{508}(\mathcal{S}_i \cdot \mathcal{S}_{i+1} )^2  \\
& +\frac{17}{4572} (\mathcal{S}_i \cdot \mathcal{S}_{i+1} )^3 
+\frac{1}{18288} (\mathcal{S}_i \cdot \mathcal{S}_{i+1} )^4 
\biggr\} \; .
\end{split}
\label{eqn:SU6-VBS}
\end{equation}

\begin{figure}[htb]
\begin{center}
\includegraphics[scale=0.6]{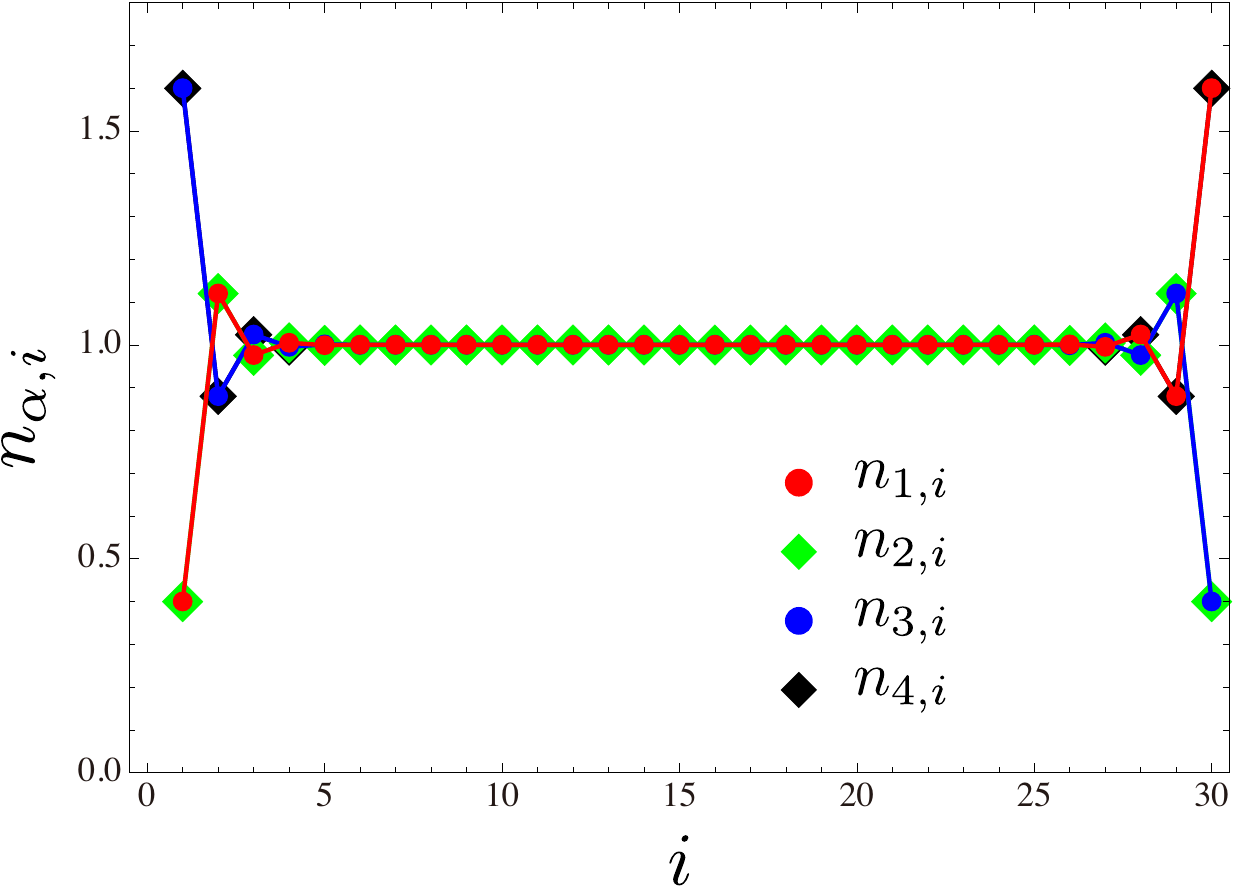}
\caption{(Color online) Local fermion density 
$n_{\alpha,i}=c_{g\alpha,i}^{\dagger}c_{g\alpha,i}+c_{e\alpha,i}^{\dagger}c_{e\alpha,i}$  
calculated for one of the 36-fold degenerate SU(4) VBS states on a finite open chain with 30 sites.  
There are six edge states as two of the four $n_{\alpha,i}$ take the same value 
(and so do the other two) around the edge.    Note that the local constraint $\sum_{\alpha=1}^{4}n_{\alpha,i}=4$ 
is satisfied at each site. 
\label{fig:local-density-SU4-VBS}}
\end{center}
\end{figure}

To be specific, let us restrict ourselves to the case of SU(4) to describe the main features of the VBS wave function.  
Most of the properties are carried over to general SU($N$) ($N$: even) with due modifications. 
The ground state is SU(4)-symmetric and featureless {\em in the bulk},  
and has the `spin-spin' correlation functions \cite{Bois-C-L-M-T-15}
\begin{equation}
\langle \mathcal{S}^{A}_{j} \mathcal{S}^{A}_{j+n} \rangle 
= 
\begin{cases}
\frac{12}{5} \left(-\frac{1}{5} \right)^n   &  n \neq 0 \\
\frac{4}{5} & n=0 
\end{cases}
\end{equation}
that are exponentially decaying 
with a very short correlation length $1/\ln 5\approx 0.6213$.   
Nevertheless, the system hosts emergent edge states  near the boundaries.  
In fact, if one measures $\VEV{\mathcal{S}^{A}_{i}}$ (with $\mathcal{S}^{A}_{i}$ being 
any three commuting generators of SU(4), or equivalently, independent linear combinations of 
local fermion densities $n_{\alpha,i}=c_{g\alpha,i}^{\dagger}c_{g\alpha,i}+c_{e\alpha,i}^{\dagger}c_{e\alpha,i}$), 
one can clearly see the structure localized around 
the two edges (Fig.~\ref{fig:local-density-SU4-VBS}.  See also the bottom panel of Fig.~\ref{fig:pbandN4} 
for a similar plot for the original fermionic model).  
At each edge, there are six different states (i.e., ${\tiny \yng(1,1)}$) distinguished by the value of the set 
of the three generators $\VEV{S^{A}_{i}}$.   This implies that $U_{g}$ in Eq.~\eqref{eqn:MPS-UAU} 
transforms like ${\tiny \yng(1,1)}$, telling us that the ground state of the VBS Hamiltonian 
\eqref{eqn:SU4-VBS} falls into the class-2 topological phase protected by the on-site PSU(4) symmetry. 

\begin{figure}[htb]
\begin{center}
\includegraphics[scale=0.7]{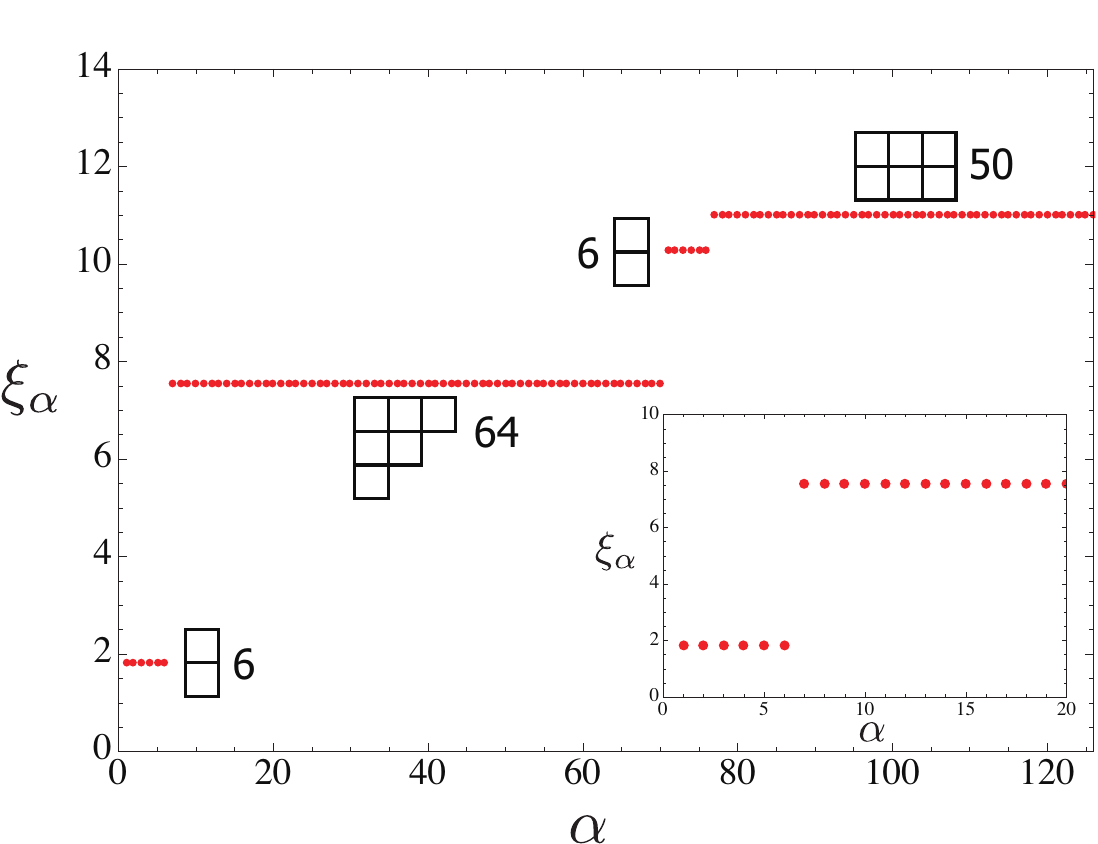}
\caption{(Color online) 
Entanglement spectrum of the SU(4) Heisenberg model \eqref{eqn:2nd-order-effective-Ham-Gorshkov}. 
From Ref.~\cite{Tanimoto-T-14}
\label{fig:ES_a00}}
\end{center}
\end{figure}

Corresponding to the physical edge states, the entanglement spectrum exhibits a particular degeneracy 
structure.  In fact, the entanglement spectrum of the SU(4) VBS state consists of a {\em single} 
six-fold degenerate level, which leads to the von Neumann (bipartite) entanglement entropy 
$S_{\text{vN}}=\ln 6$.\footnote{%
By construction, it is obvious that for general $N$ (even), the single entanglement level of 
the VBS state (Fig.~\ref{fig:VB-construction}) is $N!/[(N/2)!]^{2}$-fold degenerate.} 
This extremely simple structure is peculiar to the solvable VBS model \eqref{eqn:SU4-VBS} 
and, for a generic SU(4) Hamiltonian in the same class-2 topological phase, 
we expect the degeneracy structure compatible with 
the SU(4) irreducible representations having $n_{\text{Y}}=2$ boxes.  
Fig.~\ref{fig:ES_a00} is the plot of the entanglement spectrum of the pure SU(4) Heisenberg model 
\eqref{eqn:exch-coupling-Gorshkov} obtained \cite{Tanimoto-T-14} 
by infinite time-evolving block decimation (iTEBD) method 
\cite{Vidal-iTEBD-07,Orus-V-08}.  One can clearly see that the level structure is totally consistent with 
the topological class-2.  By linearly interpolating between the VBS model \eqref{eqn:SU4-VBS} 
and the pure Heisenberg model \eqref{eqn:exch-coupling-Gorshkov}, we can show \cite{Tanimoto-T-14} 
that the overall structure of the entanglement spectrum is maintained indicating that both 
models belong to the same unique SPT phase.  

Before concluding the discussion of the SU($N$) SPT phases, a few remarks are in order. 
One is about the other topological phases.  The group-cohomology classification tells us that 
there are two more topological classes in PSU(4)-invariant systems \cite{Duivenvoorden-Q-13}.  
In fact, the VBS states studied in Refs.~\cite{Affleck-K-L-T-88,Katsura-H-K-08,Korepin-X-10,Orus-T-11} 
correspond to these two.  The entanglement spectra calculated for these VBS states agree with 
those expected from the class-1 and 3 topological phases.  
Quite recently,  the phases of SU(3)-invariant spin chains, including 
the SU(3)-counterpart of the above two phases were investigated in Ref.~\cite{Morimoto-U-M-F-14}.   

Also interesting is the characterization of these topological phases with non-local order parameters. 
It has been known that the Haldane phase of SU(2) spin chains is characterized by a pair of 
non-local order parameters ({\em string order parameters}) 
\cite{denNijs-R-89,Kennedy-T-92-PRB,Kennedy-T-92-CMP,Oshikawa-92,Totsuka-S-mpg-95}.  
These non-local order parameters are not only useful as `working indicators' of the topological Haldane phase 
but also have intimate connection to the modern characterization of the 1D SPT phases in terms of 
the projective representation $U_g$ [see Eq.~\eqref{eqn:MPS-UAU}] \cite{Pollmann-T-12,Hasebe-T-13}.  

This concept can be generalized to other symmetries 
\cite{Duivenvoorden-Q-12,Else-B-D-13,Duivenvoorden-Q-ZnxZn-13}.  
For instance, a set of $2(N-1)$ non-local `string' order parameters distinguish between the $N$ topologically 
distinct phases predicted by group cohomology \cite{Tanimoto-T-14}.   The characterization of the SPT phases 
by non-local correlation functions seem quite interesting in view of the recent development in real-space imaging 
techniques \cite{Endres-etal-stringOP-11} as some of the string order parameters are written only 
in terms of fermion densities which are, in principle, detectable in experiments.  

Last, we comment on the relation between the trap potentials and the edge states characteristic of 
SPT phases.  In the usual setting of harmonic potentials, it is known \cite{Rigol-M-B-S-03,Schneider-etal-MIT-08}
that an island of Mott insulating region 
({\em Mott core}), that is surrounded by a compressible metallic region, is formed around the center of the trap; 
the above argument for SPT phases holds only within the Mott core.  Due to the interaction between 
the Mott core and the metallic region surrounding it, the structures around the (smeared) edges 
may be substantially reduced\footnote{%
Especially in the case of, e.g., the charge Haldane phase (see Sec.~\ref{sec:PhaseDiag-N-2-g-e} 
and Fig.~\ref{fig:4-MottPhases-SU2} for the definition) where the edge states directly couple to 
the trapping potential, the inhomogeneity of the potential causes destructive effects.} 
\cite{Kobayashi-O-O-Y-M-12,Kobayashi-O-O-Y-M-14}.   
However, thanks to the recent success in creating a box trap \cite{Gaunt2013}, one may resolve this problem.  
Combining the box trap and the technology of single-site detection (see \cite{Gross-B-review-15} for a recent review), 
quantum simulation of various SPT phases [including our SU($N$) ones] will become feasible in the near 
future.  
\subsection{Example of phase diagrams}
\label{sec:phasediagrams}
We will now turn to the presentation of some numerical phase diagrams that were obtained 
using DMRG simulations on finite chains of length $L$ with open boundary conditions.  
Since we focus on particular phases occurring at fixed densities, we do not consider the (harmonic) trapping potential. In real experiments, detection could be achieved using local quantities. Let us emphasize that we will discuss mostly gapped phases, so that they are stable with respect to small perturbations and should be more robust for observation. 
As to the simulations, let us simply mention that we have only implemented the $N$ U(1) quantum 
numbers corresponding to the separate conservation of the fermion numbers of each color, 
though it should be possible to ease simulations by implementing the full SU($N$) symmetry of the models~\cite{Weichselbaum2012}.  We refer to Ref.~\cite{Bois-C-L-M-T-15} for more technical details. 

For the sake of clarity, in this section, we restrict ourselves to the half-filled cases which already exhibit 
a rich variety of phases, including topological ones.  
Away from half-filling, one expects on general grounds the occurrence of gapless Luttinger liquid-like phases 
with (diagonal) density or superconducting (i.e. off-diagonal) correlations \cite{Bois-C-L-M-15} 
or degenerate Mott phases \cite{Szirmai-13} for some other (in)commensurate fillings.  

For simplicity, when considering the $g$-$e$ model (\ref{eqn:Gorshkov-Ham}), we will assume that 
both orbitals are equivalent, so that they have identical hoppings $t_g=t_e=t$, 
chemical potentials $\mu_g=\mu_e=\mu$, 
and interactions $U_{gg}=U_{ee}=U_{mm}$.  
Therefore, we will consider neither the case with spin-imbalance 
where spin-polarization effects may dominate~\cite{Kobayashi-O-O-Y-M-14} 
and give rise to FFLO physics, nor the strongly anisotropic case where one of the orbitals would be 
much more localized than the other (e.g., $t_e\ll t_g$) hence giving rise to physics of the SU($N$) Kondo lattice model~\cite{Coleman1983,Gorshkov-et-al-10,FossFeig2010,FossFeig2010a,Isaev2015,Nakagawa-K-15}.
\subsubsection{$N=2$ $g$-$e$ model}
\label{sec:PhaseDiag-N-2-g-e}
\begin{figure}[!htb]
\begin{center}
\centerline{
\includegraphics[width=0.5\textwidth]{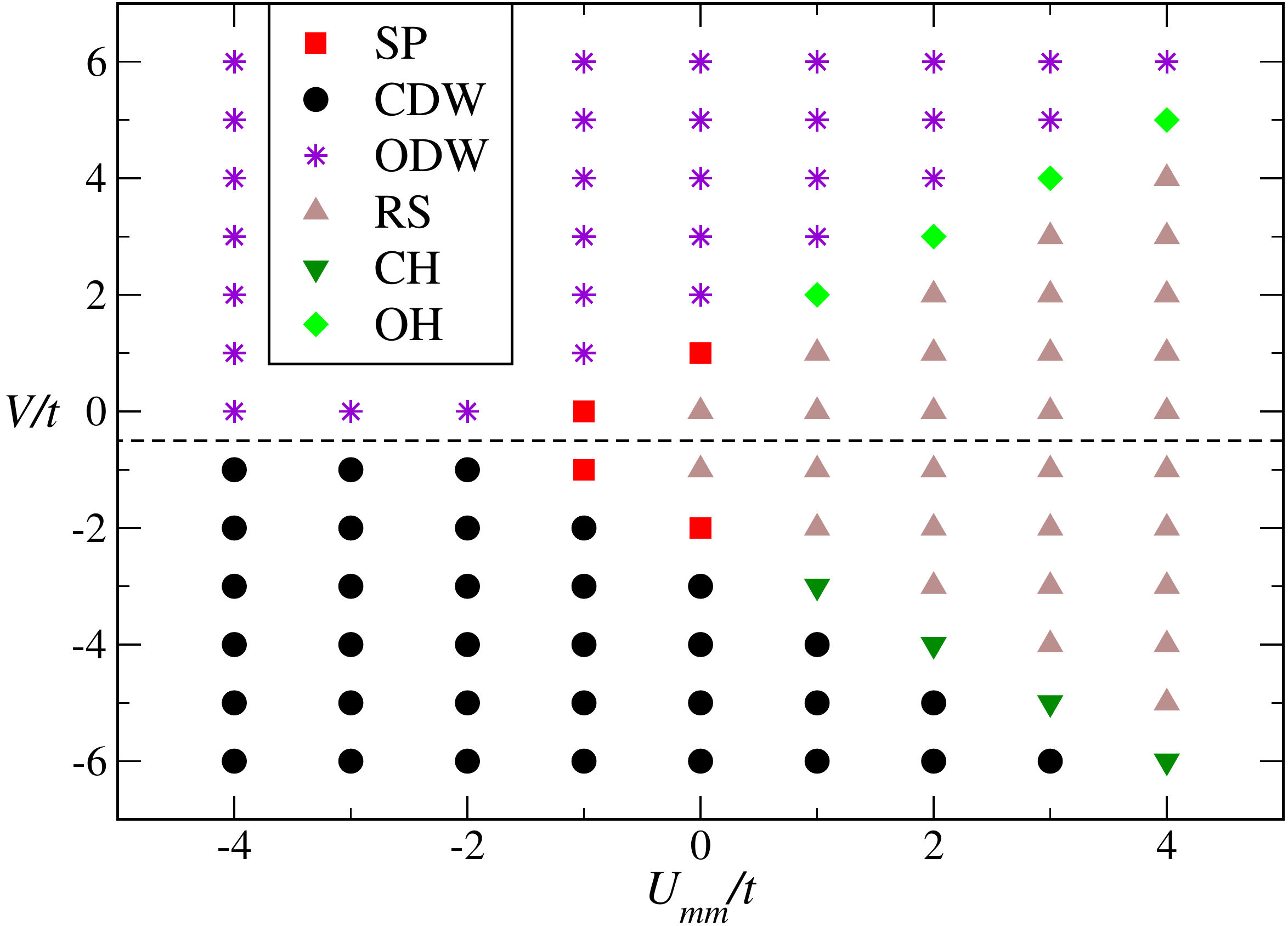}
\hspace{3mm}
\includegraphics[width=0.5\textwidth]{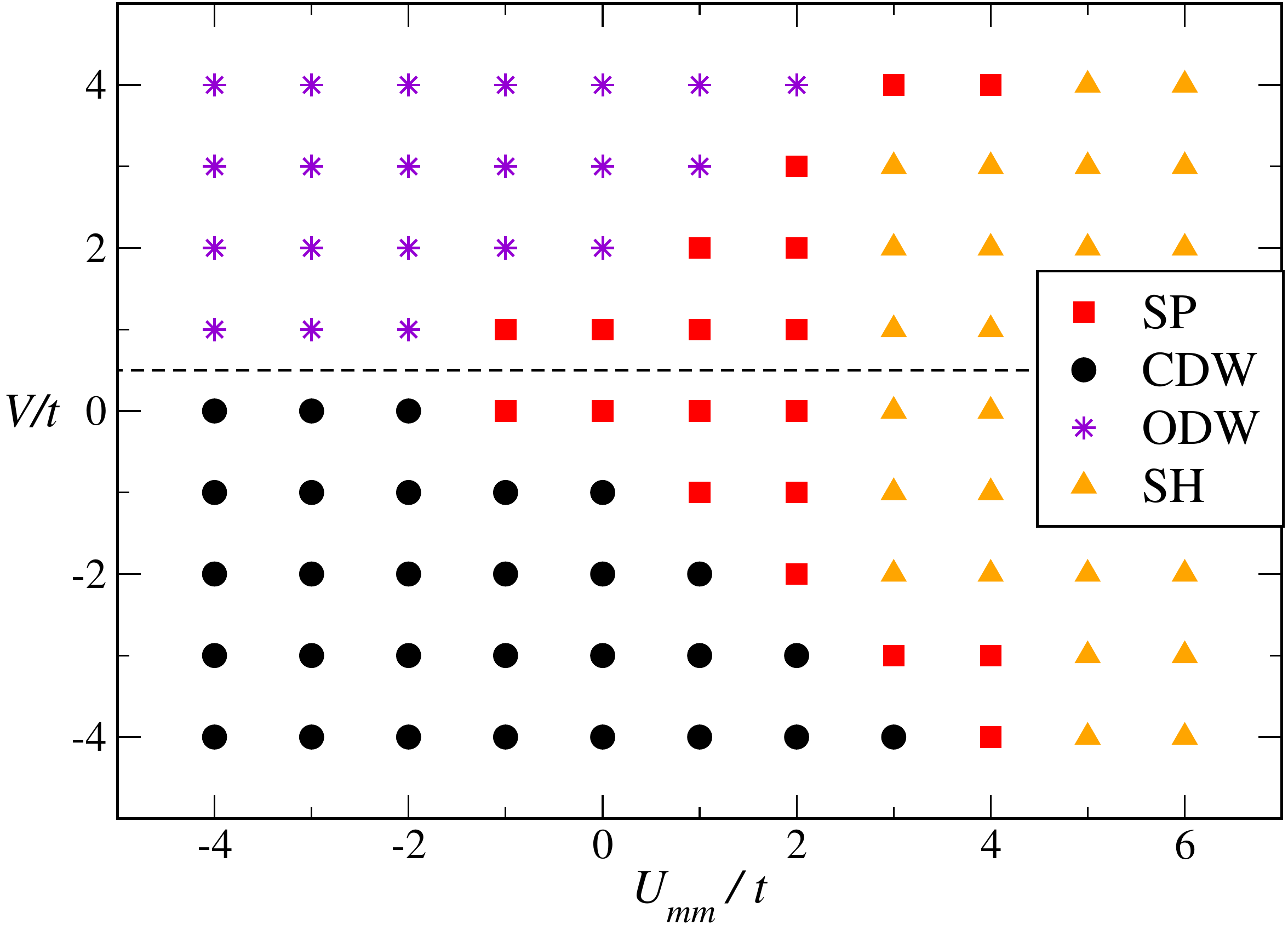}
}
\caption{(Color online) Phase diagram of the $g$-$e$ model with $N=2$ at half-filling. 
Left and right panels correspond respectively to $V_{\text{ex}}^{g\text{-}e}/t=-1$ and $V_{\text{ex}}^{g\text{-}e}/t=1$, which allows to realize various phases: spin-Peierls (SP), charge/orbital density waves (CDW/ODW), rung singlet (RS), charge-Haldane (CH), orbital-Haldane (OH) and spin-Haldane (SH). 
\label{fig:egN2}}
\end{center}
\end{figure}

\begin{figure}[!htb]
\begin{center}
\includegraphics[scale=0.85]{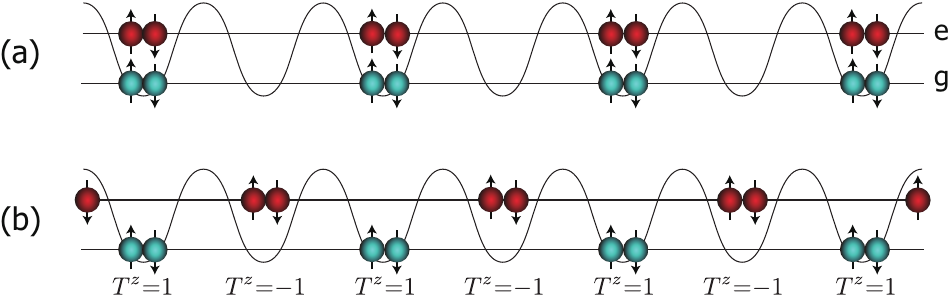}
\caption{(Color online) 
Two density-wave states for $N=2$. In-phase and out-of-phase combinations of two density waves in $g$ and $e$ orbitals respectively form (a) CDW and (b) ODW.  Adapted from Ref.~\cite{Bois-C-L-M-T-15}. 
\label{fig:2-DWPhases-SU2}}
\end{center}
\end{figure}

\begin{figure}[!htb]
\begin{center}
\includegraphics[scale=0.85]{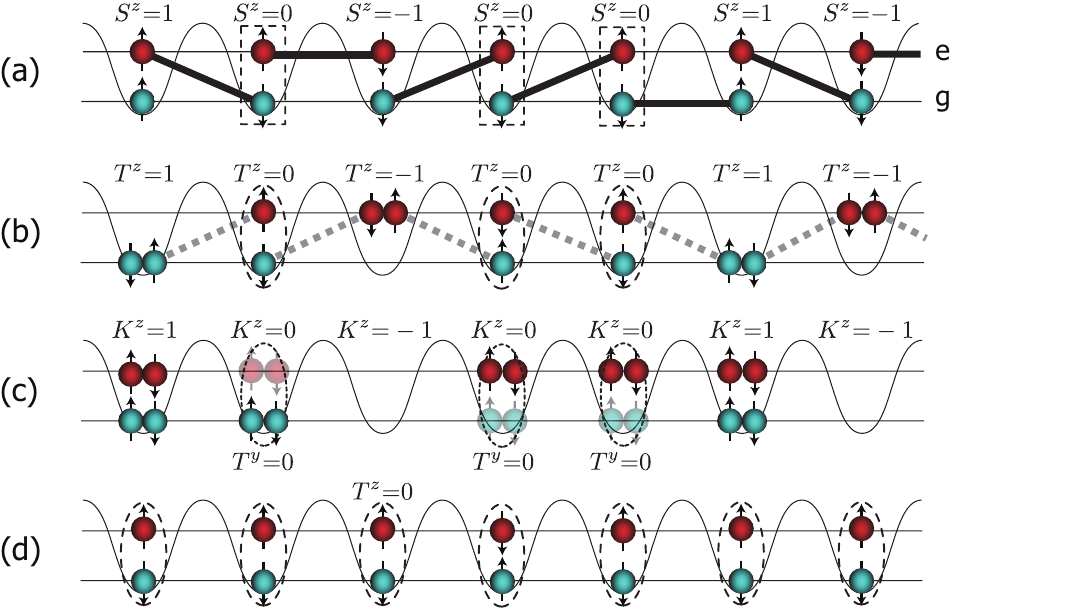}
\caption{(Color online) Four translationally invariant insulating states for
  $N=2$. (a) spin Haldane (SH), (b) orbital Haldane (OH), (c) charge
  Haldane (CH), and (d) rung-singlet (RS) phases. Singlet bonds formed
  between spins (orbital pseudo-spins) are shown by thick solid
  (dashed) lines (singlet bonds are not shown in (c)). 
  Charge pseudo-spin is defined as $K^{z}_{i}=n_{i}/2 -1$ and the $K^{z}=0$ state is a linear superposition 
  of $c_{g\uparrow}^{\dagger}c_{g\downarrow}^{\dagger}|0\rangle$ and 
  $c_{e\uparrow}^{\dagger}c_{e\downarrow}^{\dagger}|0\rangle$ (state with $T^{y}=0$).   
  Dashed ovals
  (rectangles) denote spin-singlets
  (triplets).  Adapted from Ref.~\cite{Bois-C-L-M-T-15}. 
\label{fig:4-MottPhases-SU2}}
\end{center}
\end{figure}

We present in Fig.~\ref{fig:egN2}
typical phase diagrams of the $g$-$e$ model (\ref{eqn:Gorshkov-Ham}) 
with $N=2$ which exhibit a large variety of phases: (i) charge density wave (CDW), (ii) orbital density wave (ODW), 
(iii) spin-Peierls (SP), (iv) charge Haldane (CH), (v) orbital Haldane (OH), (vi) spin Haldane (SH), 
and (vii) rung singlet (RS) (see Fig.~\ref{fig:4-MottPhases-SU2} and Table~\ref{tab:abbreviation}).    
The CH phase is the collective Haldane state formed by the (spin-singlet) charge pseudo-spin-1 states:   
$c_{g1}^{\dagger}c_{g2}^{\dagger}c_{e1}^{\dagger}c_{e2}^{\dagger}|0\rangle$ ($n=4$), 
$c_{g1}^{\dagger}c_{g2}^{\dagger}|0\rangle+c_{e1}^{\dagger}c_{e2}^{\dagger}|0\rangle$ ($n=2$), and 
$|0\rangle$ ($n=0$) [see Eq.~\eqref{eqn:cgarge-pseudo-spin} for the definition of the charge pseudo-spin] 
and is characterized by the existence of non-local string order in the charge distribution.   
Clearly, it requires strong charge fluctuations and is {\em not} a Mott insulator.  
Similar `Haldane states' in the charge sector have been found in the study of related multi-component fermions 
\cite{Nonne-L-C-R-B-10,Nonne-L-C-R-B-11,Kobayashi-O-O-Y-M-14}.
As CH is a collective insulating state with a charge gap analogous to the spin gap in the Haldane-gap systems, 
it, together with its bosonic counterpart, is also dubbed  
{\em Haldane insulator} in literatures \cite{Torre-B-A-06,Berg-T-G-A-08,Batrouni-S-R-G-13}. 
To give a simple picture of these phases, we provide, respectively in Figs.~\ref{fig:2-DWPhases-SU2} 
and \ref{fig:4-MottPhases-SU2}, a cartoon of the two-fold degenerate possible density waves (CDW and ODW) 
and the non-degenerate Mott phases  (SH,CH, and RS) that we have found. 

These very rich phase diagrams are in rather good agreement with the low-energy predictions, and they were already discussed in Refs.~\cite{Nonne-B-C-L-11,Bois-C-L-M-T-15}.  
In Fig.~\ref{fig:egN2}, one notices that the phases concerning the charge sector 
(CDW and CH) and those concerning the orbital sector (ODW and OH) appear in a very symmetric manner.  
In fact, this is a direct consequence of the following symmetry (a generalization of the Shiba duality) 
that the $N=2$ $g$-$e$ model possesses \cite{Bois-C-L-M-T-15}:
\begin{equation}
\begin{split}
& V \rightarrow -V + V_{\text{ex}}^{g\text{-}e}  \\
& V_{\text{ex}}^{g\text{-}e} \rightarrow V_{\text{ex}}^{g\text{-}e} \, , \;\;
 U_{mm} \rightarrow  U_{mm}  \; ,
\end{split}
\end{equation}
that swaps a phase related to charge and the corresponding orbital phase. 
\subsubsection{$N=2$ $p$-band model}
We present here some data for the $p$-band model \eqref{eqn:p-band} with $N=2$ at half-filling.  
We refer the readers to Ref.~\cite{Bois-C-L-M-T-15} for the full phase diagram.  
We focus here along the line $U_1=3U_2$ which could be realized using the harmonic trapping 
[precisely, traps that are axially symmetric with respect to the chain direction; 
see Eq.\eqref{eqn:U1-U2-axially-symmetric}] 
and we simply plot some local quantities, that would be most easily measured experimentally using 
the existing quantum-optical techniques, namely: local densities and kinetic energy
\begin{equation}
\begin{split}
& n_{\alpha}(i) =  \left\langle \sum_{m=p_x,p_y}  c_{m\alpha,i}^{\dagger}c_{m\alpha,i} \right\rangle \\ 
& E_{\text{kin}}(i) =  
\left\langle \sum_{\alpha=\uparrow,\downarrow}\sum_{m=p_x,p_y}  c_{m\alpha,i}^{\dagger}c_{m\alpha,i+1} + \text{H.c.} \right \rangle \; .
\end{split}
\label{eq:def_local}
\end{equation}

\begin{figure}[!htb]
\begin{center}
\centerline{
\includegraphics[width=0.5\textwidth]{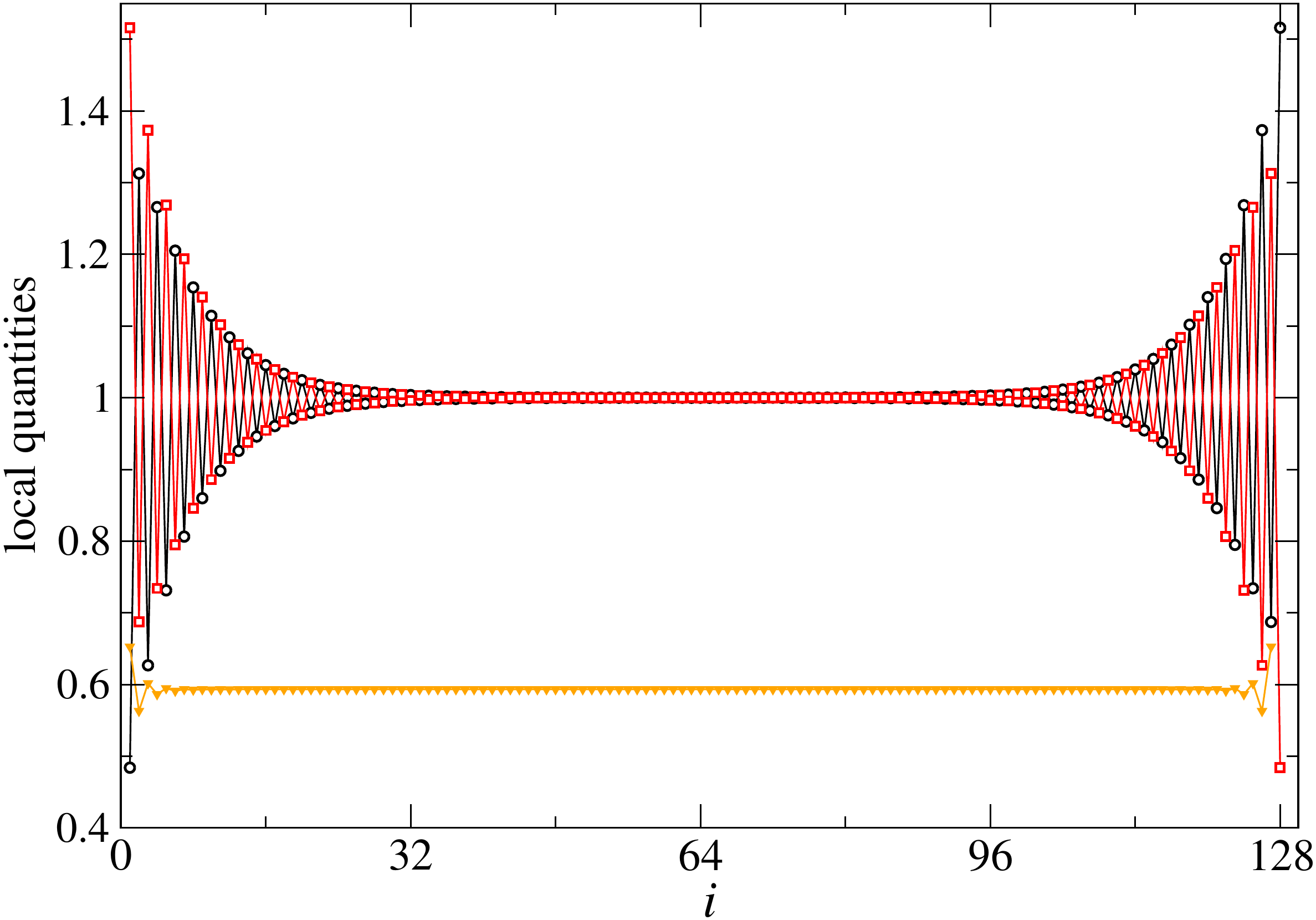}
\hspace{3mm}
\includegraphics[width=0.5\textwidth]{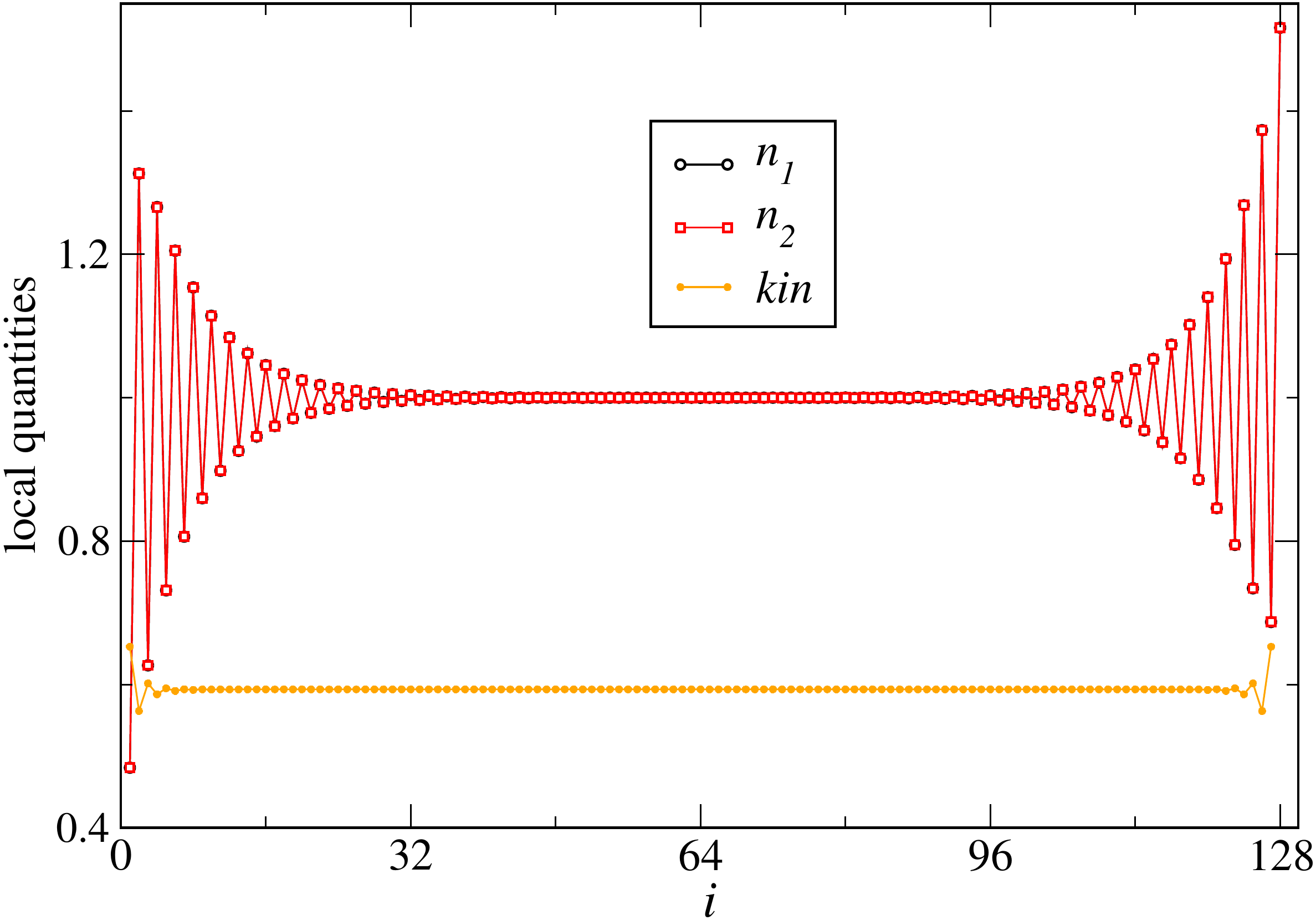}
}
\caption{(Color online) Local quantities (fermion density and kinetic energy) 
for the $p$-band model at half-filling with $N=2$ obtained on $L=128$ chain using DMRG.
Left panel: $(U_1/t,U_2/t)=(12,4)$ exhibits spin edge states characteristics of SH phase.  
Right panel:  $(U_1/t,U_2/t)=(-12,-4)$ exhibits charge edge states characteristics of CH phase. 
\label{fig:pbandN2}}
\end{center}
\end{figure}

In Fig.~\ref{fig:pbandN2}, one can clearly identify the edge states both for repulsive 
($U_1=3U_2 >0$) and attractive ($U_1=3U_2 <0$) interactions.  
They correspond respectively to the induced edge states in the spin and the charge Haldane phases (SH and CH; 
see Fig.~\ref{fig:4-MottPhases-SU2}) 
and are simple manifestations of the topological properties in the bulk (see the discussion in Sec.~\ref{sec:SPT}).  
The existence of the CH phase in fermionic systems was pointed out 
in Refs.~\cite{Nonne-L-C-R-B-10,Nonne-L-C-R-B-11} for a related but different multi-component Fermi system.  
The CH phase in the $p$-band model was first found in Ref. \cite{Kobayashi-O-O-Y-M-14}.

Let us remind the readers that, for this model, the sign of the interaction can be flipped 
formally [$(U_1,U_2) \to (-U_1,-U_2)$]  
using spin-charge interchange transformation \cite{Kobayashi-O-O-Y-M-14} 
similar to the one used in Sec.~\ref{sec:halfilling} to discuss the attractive ($U<0$) side of 
the half-filled single-band SU(2) Hubbard chain (see also the discussion in Sec.~\ref{sec:p-band-hamiltonian}).  
Because of this spin-charge symmetry in the $N=2$ $p$-band model, SH and CH appear 
in a symmetric manner in the phase diagram (see Fig.~13 of Ref.~\cite{Bois-C-L-M-T-15}).  
\subsubsection{$N=4$ $g$-$e$ model}
\begin{figure}[!htb]
\begin{center}
\centerline{
\includegraphics[width=0.5\textwidth]{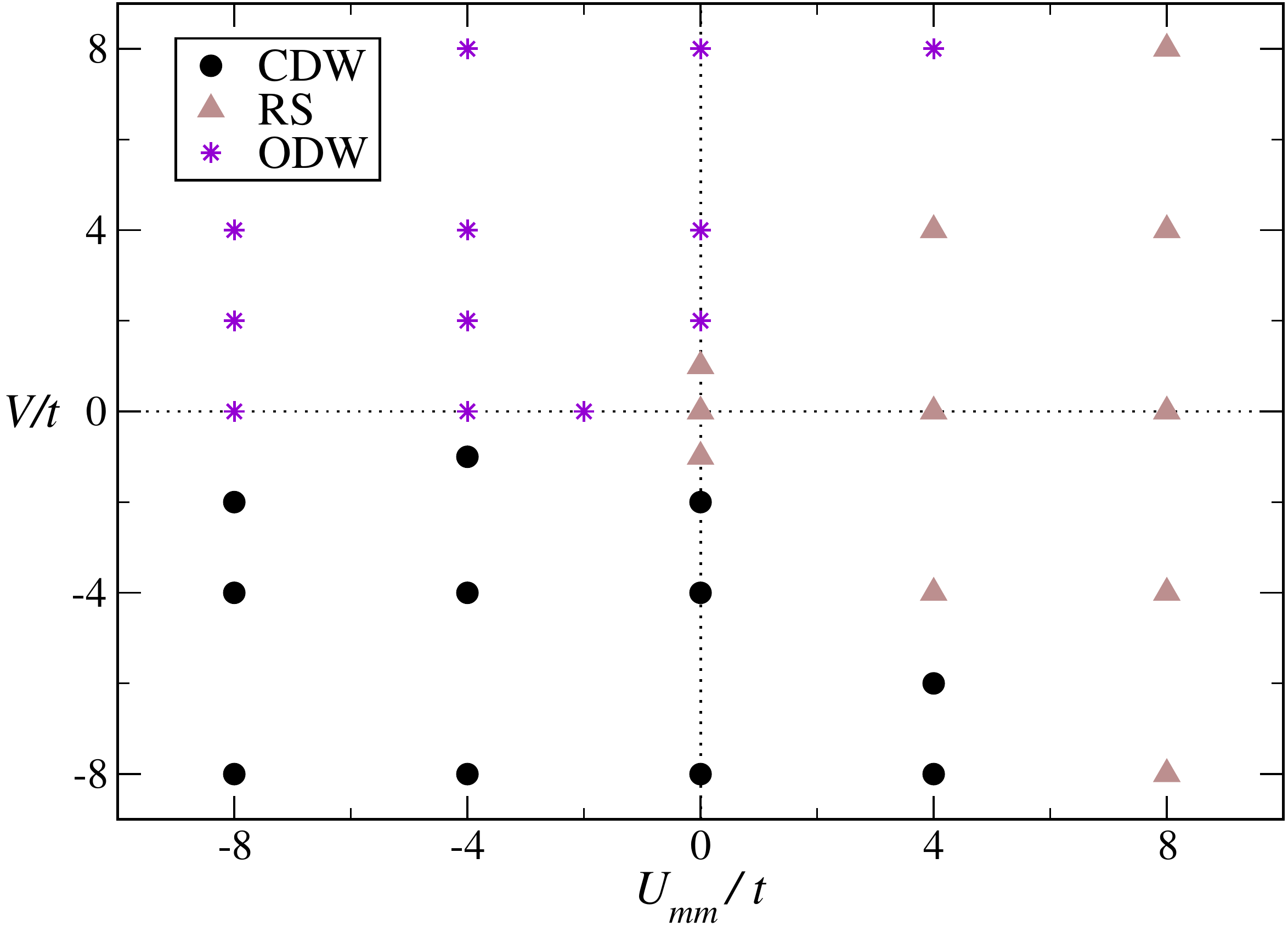}
\hspace{3mm}
\includegraphics[width=0.5\textwidth]{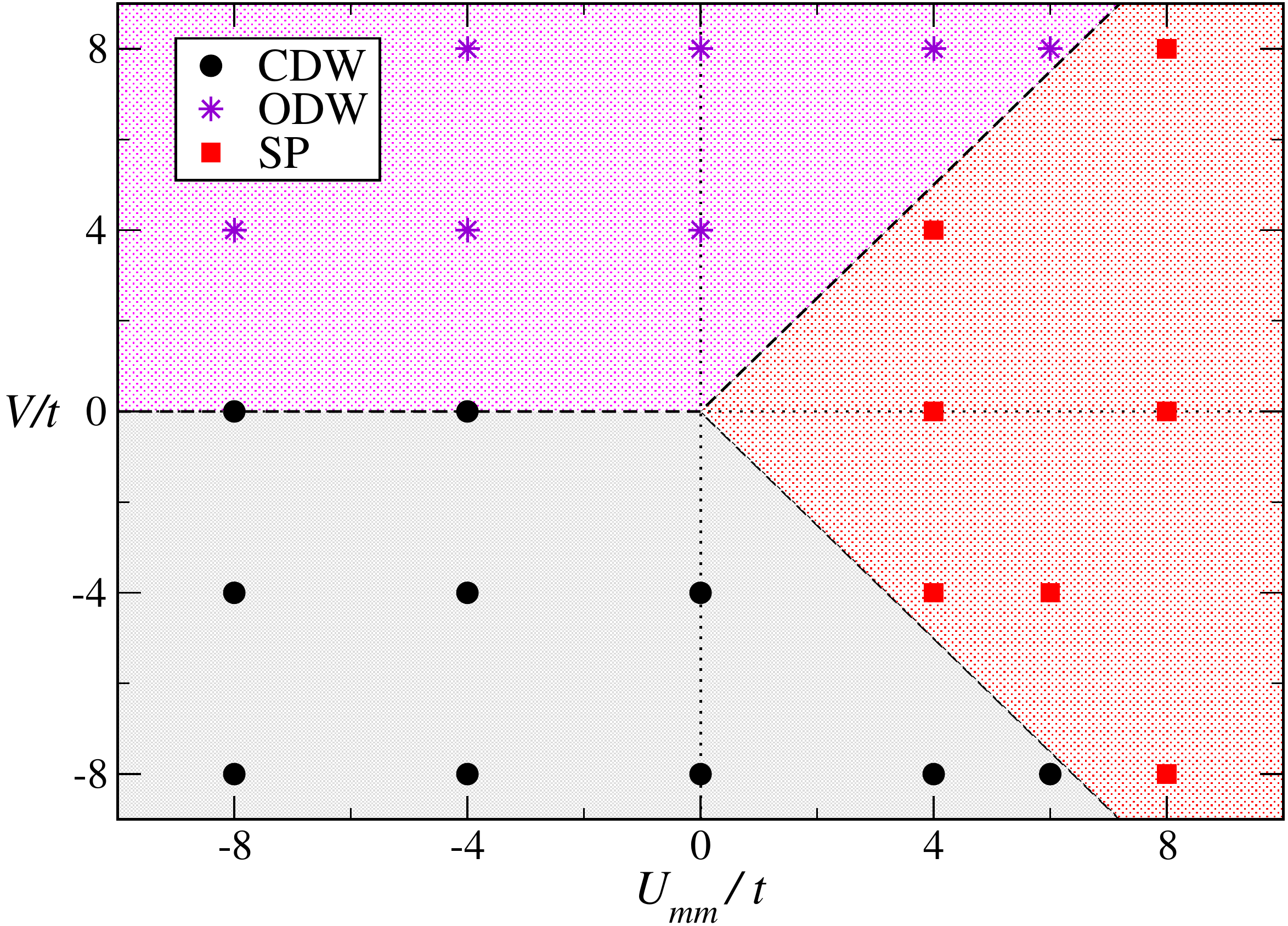}
}
\vspace{5mm}
\includegraphics[width=0.5\textwidth]{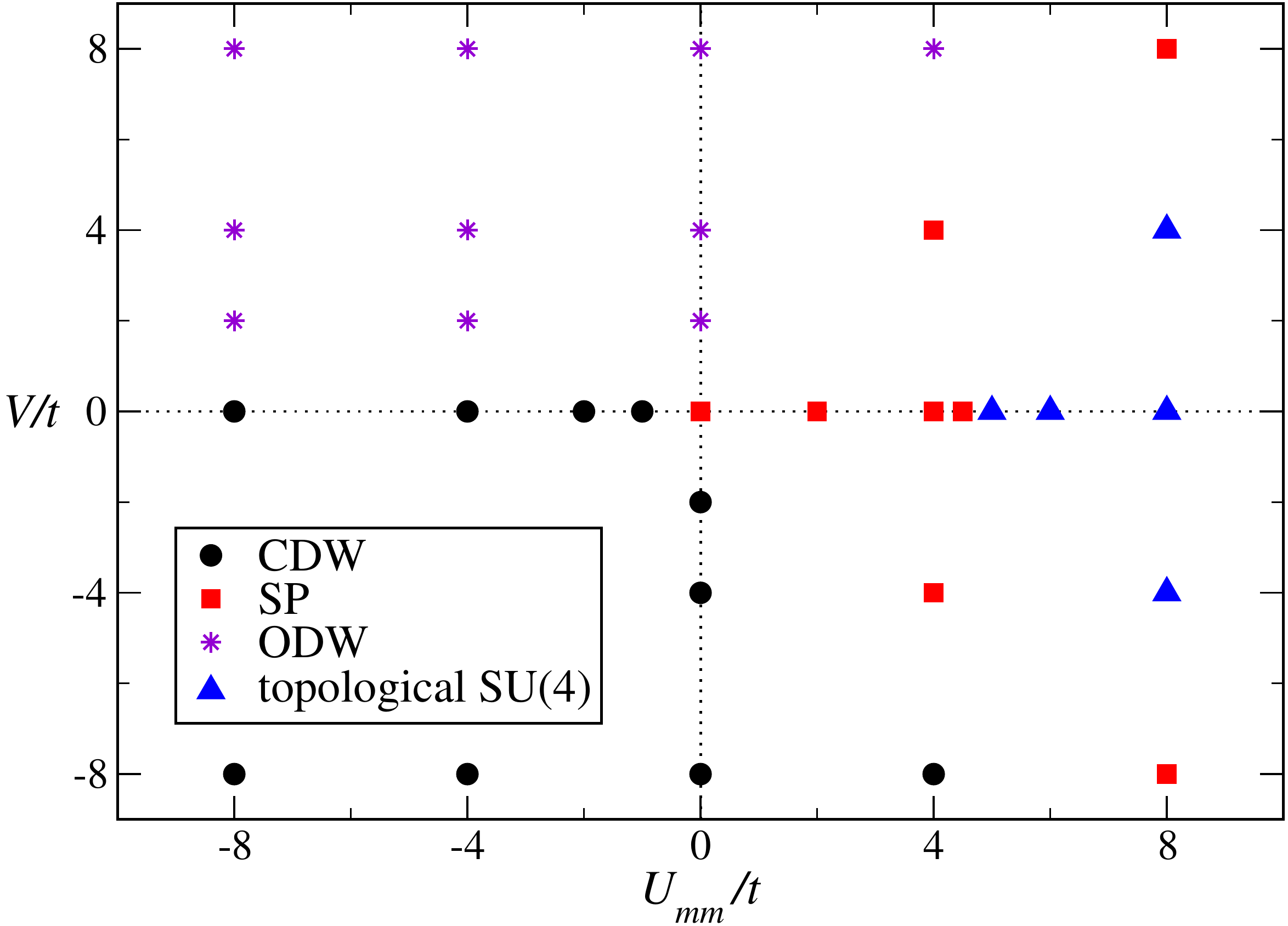}
\caption{(Color online) 
Phase diagram of the $g$-$e$ model with $N=4$ at half-filling. From top-left to bottom,
we have fixed $V_{\text{ex}}^{g\text{-}e}/t=-1$, $0$ and $1$, which allows to realize conventional phases (SP, CDW and ODW) as well as the topological SU(4) one.  The symmetry with respect to $V=0$ in the second plot 
is a direct consequence of the orbital-charge interchange that maps CDW into ODW.   
See Table~\ref{tab:abbreviation} for the meaning of the abbreviations `ODW', `RS', etc. 
\label{fig:egN4}}
\end{center}
\end{figure}

We now turn to the $N=4$ $g$-$e$ model (\ref{eqn:Gorshkov-Ham}) and present its phase diagrams 
for several fixed values of $V_{\text{ex}}^{g\text{-}e}/t$ in Fig.~\ref{fig:egN4}.  
Due to the existence of the symmetry $V \to -V$ 
that swaps charge and orbital when $V_{\text{ex}}^{g\text{-}e}=0$ \cite{Bois-C-L-M-T-15}, 
the phase diagram shown in the second panel is symmetric with respect to $V=0$.    

In the weak-coupling regime, we only find conventional degenerate phases with broken translation symmetry 
(CDW, ODW and SP; See Table~\ref{tab:abbreviation} for the meanings of the abbreviations), 
that are in good agreement with the low-energy predictions~\cite{Bois-C-L-M-T-15}.  
However, in the intermediate and strong-coupling regions, we find the SU(4) topological phase 
as has been predicted by the strong-coupling argument (see Secs.~\ref{sec:strong-coupling} 
and \ref{sec:solvable-Ham}).  
Its signatures are again given by the existence of non-trivial edge states 
[see Fig.~15(a) of Ref.~\cite{Bois-C-L-M-T-15}, which looks similar to Fig.~\ref{fig:pbandN4}(c)]  
or could also be probed by computing the degeneracy of the entanglement spectrum~\cite{Tanimoto-T-14} 
[see Fig.~\ref{fig:ES_a00} for the entanglement spectrum of the effective model 
\eqref{eqn:2nd-order-effective-Ham-Gorshkov} that describes the topological phase]. 
There is thus a quantum phase transition between (weak-coupling) SP and the SU(4) topological phase 
(class-2 SPT phase) in the strong-coupling region. The nature of the universality
class of the transition is, however, difficult to determine numerically.  
In Ref. \cite{Nonne-M-C-L-T-13}, it was conjectured 
that the quantum phase transition is governed by an SU(4)$_2$ CFT with the central charge $c=5$. The high value
of the central charge  calls for large-scale numerical simulations.
\subsubsection{$N=4$ $p$-band model}
\begin{figure}[!htb]
\begin{center}
\centerline{
\includegraphics[width=0.5\textwidth]{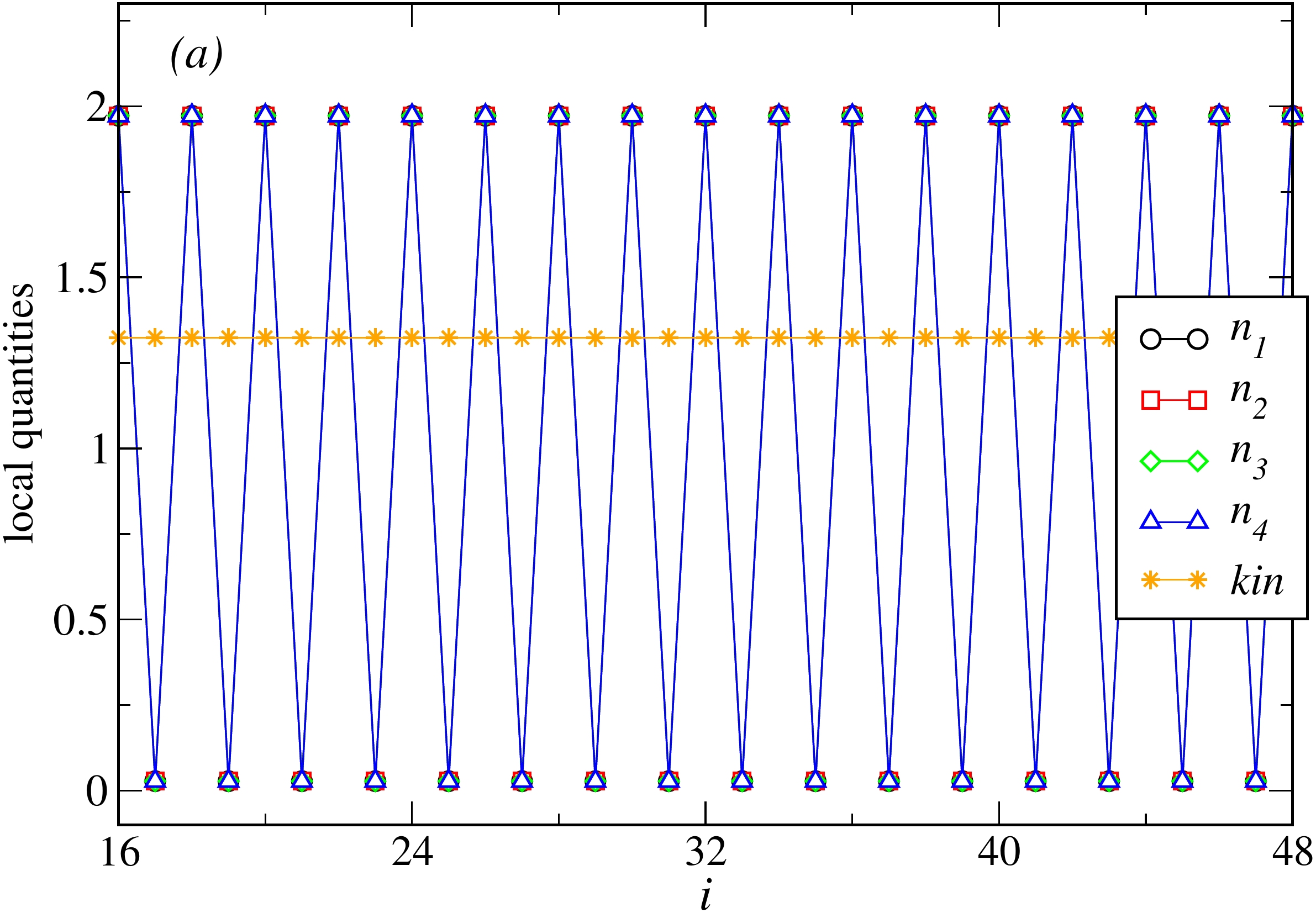}
\hspace{3mm}
\includegraphics[width=0.5\textwidth]{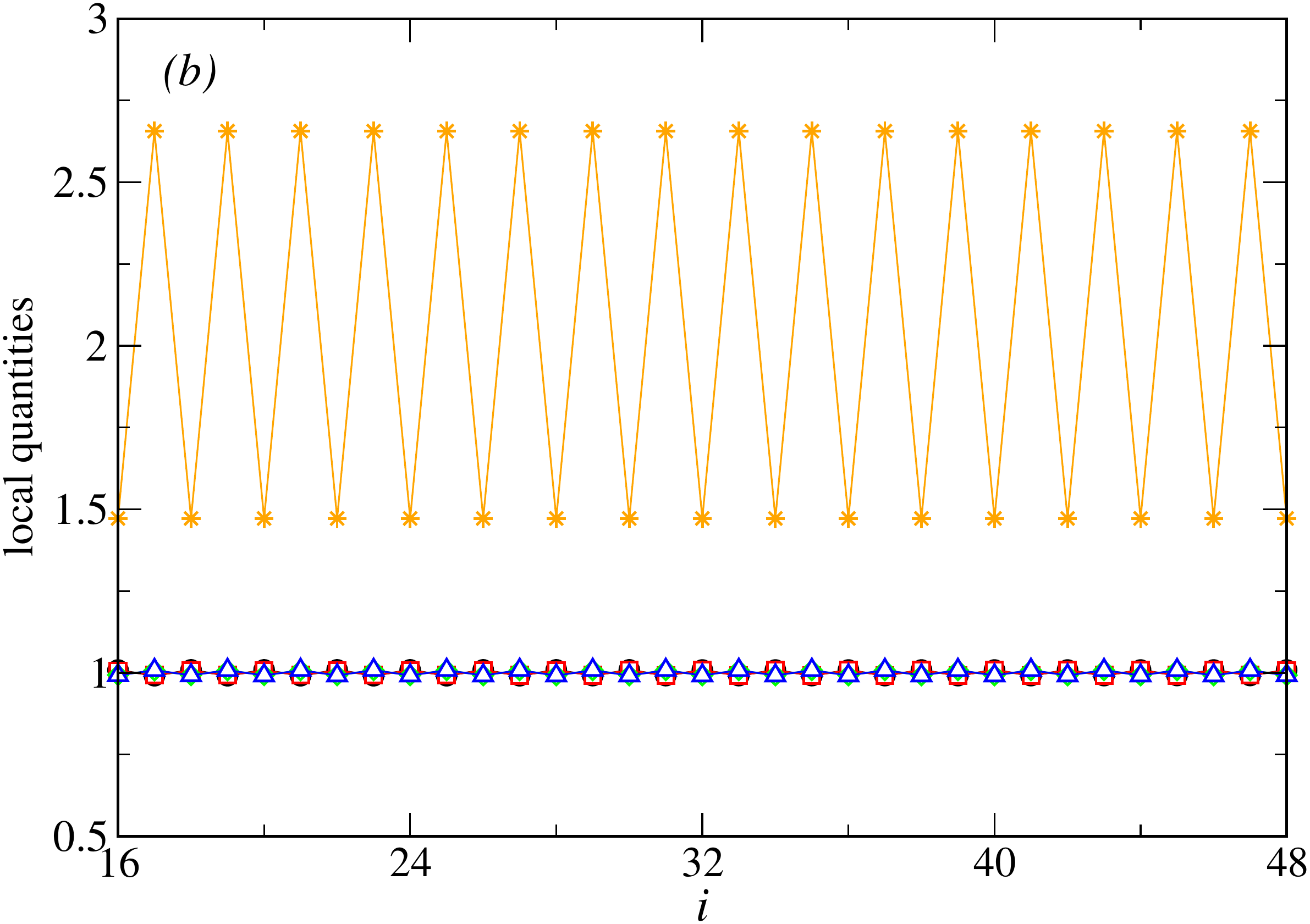}
}
\vspace{5mm}
\includegraphics[width=0.5\textwidth]{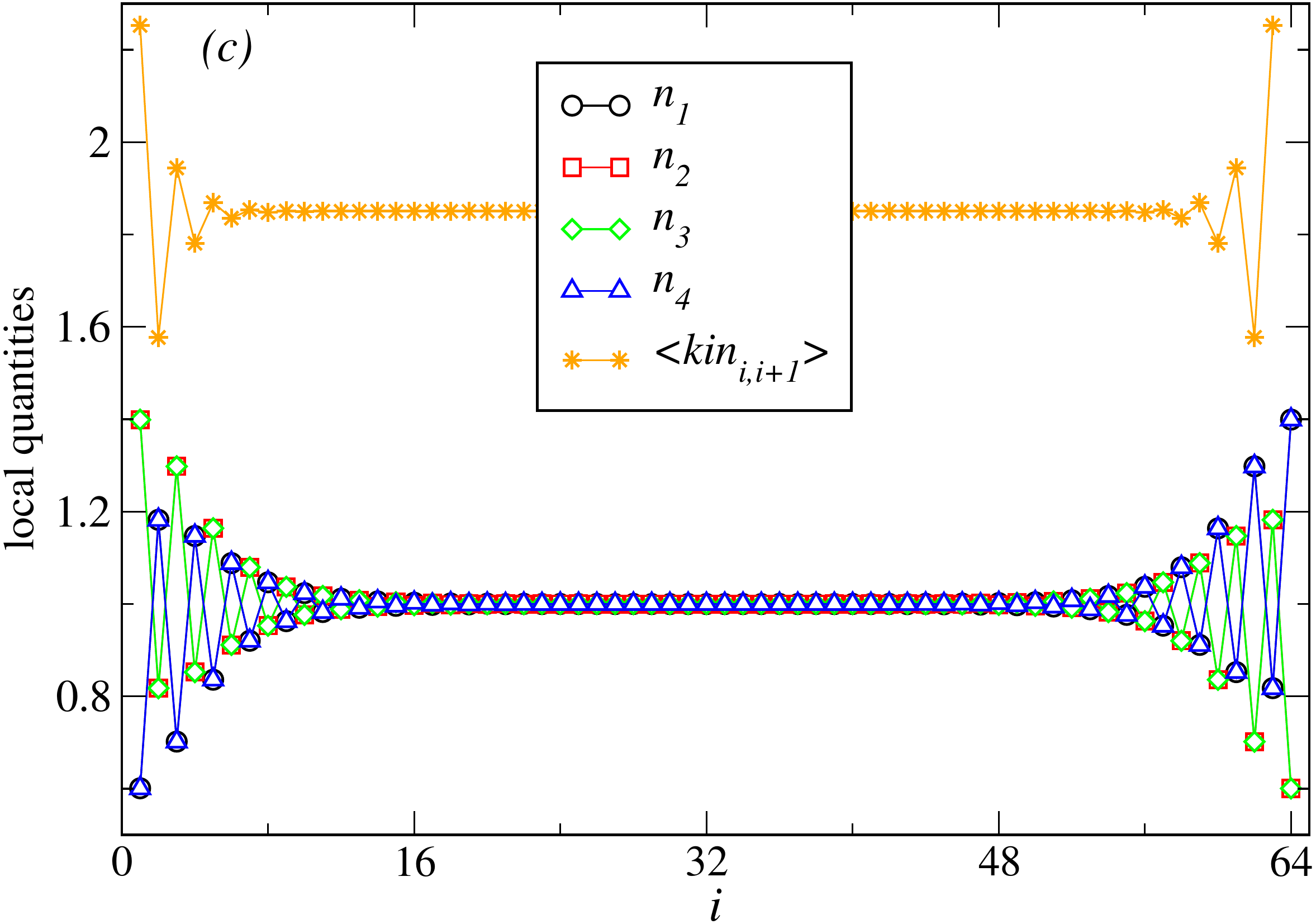}
\caption{(Color online) 
Local quantities for the $p$-band model at half-filling with $N=4$ obtained on $L=64$ chain using DMRG. 
The three panels correspond to several points relevant for harmonic trapping: 
(a) $(U_1/t,U_2/t)=(-3,-1)$ corresponding to a CDW phase; (b) $(U_1/t,U_2/t)=(3,1)$ exhibits strong dimerization, as expected for a SP phase; (c) $(U_1/t,U_2/t)=(9,3)$ shows non-trivial edge states characteristics of the topological SU(4) phase. Note that in the first two panels, we only plot data in the bulk of the chain for readability.  
\label{fig:pbandN4}}
\end{center}
\end{figure}

\begin{figure}[!htb]
\centering
\includegraphics[width=0.7\textwidth]{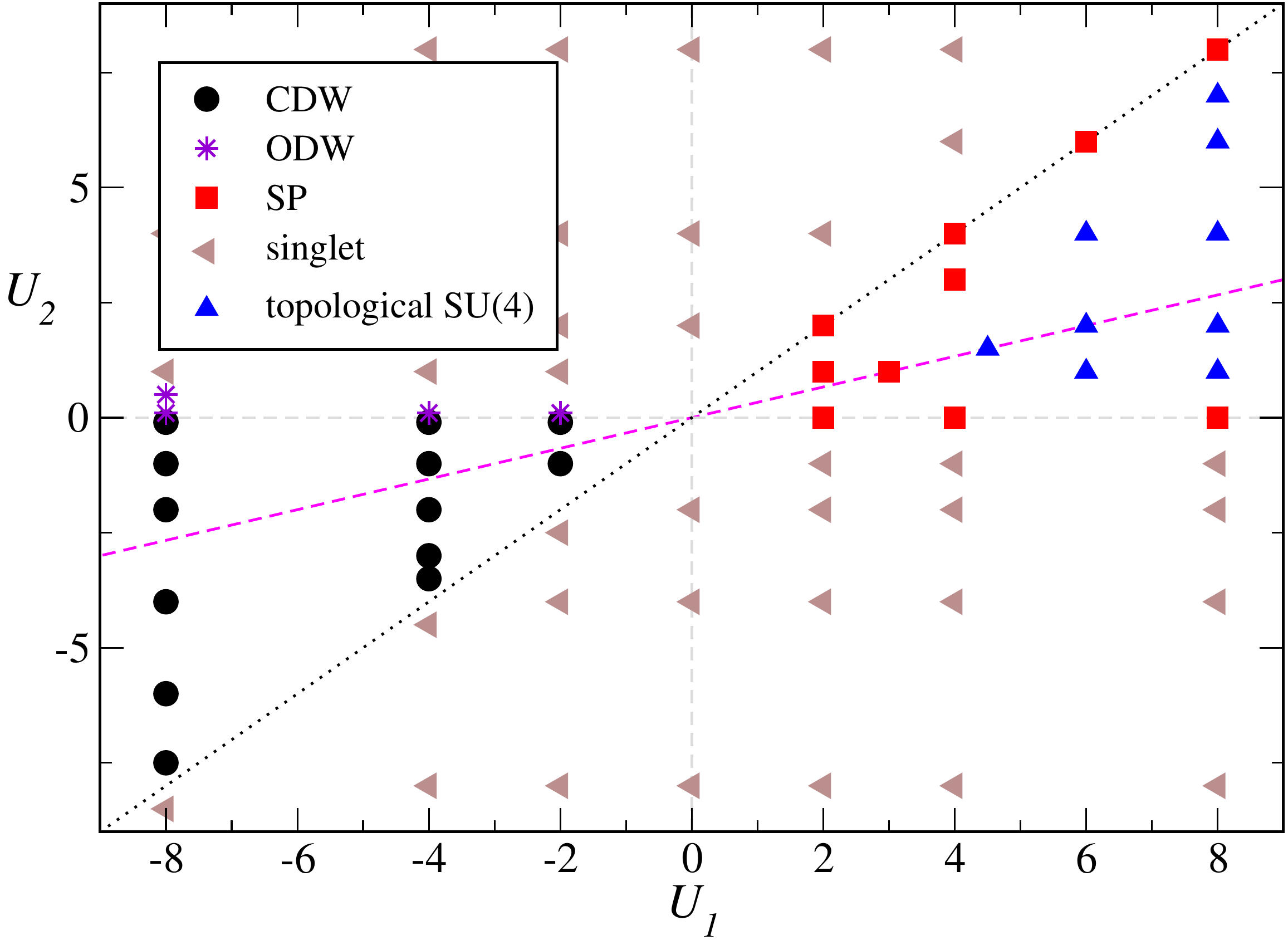}
\caption{(Color online) 
Phase diagram for half-filled $N=4$ $p$-band model (\ref{eqn:p-band}) obtained by DMRG on $L=32$. 
Dashed line corresponds to the condition $U_1=3U_2$ satisfied for an axially symmetric trap. 
$U_2=0$ correspond to two decoupled SU(4) Hubbard.  Adapted from Ref.~\cite{Bois-C-L-M-T-15}.  
\label{fig:phasediag_pband_N4}}
\end{figure}

When considering the $N=4$ $p$-band model, we restrict ourselves to the case with half-filling.   
Here we only plot relevant local quantities (local fermion densities and kinetic-energy density) 
in Fig.~\ref{fig:pbandN4} for several couplings along the line $U_1=3U_2$ (harmonic potential). 
The full $U_1$-$U_2$ phase diagram obtained 
in Ref.~\cite{Bois-C-L-M-T-15} is also shown in Fig.~\ref{fig:phasediag_pband_N4}. 
The definitions of the local quantities are similar to Eq.~(\ref{eq:def_local}) 
with the only difference that $\alpha=1,\ldots,4$ for $N=4$.  At weak-coupling, Figs.~\ref{fig:pbandN4}(a,b) indicate the existence of CDW phase 
for attractive interactions ($U_1<0$) and SP in the repulsive ($U_1>0$) case, as predicted using the low-energy 
field theories~\cite{Bois-C-L-M-T-15}.  
For strong repulsive interactions, Fig.~\ref{fig:pbandN4}(c) shows a very different behavior 
since non-trivial edge states appear while the bulk becomes featureless, 
and this is in agreement with the property of the proposed 
topological SU(4) phase (see Fig.~\ref{fig:local-density-SU4-VBS}).  
Indeed, each edge state is 6-fold degenerate as it results from choosing 2 colors among 4, 
and this degeneracy can be directly seen in the entanglement spectrum (see the lowest level 
in Fig.~\ref{fig:ES_a00}). 
As in the $N=4$  $g$-$e$ model, we expect a quantum phase transition in the SU(4)$_2$  universality class 
between SP and SU(4) topological phases.
\begin{table}[!htb]
\begin{center}
\begin{tabular}{lccc}
\hline\hline
phases &   abbreviation & SU($N$) & orbital ($T$) \\
\hline
spin-Haldane & SH & $S=1$ & local singlet \\
orbital-Haldane & 
OH & local singlet & $N/2$ \\
charge-Haldane & 
CH & local singlet & $-$
\\
orbital large-$D_{x,y}$ & 
OLD$_{x,y}$ & local singlet & $N/2$\\
rung-singlet (OLD$_{z}$) & 
RS & local singlet &  $N/2$
\\
spin-Peierls  & SP & $-$ & $N/2$ 
\\
charge-density wave & CDW & local singlet & local singlet
\\
orbital-density wave & ODW & local singlet & $N/2$ \\
\hline\hline
\end{tabular}
\caption{List of dominant phases and their abbreviations. 
Local SU($N$)/orbital degrees of freedom are shown, too. 
(see also Fig.~\ref{fig:4-MottPhases-SU2})  
`SH' and `OH' appear only in the $N=2$ case.  
`RS' and `ODW' respectively are a `large-$D$' state $\otimes | T^{z}=0 \rangle$ and 
the `N\'{e}el-ordered' state of $T=N/2$ chains.    
\label{tab:abbreviation}}
\end{center}
\end{table}
\clearpage
\section{Concluding remarks}
\label{sec:conclusion}
Alkaline-earth and ytterbium cold atomic gases allow the realization of fermionic cold gases 
with SU($N$) symmetric interactions in a very controlled way.  
On top of the almost perfect SU($N$) symmetry, one has the advantage of being able to use 
two orbital levels so that we can play with spin-orbital coherent exchange interactions, 
as demonstrated experimentally~\cite{Zhang-et-al-14,Scazza-et-al-14,Cappellini-et-al-14}.  
Both ingredients are expected to lead to a variety of states of matter and phenomena depending 
on the microscopic parameters, the dimensionality and so on~\cite{Gorshkov-et-al-10}.  
In such a context, our review has focused on the one-dimensional case, where powerful analytical and numerical 
techniques are available.  
 
Our review has covered, in depth, the case of single-band SU($N$) Fermi-Hubbard model 
which ultracold alkaline-earth fermions loaded into a one-dimensional optical lattice can simulate 
up to $N$ as large as 10.   The physics of the single-band Fermi-Hubbard model with a single contact interaction $U$
is very rich for general fillings and turns out to be very different between $N=2$ and $N\geq 3$. 
One fascinating aspect of the model 
is the possible occurrence of the (finite-$U$) Mott transition for commensurate fillings when $N>2$ in
sharp contrast to the $N=2$ case where the critical value is $U_{\text{c}}=0$ and additional nearest-neighbor interactions 
are necessary to have a finite $U_{\text{c}}$.  
For one atom per site, a filling at which we have least three-body losses, the low-energy physics of 
the Mott-insulating phase is described by the integrable Sutherland model with $N-1$ gapless relativistic modes. 
For incommensurate fillings and in the low-density regime, a new 
Luther-Emery liquid with a gap in the SU($N$) sector 
and one gapless bosonic (charge) mode emerges in the attractive $U<0$ case. 
The instability toward a paired fermionic superfluid (an analogue of the BCS superconductivity) 
is completely suppressed in this phase and the dominant superfluid instability occurs in the $N$-particle channel, 
e.g., the trionic and quartetting instabilities when $N=3$ and $N=4$, respectively. 
Finally, $N>2$ systems support fully gapped Mott-insulating phases for special commensurate fillings;  
these phases display a bond-ordering with ground-state degeneracies as in a SP phase at half filling.

All these results obtained for the single-band case clearly imply a rich variety of physics when $N>2$ 
which cannot be realized in the SU(2) two-component Fermi gas with a contact interaction.  
However, as already seen, all the fully gapped SU($N$) Mott-insulating phases occurring in the single-band 
model spontaneously break the translation symmetry.  
In this respect, many interesting fermionic or bosonic SPT phases are beyond the scope 
of the single-band SU($N$) Fermi-Hubbard model.  
The realization of these important phases thus necessitates the generalization 
of the model by introducing, for instance, additional degrees of freedom.  
To this end, we have presented two microscopic models which describe the low-energy physics of 
the short-range interacting two-orbital SU($N$) cold fermions on a lattice, 
namely the $g$-$e$ model and the $p$-band one.   Both models can be realized
with ultracold alkaline-earth or ytterbium atoms.  The zero-temperature phase diagrams of these models are very rich and 
harbor very different Mott and collective insulating phases. 
On the basis of low-energy field-theory approaches as well as rigorous constructions of the ground states in some strong-coupling regime,  a variety of phases have been found that are either degenerate (ODW or CDW, SP etc.) 
or non-degenerate (trivial singlet, various Haldane-like phases, as well as other SPT phases).  
All these results can be confirmed numerically by mapping out various phase diagrams, 
as we have presented for even $N=2$ and $4$ at half-filling for both models. 

Let us emphasize that in the SPT phases, the edge states are protected 
(and thus cannot be removed without closing a gap) as long as a given symmetry is present.  
This is precisely the case, for instance, for the SU($N$)-protected topological phase 
that we have discussed here and our observation provides 
one microscopic realization of one of the $N-1$ possible SPT phases predicted in Ref.~\cite{Duivenvoorden-Q-13}.  
Therefore, its experimental observation using, for instance, 
polarization measurements~\cite{Partridge-L-K-L-H-06,Liao-et-al-10}, site-resolved imaging technologies 
\cite{Endres-etal-stringOP-11},  
and spin-selective detection~\cite{Weitenberg2011} 
looks like an exciting possibility, although the edge states may be suppressed or even absent 
if one takes into account a harmonic trap~\cite{Kobayashi-O-O-Y-M-12, Kobayashi-O-O-Y-M-14}.  
In order to circumvent this difficulty, an interesting framework would be to use a box-shaped trapping 
potential~\cite{Gaunt2013} where presumably edge states should be more visible.  

Of course, there are many important topics which have not been covered in this review. 
For instance, when discussing the two-orbital models, 
we have assumed that both orbitals are not very different from each other, i.e., atoms have similar hoppings and chemical 
potentials for both $g$ and $e$ orbitals, 
as can be achieved experimentally using, for instance, the $p$-band levels discussed 
in Sec.~\ref{sec:p-band-hamiltonian}.  Therefore, we have not touched upon, e.g.,  
the fascinating Kondo physics in the context of ultracold alkaline-earth atoms~\cite{FossFeig2010,FossFeig2010a,Isaev2015,Nakagawa-K-15}.  
Also, we have not discussed interesting phases that would appear away from half-filling.  
For instance, as in the single-band case, we may expect various superfluid instabilities to occur 
for incommensurate fillings.  This problem is yet to be fully understood except for 
the $N=2$ case where we found several competing pairing phases as well as Mott ones \cite{Bois-C-L-M-15}.

Recently, it has also been proposed that using synthetic gauge field on one-dimensional SU($N$) cold atoms with contact interactions may be related to 2D Chern insulators~\cite{Goldman2013} or quantum Hall phases~\cite{Barbarino2015}, which also paves the way to realize exotic phases of matter with these systems. 
In the light of the recent experimental achievements with  alkaline-earth cold fermionic gases,
we hope that it will be possible in the future to unveil part of the richness that we highlighted in this review.

\section*{Acknowledgements}
The authors are very grateful to P. Azaria, V. Bois, A. Bolens, E. Boulat, P. Fromholz, T. Koffel, 
M. Moliner, H. Nonne, G. Roux, K. Tanimoto, A. M. Tsvelik, and S. R. White for collaborations on this topic over the years. 
Numerical simulations have been performed using HPC resources from GENCI--TGCC, GENCI--IDRIS (Grant 2015050225) and CALMIP. The authors would like to thank CNRS for financial support (PICS grant).  
One of the authors (K.T.) was supported in part by JSPS KAKENHI Grant No.~24540402 and No.~15K05211.  

\bibliographystyle{elsarticle-num}
\bibliography{alkaline-earth}
\end{document}